\documentclass[11pt]{article}
\pdfoutput=1 

\usepackage[letterpaper,margin=1in]{geometry}
\usepackage{titling}
\usepackage{amsmath}
\usepackage{amsfonts}
\usepackage{amssymb}

\usepackage[authoryear,sort&compress,round]{natbib}
\usepackage{multibib}
\newcites{App}{Online appendix references}
\bibliographystyleApp{aea}
\usepackage{url}

\usepackage{enumerate}

\usepackage[capposition=top,footfont=small,subfloatrowsep=none]{floatrow}

\usepackage{booktabs}
\usepackage{multirow}
\usepackage{rotating}
\usepackage{afterpage} 
\usepackage{makecell}
\usepackage{pdflscape}
\usepackage{longtable}
\usepackage{color,colortbl}
\usepackage{dcolumn}
\newcolumntype{d}[1]{D{.}{.}{#1}}
\usepackage{changepage}   % for the adjustwidth environment
\usepackage{siunitx}
\usepackage{array}
\newcolumntype{$}{>{\global\let\currentrowstyle\relax}}
\newcolumntype{^}{>{\currentrowstyle}}

\newcolumntype{C}[1]{>{\centering\arraybackslash}b{#1}}

\usepackage{graphicx}
\usepackage[position=bottom]{subfig}
\usepackage{tikz}
\usetikzlibrary{backgrounds,fit,positioning}
\usepackage[edges]{forest}

\providecommand{\Pr}{\mathrm{Pr}}

\providecommand{\e}{\varepsilon}

\providecommand{\Var}{\mathrm{Var}}
\providecommand{\Cov}{\mathrm{Cov}}
\usepackage{appendix}

\usepackage{chngcntr} % for numbering Appendix figures and tables within sections
\usepackage{setspace}

\title{Recruitment, effort, and retention effects of performance contracts for civil servants: Experimental evidence from Rwandan primary schools}
\author{Clare Leaver, Owen Ozier, Pieter Serneels, and Andrew Zeitlin\thanks{Leaver: Blavatnik School of Government, University of Oxford and CEPR (email: clare.leaver@bsg.ox.ac.uk). Ozier: Department of Economics, Williams College, World Bank Development Research Group, BREAD, and IZA (email: owen.ozier@williams.edu). Serneels: School of International Development, University of East Anglia, EGAP, and IZA (email: p.serneels@uea.ac.uk). Zeitlin: McCourt School of Public Policy, Georgetown University, and CGD (email: andrew.zeitlin@georgetown.edu). We thank counterparts at REB and MINEDUC for advice and collaboration, and David Johnson for help with the design of student and teacher assessments. We are grateful to the three anonymous referees, Katherine Casey, Jasper Cooper, Ernesto Dal B\'o, Erika Deserranno, David Evans, Dean Eckles, Frederico Finan, James Habyarimana, Caroline Hoxby, Macartan Humphreys, Pamela Jakiela, Julien Labonne, David McKenzie, Ben Olken, Berk \"{O}zler, Cyrus Samii, Kunal Sen, Martin Williams, and audiences at BREAD, DfID, EDI, NBER,  SIOE, and SREE for helpful comments. IPA staff members Kris Cox, Stephanie De Mel, Olive Karekezi Kemirembe, Doug Kirke-Smith, Emmanuel Musafiri, and Phillip Okull, and research assistants Claire Cullen, Robbie Dean, Ali Hamza, Gerald Ipapa, and Saahil Karpe all provided excellent support.  Financial support was provided by the U.K. Department for International Development (DfID) via the International Growth Centre and the Economic Development and Institutions Programme, by Oxford University's John Fell Fund, and by the World Bank's SIEF and REACH trust funds. Leaver is grateful for the hospitality of the Toulouse School of Economics, 2018--2019. Research was conducted under Rwanda Ministry of Education permit number MINEDUC/S\&T/308/2015 and received IRB approval from the Rwanda National Ethics Committee (protocol 00001497) and from Innovations for Poverty Action (protocol 1502). This study is registered as AEA RCT Registry ID AEARCTR-0002565 \citep{LeaOziSerZei18rctregistry}. The findings in this paper are the opinions of the authors, and do not represent the opinions of the World Bank, its Executive Directors, or the governments they represent. All errors and omissions are our own.}}
\date{January 31, 2021}

\begin{document}

\maketitle

\begin{abstract}
This paper reports on a two-tiered experiment designed to separately identify the selection and effort margins of pay-for-performance (P4P). At the recruitment stage, teacher labor markets were randomly assigned to a ‘pay-for-percentile’ or fixed-wage contract. Once recruits were placed, an unexpected, incentive-compatible, school-level re-randomization was performed, so that some teachers who applied for a fixed-wage contract ended up being paid by P4P, and vice versa. By the second year of the study, the within-year effort effect of P4P was 0.16 standard deviations of pupil learning, with the total effect rising to 0.20 standard deviations after allowing for selection.
\end{abstract}

\thispagestyle{empty}
\setcounter{page}{0}
\clearpage 
\onehalfspacing

The ability to recruit, elicit effort from, and retain civil servants is a central issue for any government. This is particularly true in a sector such as education where people---that is, human rather than physical resources---play a key role. Effective teachers generate private returns for students through learning gains, educational attainment, and higher earnings  \citep{CheFriRoc14aer1,CheFriRoc15aer2}, as well as social returns through improved labor-market skills that drive economic growth \citep{HanWoe12jeg}. And yet in varying contexts around the world, governments struggle to maintain a skilled and motivated teacher workforce \citep{Bol17jep}.

One policy option in this context is \emph{pay-for-performance}. These compensation schemes typically reward  teacher inputs such as presence and conduct in the classroom, teacher value added based on student learning, or both \citep[see, e.g.,][]{MurSun11jpe}. In principle, they can address the difficulty of screening for teacher quality ex ante \citep{StaRoc10jep}, as well as the limited oversight of teachers on the job \citep{ChaHamKreMurRog06jep}.

Yet pay-for-performance divides opinion. Critics, drawing upon public administration, social psychology, and behavioral economics, argue that pay-for-performance could dampen the effort of workers \citep{DeciRyan85,Krepps97,BenTir03res}. Concerns are that such schemes may: recruit the wrong types, individuals who are ``in it for the money''; lower effort by eroding intrinsic motivation; and fail to retain the right types because good teachers become de-motivated and quit. By contrast, proponents point to classic contract theory \citep{Laz03sepr,Rot15aer} and evidence from private-sector jobs with readily measurable output \citep{Laz00aer} to argue that pay-for-performance will have positive effects on both compositional and effort margins. Under this view, such schemes: recruit the right types, individuals who anticipate performing well in the classroom; raise effort by strengthening extrinsic motivation; and retain the right types because good teachers feel rewarded and stay put.

This paper conducts the first prospective, randomized controlled trial designed to identify both the compositional and effort margins of pay-for-performance. A novel, two-tiered experiment separately identifies these effects. This is combined with detailed data on applicants to jobs, the skills and motivations of actual hires, and their performance over two years on the job, to evaluate the effects of pay-for-performance on the recruitment, effort, and retention of civil servant teachers.

At the center of this study is a pay-for-performance (hereafter P4P) contract, designed jointly with the Rwanda Education Board and Ministry of Education. Building on extensive consultations and a pilot year, this P4P contract rewards the top 20 percent of teachers with extra pay using a metric that equally weights learning outcomes in teachers' classrooms alongside three measures of teachers' inputs into the classroom (presence, lesson planning, and observed pedagogy). The measure of learning used was based on a \emph{pay-for-percentile} scheme that makes student performance at all levels relevant to teacher rewards \citep{BarNea12aer}.  The tournament nature of this contract allows us to compare it to a fixed-wage (hereafter FW) contract that is equal in expected payout.

Our two-tiered experiment first randomly assigns labor markets to either P4P or FW \emph{advertisements}, and then uses a surprise re-randomization of \emph{experienced} contracts at the school level to enable estimation of pure compositional effects within each realized contract type. The first stage was undertaken during recruitment for teacher placements for the 2016 school year. Teacher labor markets are defined at the district by subject-family level. We conducted the experiment in six districts (18 labor markets) which, together, cover more than half the upper-primary teacher hiring lines for the 2016 school year. We recruited into the study all primary schools that received such a teacher to fill an upper-primary teaching post (a total of 164 schools). The second stage was undertaken once 2016 teacher placements had been finalized. Here, we randomly re-assigned each of these 164 study schools in their entirety to either P4P or FW contracts; all teachers who taught core-curricular classes to upper-primary students, including both newly placed recruits and incumbents, were eligible for the relevant contracts. We offered a signing bonus to ensure that no recruit, regardless of her belief about the probability of winning, could be made worse off by the re-randomization and, consistent with this, no one turned down their (re-)randomized contract. As advertised at the time of recruitment, incentives were in place for two years, enabling us to study retention as well as to estimate higher-powered tests of effects using outcomes from both years.

Our three main findings are as follows. First, on recruitment, advertised P4P contracts did not change the distribution of measured teacher skill either among applicants in general or among new hires in particular. This is estimated sufficiently precisely to rule out even small negative effects of P4P on measured skills. Advertised P4P contracts did, however, select teachers who contributed less in a framed Dictator Game played at baseline to measure intrinsic motivation. In spite of this, teachers recruited under P4P were at least as effective in promoting learning as were those recruited under FW (holding experienced contracts constant). 

Second, in terms of incentivizing effort, placed teachers working under P4P contracts elicited better performance from their students than teachers working under FW contracts (holding advertised contracts constant). Averaging over the two years of the study, the within-year effort effect of P4P was 0.11 standard deviations of pupil learning and for the second year alone, the within-year effort effect of P4P was 0.16 standard deviations. There is no evidence of a differential impact of experienced contracts by type of advertisement. 

In addition to teacher characteristics and student outcomes, we observe a range of teacher behaviors. These behaviors corroborate our first finding: P4P recruits performed no worse than the FW recruits in terms of their presence, preparation, and observed pedagogy. They also indicate that the learning gains brought about by those experiencing P4P contracts may have been driven, at least in part, by improved teacher presence and pedagogy. Teacher presence was 8 percentage points higher among recruits who experienced the P4P contract compared to recruits who experienced the FW contract. This is a sizeable impact given that baseline teacher presence was close to 90 percent. And teachers who experienced P4P were more effective in their classroom practices than teachers who experienced FW by 0.10 points, as measured on a 4-point scale. 

Third, on retention, teachers working under P4P contracts were no more likely to quit during the two years of the study than teachers working under FW contracts. There was also no evidence of differential selection-out on baseline teacher characteristics by experienced contract, either in terms of skills or measured motivation. On the retention margin, we therefore find little evidence to support claims made by either proponents or opponents of pay-for-performance. 

To sum up, by the second year of the study, we estimate the within-year effort effect of P4P to be 0.16 standard deviations of pupil learning, with the total effect rising to 0.20 standard deviations after allowing for selection. Despite evidence of lower intrinsic motivation among those recruited under P4P, these teachers were at least as effective in promoting learning as were those recruited under FW. These results support the view that pay-for-performance can improve effort while also allaying fears of harmful effects on selection. Of course, we have studied a two-year intervention---impacts of a long-term policy might be different, particularly if P4P influences individuals' early-career decisions to train as a teacher.   

Our findings bring new experimental results on pay-for-performance to the literature on the recruitment of civil servants in low- and middle-income countries. Existing papers have examined the impact of advertising higher \emph{unconditional} salaries and career-track motivations, with mixed results.  In Mexico, \citet{DalFinRos13qje} find that higher base salaries attracted both skilled and motivated applicants for civil service jobs. In Uganda, \citet{Des19aejapp} finds that the expectation of higher earnings discouraged pro-social applicants for village promoter roles, resulting in lower effort and retention. And in Zambia, \citet{AshBanDavLee19wp} find that emphasis on career-track motivations for community health work, while attracting some applicants who were less pro-social, resulted in hires of equal pro-sociality and greater talent overall, leading to improvements in a range of health outcomes.
By studying pay-for-performance and by separately manipulating advertised and experienced contracts, we add evidence on the compositional and effort margins of a different, and widely debated, compensation policy for civil servants.

How the teaching workforce changes in response to pay-for-performance is of interest in high-income contexts as well.  In the United States, there is a large (but chiefly observational) literature on the impact of compensation on who enters and leaves the teaching workforce. Well-known studies have simulated the consequences of dismissal policies \citep{Nea11hbk,Rot15aer,CheFriRoc15aer2} or examined the role of teachers' outside options in labor supply \citep{ChiWes12efp}. Recent work has examined the District of Columbia's teacher evaluation system, where financial incentives are linked to measures of teacher performance (including student test scores): \citet{DeeWyc15jpam} use a regression discontinuity design to show that low-performing teachers were more likely to quit voluntarily, while \citet{AdnDeeKatWyc17eepa} confirm that these `quitters' were replaced by higher-performers. In Wisconsin, a reform permitted approximately half of
the state's school districts to introduce flexible salary schemes that allow pay to vary with performance.  In that setting, \citet{Bia19nber} finds that high-value-added teachers were more likely to move to districts with flexible pay, and were less likely to quit, than their low-value-added counterparts. Our prospective, experimental study of pay-for-performance  contributes to this literature methodologically but also substantively since the Rwandan labor market shares important features with high-income contexts.\footnote{Notably, there is no public sector pay premium in Rwanda, which is unusual for a low-income country and more typical of high-income countries \citep{FinOlkPan17hbk}. The 2017 Rwanda Labour Force Survey includes a small sample of recent Teacher Training College graduates (aged below age 30). Of these, 37 percent were in teaching jobs earning an average monthly salary of 43,431 RWF, while 15 percent were in non-teaching jobs earning a higher average monthly salary of 56,347 RWF---a \emph{private sector} premium of close to 30 percent \citep{nisr17laborsurvey}.}

While our paper is not the first on the broader topic of incentive-based contracts for teachers,\footnote{See, e.g., \citet{Imb15iza} and \citet{JacRocSta14annrev} who provide a review.} we go to some length to address two challenges thought to be important for policy implementation at scale. One is that the structure of the incentive should not unfairly disadvantage any particular group \citep{BarNea12aer}; the other is that the incentive should not be inappropriately narrow \citep{Stecheretal}. We address the first issue by using a measure of learning based on a pay-for-percentile scheme that makes student performance at all levels relevant to teacher rewards, and the second by combining this with measures of teachers' inputs into the classroom to create a broad, composite metric. There is a small but growing literature studying pay-for-percentile schemes in education: \cite{LoyalkaJoLE} in China, \citet{GilKarKasLucNeaXXjhr} in Uganda, and \citet{Mbitietalwp} in Tanzania. Our contribution is to compare the effectiveness of contracts, P4P versus FW, that are based on a composite metric and are budget neutral in salary.

A final, methodological contribution of the paper, in addition to the experimental design, is the way in which we develop a pre-analysis plan. In our registered plan \citep{LeaOziSerZei18rctregistry}, we pose three questions. What outcomes to study? What hypotheses to test for each outcome? And how to test each hypothesis? We answered the `what' questions on the basis of theory, policy relevance, and available data. With these questions settled, we then answered the `how' question using blinded data. Specifically, we used a blinded dataset that allowed us to learn about a subset of the statistical properties of our data without deriving hypotheses from realized treatment responses, as advocated by, e.g.,  \citet{Olk15jep}.\footnote{We have not found prior examples of such blinding in economics.  \citet{HumSanVan13politanalysis} argue for, and undertake, a related approach with partial endline data in a political science application.} This approach achieves power gains by choosing from among specifications and test statistics on the basis of simulated power, while protecting against the risk of false positives that could arise if specifications were chosen on the basis of their realized statistical significance. The spirit of this approach is similar to recent work by \citet{AndMag17nberwp} and \citet{FafLab17polanalysis}.\footnote{In contrast to those two papers, we forsake the opportunity to undertake exploratory analysis because our primary hypotheses were determined \emph{a priori} by theory and policy relevance. In return, we avoid having to discard part of our sample, with associated power loss.} For an experimental study in which one important dimension of variation occurs at the labor-market level, and so is potentially limited in power, the gains from these specification choices are particularly important. The results reported in our pre-analysis plan demonstrate that, with specifications appropriately chosen, the study design is well powered, such that even null effects would be of both policy and academic interest.

In the remainder of the paper, Sections \ref{s:design} and \ref{s:data} describe the study design and data, Sections \ref{s:results} and \ref{s:robustness} report and discuss the results, and Section \ref{s:conclusion} concludes.

\section{Study design}\label{s:design}

\subsection{Setting}

The first tier of the study took place during the actual recruitment for civil service teaching jobs in upper primary in six districts of Rwanda in 2016.\footnote{Upper primary refers to grades 4, 5, and 6; schools typically include grades 1 through 6.} To apply for a civil service teaching job, an individual needs to hold a Teacher Training College (TTC) degree. Eligibility is further defined by specialization. Districts solicit applications at the district-by-subject-family level, aggregating curricula subjects into three `families' that correspond to the degree types issued by TTCs: math and science (TMS); modern languages (TML); and social studies (TSS). Districts invite applications between November and December, for the academic year beginning in late January/early February. Individuals keen to teach in a particular district submit one application and are then considered for all eligible teaching posts in that district in that hiring round. 

Given this institutional setting, we can think of district-by-subject-family pairs as \emph{labor markets}. The subject-family boundaries of these labor markets are rigid; within each district, TTC degree holders are considered for jobs in pools alongside others with the same qualification. The district boundaries may be more porous, though three quarters of the new teaching jobs in our study were filled by recruits living within the district at the time of application. Since this is the majority of jobs, we proceed by treating these labor markets as distinct for our primary analysis and provide robustness checks for cross-district applications in Online Appendix C.\footnote{As we note in the appendix, cross-district applications would not lead us to find a selection effect where none existed but we might overstate the magnitude of any selection effect.}

There are 18 such labor markets in our study.\footnote{Inference based on asymptotics could easily be invalid with 18 randomizable markets.  We address this risk by committing to randomization inference for all aspects of statistical testing.}
This is a small number in terms of statistical power (as we address below) but not from a system-scale perspective. The study covers more than 600 hiring lines constituting over 60 percent of the country's planned recruitment in 2016. Importantly, it is not a foregone conclusion that TTC graduates will apply for these civil service teaching jobs. Data from the 2017 Rwanda Labour Force Survey indicate that only 37 percent of recent TTC graduates were in teaching jobs, with 15 percent in non-teaching, salaried employment \citep{nisr17laborsurvey}. This is not because the teacher labor market is tight; nationwide close to a quarter of vacancies created by a teacher leaving a school remain unfilled in the following school year \citep{Zei20turnover}. A more plausible explanation is that the recent graduates in the outside sector earned a premium of close to 30 percent, making occupational choice after TTC a meaningful decision.  

\subsection{Experiment}

\paragraph{Contract structure}
The experiment was built around the comparison of two contracts paying a bonus on top of teacher salaries in each of the 2016 and 2017 school years, and was managed by Innovations for Poverty Action (IPA) in coordination with REB. The first of these was a P4P contract, which paid RWF 100,000  (approximately 15 percent of annual salary) to the top 20 percent of upper-primary teachers within a district, as measured by a composite performance metric.\footnote{The exchange rate on January 1, 2016 was 734 RWF to 1 USD, so the RWF 100,000 bonus was worth roughly 136 USD.} This metric equally weighted student learning alongside three measures of teachers' inputs into the classroom (presence, lesson preparation, and observed pedagogy). The measure of learning was based on a pay-for-percentile scheme that makes student performance at all levels relevant to teacher rewards \citep{BarNea12aer}.\footnote{Student learning contributed to an individual teacher's score via percentiles within student-based brackets so that 
a teacher with a particular mix of low-performing and high-performing students was, in effect, competing with other teachers with similar mixes of students. The data used to construct this measure, and the measures of teachers' inputs, are described in Sections \ref{ss:data_assessments} and \ref{ss:data_inputs} respectively, and we explain the adaptation of the Barlevy-Neal measure of learning outcomes to a repeated cross-section of pupils in Online Appendix D.} The 2016 performance award was conditional on remaining in post during the entire 2016 school year, and was to be paid early in 2017. Likewise, the 2017 performance award was conditional on remaining in post during the entire 2017 school year, and was to be paid early in 2018. The second was a fixed-wage (FW) contract that paid RWF 20,000 to all upper-primary teachers. This bonus was paid at the same time as the performance award in the P4P contract. 

Although P4P contracts based on a composite metric of teacher inputs and student performance have been used in a number of policy settings in the US \citep{Stecheretal, Imb15iza}, such contracts have been relatively less studied in low- and middle-income countries. In their comprehensive review, \citet{GleMur16chapter} discuss several evaluations of teacher incentives based on student test scores or attendance checks, but none based on a combination of both. After extensive discussions with REB about what would be suitable in this policy setting, a decision was made to use the P4P contract described above, based on a composite metric.

\paragraph{Design overview}
The design, summarized visually in Figure A.1 in Online Appendix A,
draws on a two-tiered experiment, as used elsewhere (see \cite{KarZin09ecta}, \cite{AshBerSha10aer}, and \cite{CohDup10qje} in credit-market and public-health contexts). Both tiers employ the contract variation described above.

Potential applicants, not all of whom were observed, were assigned to either advertised FW or advertised P4P contracts, depending on the labor market in which they resided. Those who actually applied, and were placed into schools, fall into one of the four groups summarized in Figure \ref{f:twobytwo}. For example, group $a$ denotes teachers who applied to jobs advertised as FW, and who were placed in schools assigned to FW contracts, while group $c$ denotes teachers who applied to jobs advertised as FW and who were then placed in schools re-randomized to P4P contracts. Under this experimental design, comparisons between groups $a$ and $b$, and between groups $c$ and $d$, allow us to learn about a pure compositional effect of P4P contracts on teacher performance, whereas comparisons along the diagonal of $a$--$d$ are informative about the total effect of such contracts, along
both margins.

\begin{figure}[htbp]
    \begin{center}
      \begin{tabular}{l l c c}
        & & \multicolumn{2}{c}{Advertised} \\
            &  & FW & P4P \\
        \multirow{ 2}{*}{Experienced}& FW & $a$ & $b$ \\
        &  P4P & $c$ & $d$ \\
      \end{tabular}
    \end{center}
\caption{Treatment groups among recruits placed in study schools}\label{f:twobytwo}
\end{figure}

\paragraph{First tier randomization: Advertised contracts}\label{ss:2ndtx}

Our aim in the first tier was to randomize the 18 distinct labor markets to contracts, `treating' all potential applicants in a given market so that we could detect the supply-side response to a particular contract. The result of the randomized assignment is that 7 of these labor markets can be thought of as being in a `P4P only' advertised treatment, 7 in a `FW only' advertised treatment, and 4 in a `Mixed' advertised treatment.\footnote{This randomization was performed in MATLAB by the authors. The Mixed advertised treatment arose due to logistical challenges detailed in the pre-analysis plan: the first-tier randomization was carried out at the level of the subject rather than the subject-family. An example of a district-by-subject-family assigned to the Mixed treatment is Ngoma-TML. An individual living in Ngoma with a TML qualification could have applied for an advertised Ngoma post in English on a FW contract, or an advertised Ngoma post in Kinyarwanda on a P4P contract. In contrast, Kirehe-TML is in the P4P only treatment. So someone in Kirehe with a TML qualification could have applied for either an English or Kinyarwanda post, but both would have been on a P4P contract.}
Empirically, we consider the Mixed treatment as a separate arm; we estimate a corresponding advertisement effect only as an incidental parameter.

This first-tier randomization was accompanied by an advertising campaign to increase awareness of the new posts and their associated contracts.\footnote{Details of the promotional materials used in this campaign are provided in Online Appendix E.} In November 2015, as soon as districts revealed the positions to be filled, we announced the advertised contract assignment. In addition to radio, poster, and flyer advertisements, and the presence of a person to explain the advertised contracts at District Education Offices, we also held three job fairs at TTCs to promote the interventions. These job fairs were advertised through WhatsApp networks of TTC graduates. All advertisements emphasized that the contracts were available for recruits placed in the 2016 school year and that the payments would continue into the 2017 school year. Applications were then submitted in December 2015. In January 2016, all districts held screening examinations for potential candidates. Successful candidates were placed into schools by districts during February--March 2016, and were then assigned to particular grades, subjects, and streams by their head teachers. 

\paragraph{Second-tier randomization: Experienced contracts}

Our aim in the second tier was to randomize the schools to which REB had allocated the new posts to contracts. A school was included in the sample if it had at least one new post that was filled and assigned to an upper-primary grade.\footnote{Because schools could receive multiple recruits, for different teaching specializations, it was possible for enrolled schools to contain two recruits who had experienced distinct advertised treatments. Recruits hired under the mixed advertisement treatment, and the schools in which they were placed, also met our enrollment criteria.  These were similarly re-randomized to either experienced P4P or experienced FW in the second-tier randomization.} Following a full baseline survey, schools were randomly assigned to either P4P or FW. Of the 164 schools in the second tier of the experiment, 85 were assigned to P4P and 79 were assigned to FW contracts.

All upper-primary teachers---placed recruit or incumbent---within each school received the new contract. At individual applicant level, this amounted to re-randomization and hence a change to the initial assignment for some new recruits. A natural concern is that individuals who applied under one contract, but who were eventually offered another contract, might have experienced disappointment (or other negative feelings) which then had a causal impact on their behavior. To mitigate this concern,  all new recruits were offered an \emph{end-of-year retention bonus} of RWF 80,000 on top of their school-randomized P4P or FW contract. An individual who applied under advertised P4P in the hope of receiving RWF 100,000 from the scheme, but who was subsequently re-randomized to experienced FW, was therefore still eligible to receive RWF 100,000 (RWF 20,000 from the FW contract plus RWF 80,000 as a retention bonus). Conversely, an individual who applied under advertised FW safe in the knowledge of receiving RWF 20,000 from the scheme, but who was subsequently re-randomized to experienced P4P, was still eligible for at least RWF 80,000. None of the recruits objected to the (re)randomization or turned down their re-randomized contract. 

Of course, surprise effects, disappointment or otherwise, may still be present in on-the-job performance. When testing hypotheses relating to student learning, we include a secondary specification with an interaction term to allow the estimated impact of experienced P4P to differ by advertised treatment. We also explore whether surprise effects are evident in either retention or job satisfaction. We find no evidence for any surprise effect. 
To ensure that teachers in P4P schools understood the new contract, we held a compulsory half-day briefing session in every P4P school to explain the intervention. This session was conducted by a team of qualified enumerators and District Education Office staff, who themselves received three days of training from the Principal Investigators in cooperation with IPA. Online Appendix E reproduces an extract of the English version of the enumerator manual, which was piloted before use. The sessions provided ample space for discussion and made use of practical examples. Teachers' understanding was tested informally at the end of the session. We also held a comparable (but simpler) half-day briefing session in every FW school.  

\subsection{Hypotheses}\label{ss:hypotheses}

Pre-commitment to an analytical approach can forestall $p$-hacking, but requires clear specification of both what to test and how to test it; this presents an opportunity, as we now discuss.
A theoretical model, discussed briefly below, and included in our pre-analysis plan and Online Appendix B, guides our choice of \emph{what} hypotheses to test. However, exactly \emph{how} to test these hypotheses in a way that maximizes statistical power is not fully determined by theory, as statistical power may depend on features of the data that could not be known in advance: the distribution of outcomes, their relationships with possible baseline predictors, and so on. We used blinded data to help decide how to test the hypotheses. In what follows we first briefly describe the theoretical model, and then discuss our statistical approach. 

\paragraph{Theory}

The model considers a fresh graduate from teacher training who decides whether to apply for a state school teaching post, or a job in another sector (a composite `outside sector'). The risk neutral individual cares about compensation $w$ and effort $e$. Her payoff is sector specific: in teaching it is $w - (e^2 - \tau \, e)$, while in the outside sector it is $w - e^2$. The parameter $\tau \geq 0 $ captures the individual's \emph{intrinsic motivation} to teach, which is perfectly observed by the individual herself but not by the employer at the time of hiring.\footnote{See  \citet{DelDur07ej} for a related approach to modeling differential worker motivation across sectors.} Effort generates a performance metric $m = e\, \theta + \e$, where $\theta \geq 1$ represents her \emph{ability}, which is also private information at the time of hiring. Compensation corresponds to one of the four cells in Figure \ref{f:twobytwo}. The timing is as follows. Teacher vacancies are advertised as either P4P or FW. The individual, of type $(\tau,\theta)$, applies either to a teaching job or to an outside job. Employers hire, at random, from the set of $(\tau,\theta)$ types that apply. Thereafter, contracts are re-randomized. If the individual applies to, and is placed in a school, she learns about her experienced contract and chooses her effort level, which results in performance $m$ at the end of the year. Compensation is paid according to the experienced contract. 

This model leads to the following hypotheses, as set out in our pre-analysis plan:
 
\begin{enumerate}[I.]
    \item Advertised P4P induces differential \emph{application} qualities;
    \item Advertised P4P affects the observable skills of recruits \emph{placed} in schools;
    \item Advertised P4P induces differentially intrinsically motivated recruits to be  \emph{placed} in schools;
    \item Advertised P4P induces the supply-side selection-in of higher- (or lower-) performing teachers, as measured by the learning outcomes of their students; 
    \item Experienced P4P creates effort incentives which contribute to higher (or lower) teacher performance, as measured by the learning outcomes of their students;
    \item These selection and incentive effects are apparent in the composite performance metric.
\end{enumerate}

\noindent The model predicts that the set of $(\tau,\theta)$ types preferring a teaching job advertised under P4P to a job in the outside sector is different from the set of types preferring a teaching job advertised under FW to a job in the outside sector. This gives Hypothesis I. Since the model abstracts from labor demand effects (by assuming employers hire at random from the set of $(\tau,\theta)$ types that apply), this prediction simply maps through to placed recruits; i.e. to Hypothesis II via $\theta$, Hypothesis III via $\tau$, and Hypotheses IV to VI via the effect of $\theta$ and $\tau$ on performance.\footnote{When mapping the theory to our empirical context, we distinguish between these hypotheses for two reasons: we have better data for placed recruits because we were able to administer detailed survey instruments to this well-defined sub-sample; and for placed recruits we can identify the advertised treatment effect from student learning outcomes, avoiding the use of proxies for $(\tau,\theta)$. A further consideration is that the impact of advertised treatment might differ between placed recruits and applicants due to labor-demand effects. We discuss this important issue in Section \ref{s:robustness}.} The model also predicts that any given $(\tau,\theta)$ type who applies to, and is placed in, a teaching job will exert more effort under experienced P4P than experienced FW. This gives Hypotheses V and VI via the effect of $e$ on performance.

\paragraph{Analysis of blinded data}
Combining several previously-known insights, we used blinded data to maximize statistical power for our main hypothesis tests.

The first insights, pertaining to simulation, are due to \citet{HumSanVan13politanalysis} and \citet{Olk15jep}.  Researchers can use actual outcome data with the treatment variable scrambled or removed to estimate specifications in `mock' data.  This permits navigation of an otherwise intractable `analysis tree'. They can also improve statistical power by simulating treatment effects and choosing the specification that minimizes the standard error.
Without true treatment assignments, the influence of any decision over eventual treatment effect estimates is unknown; thus, these benefits are garnered without risk of $p$-hacking.\footnote{This would not be true if, for example, an outcome in question was known to have different support as a function of treatment, allowing the `blinded' researcher to infer treatment from the outcome variable.  For our blinded pre-analysis, we only consider outcomes (TTC score, and student test scores) that are nearly continuously distributed and which we believe are likely to have the same support in all study arms.  To make this analysis possible, we drew inspiration from \citet{FafLab17polanalysis}, who suggest dividing labor within a research team. In our case, IPA oversaw the data-blinding process.
Results of the blinded analysis (for which IPA certified that we used only blinded data) are in our pre-analysis plan. Our RCT registry entry \citep[][AEARCTR-0002565]{LeaOziSerZei18rctregistry} is accompanied by IPA's letter specifying the date after which treatment was unblinded.}

The second set of insights pertain to randomization inference. Since the market-level randomization in our study involves 18 randomizable units, asymptotic inference is unsuitable, so we use randomization inference. It is known that any scalar function of treatment and comparison groups is a statistic upon which a (correctly-sized) randomization-inference-based test of the sharp null hypothesis could be built, but also that such statistics may vary in their statistical power in the face of any particular alternative hypothesis \citep{ImbRub15book}. We anticipated that, even with correctly-sized tests, the market-level portion of our design may present relatively low statistical power. Consequently, we conducted blinded analysis to choose, on the basis of statistical power, among testing approaches for several hypotheses: Hypothesis I, and a common framework for Hypotheses IV and V.\footnote{Hypotheses II and III employ data that our team collected, so did not have power concerns associated with them; Hypothesis VI offered fewer degrees of freedom.}

Hypothesis I is the test of whether applicants to different contracts vary in their TTC scores.  Blinded analysis, in which we simulated additive treatment effects and calculated the statistical power under different approaches, suggested that ordinary least squares regression (OLS) would yield lower statistical power than would a Kolmogorov-Smirnov (KS) test of the equality of two distributions.
Over a range of simulations, the KS test had between one and four times the power of OLS. 
We therefore committed to KS (over OLS and two other alternatives) as our primary test of this hypothesis.\footnote{This refers to \citet{LeaOziSerZei18rctregistry}, Table C.1, comparing the first and third rows.}
This prediction is borne out in Table C.1 in Online Appendix C.\footnote{The confidence interval for the KS test is roughly half the width of  the corresponding OLS confidence interval: a gain in precision commensurate with more than tripling the sample size.}

Hypotheses IV and V relate to the effects of advertised and experienced contracts on student test scores. Here, with the re-randomization taking place at the school level, we had many possible specifications to choose from.  We examined 14 specifications (modeling random effects or fixed effects at different levels), and committed to one with the highest power. Simulations suggested that this could produce a 20 to 25 percent narrower confidence interval than in a simple benchmark specification.\footnote{This refers to \citet{LeaOziSerZei18rctregistry}, Table C.3, comparing row 12 to row 1.} Comparing Table \ref{t:learning_fx} to Table A.4 in Online Appendix A, this was substantively borne out.\footnote{For the pooled advertised treatment effect, the pre-committed random effects model yields a confidence interval that is 67 percent as wide as the interval from OLS: a gain in precision commensurate with increasing the sample size by 125 percent. The gain in precision for the pooled experienced treatment effect is smaller and commensurate with increasing sample size by 22 percent.} 

On the basis of this theory and analysis of blinded data, we settled on six primary tests: an outcome, a sample, a specification and associated test statistic, and an inference procedure for each of Hypotheses I-VI, as set out in Table A.1 in Online Appendix A. We also included a small number of secondary tests based on different outcomes, samples, and/or specifications. In Section \ref{s:results}, we report results for every primary test; secondary tests are in Section \ref{s:results} or in an appendix. To aid interpretation, we also include a small amount of supplementary analysis that was not discussed in the pre-analysis plan---e.g. impacts of advertised P4P on teacher attributes beyond observable skill and intrinsic motivation, and estimates from a teacher value added model---but are cautious and make clear when this is post-hoc.

\section{Data}\label{s:data}

The primary analyses make use of several distinct types of data. Conceptually, these trace out the causal chain from the advertisement intervention to a sequence of outcomes: that is, from the candidates' application decisions, to the set (and attributes) of candidates hired into schools, to the learning outcomes that they deliver, and, finally, to the teachers' decisions to remain in the schools.  In this section, we describe the administrative, survey, and assessment data available for each of these steps in the causal chain.\footnote{All data generated by the study and used in this paper are made available in the replication materials \citep{LeaOziSerZei20icpsr}.} Our understanding of these data informs our choices of specification for analysis, as discussed in detail in the pre-analysis plan.

\subsection{Applications}\label{ss:data_applications}

Table \ref{t:applications} summarizes the applications for the newly advertised jobs, submitted in January 2016, across the six districts.\footnote{These data were obtained from the six district offices and represent a census of applications for the new posts across these districts.}  Of the 2,184 applications, 1,962 come from candidates with a TTC degree---we term these \emph{qualified} since a TTC degree is required for the placements at stake. In the table, we present TTC scores, genders, and ages---the other observed CV characteristics---for all qualified applicants. Besides these two demographic variables, TTC scores are the only consistently measured characteristics of all applicants. 

\begin{table}[!htbp]
    \caption{Application characteristics, by district}
    \label{t:applications}
    \begin{footnotesize}
    \begin{tabular}{l c c c c c c c}
\toprule
 & Gatsibo  & Kayonza  & Kirehe  & Ngoma  & Nyagatare  & Rwamagana  & All  \\  
\midrule 
Applicants  & 390 & 310 & 462 & 380 & 327 & 315 & 2,184 \\  
Qualified  & 333 & 258 & 458 & 364 & 272 & 277 & 1,962 \\  
Has~TTC~score  & 317 & 233 & 405 & 337 & 260 & 163 & 1,715 \\  
Mean~TTC~score  & 0.53 & 0.54 & 0.50 & 0.53 & 0.54 & 0.55 & 0.53 \\  
SD~TTC~score  & 0.14 & 0.15 & 0.19 & 0.15 & 0.14 & 0.12 & 0.15 \\  
Qualified~female  & 0.53 & 0.47 & 0.45 & 0.50 & 0.44 & 0.45 & 0.48 \\  
Qualified~age  & 27.32 & 27.78 & 27.23 & 27.25 & 26.98 & 27.50 & 27.33 \\  
\bottomrule
\end{tabular}

    \end{footnotesize}
\end{table}

The 2,184 applications come from 1,424 unique individuals, of whom 1,246 have a TTC qualification. The majority (62 percent) of qualified applicants complete only one application, with 22 percent applying to two districts and 16 percent applying to three or more. Multiple applications are possible but not the norm, most likely because each district requires its own exam. Of those applying twice, 92 percent applied to adjacent pairs of districts. In Online Appendix C, we use this geographical feature of applications to test for cross-district labor-supply effects and fail to reject the null that these effects are zero. 

\subsection{Teacher attributes}\label{ss:data_teachers}

During February and March 2016, we visited schools soon after they were enrolled in the study to collect baseline data using surveys and `lab-in-the-field' instruments. School surveys were administered to head teachers or their deputies, and included a variety of data on management practices---not documented here---as well as administrative records of teacher attributes, including age, gender, and qualifications.  The data cover all teachers in the school, regardless of their eligibility for the intervention. Teacher surveys were administered to all teachers responsible for at least one upper-primary, core-curricular subject and included questions about demographics, training, qualifications and experience, earnings, and other characteristics. 

The `lab-in-the-field' instruments were administered to the same set of teachers, and were intended to measure the two characteristics introduced in the theory: intrinsic motivation and ability. In the model, more intrinsically motivated teachers derive a higher benefit (or lower cost) from their efforts to promote learning. To capture this idea of other-regarding preferences towards students, taking inspiration from the work of \citet{AshBanJack14jpube}, we used a framed version of the \emph{Dictator Game} \citep{EckGross96geb}.\footnote{Previous work shows the reliability of the DG as a measure of other-regarding preferences related to intrinsic motivation \citep{BrockLanLeon16jhr, BanKeef16eer, Des19aejapp}.} Teachers were  given 2,000 Rwandan francs (RWF) and asked how much of this money they wished to allocate to the provision of school supply packets for students in their schools, and how much they wished to keep for themselves. Each packet contained one notebook and pen and was worth 200 RWF. Teachers could decide to allocate any amount, from zero to all 2,000 RWF, which would supply ten randomly chosen students with a packet. 

We also asked teachers to undertake a \emph{Grading Task} which measured their mastery of the curriculum in the main subject that they teach.\footnote{See \cite{Bol17jep} who use a similar approach to assess teacher content knowledge.} Teachers were asked to grade a student examination script, and had 5 minutes to determine if a series of student answers were correct or incorrect. They received a fixed payment for participation. Performance on this task was used to estimate a measure of teacher skill based on a two-parameter item response theory (IRT) model.%

\subsection{Student learning}\label{ss:data_assessments}

Student learning was measured in three rounds of assessment: baseline, the end of the 2016 school year, and the end of the 2017 school year (indexed by $r=0,1,2$). These student assessments play a dual role in our study:  they provide the primary measure of learning for analysis of program impacts, and they were used in the experienced P4P arm for purposes of performance awards.

Working with the Ministry of Education, we developed comprehensive subject- and grade-specific, competency-based assessments for grades 4, 5 and 6. These assessments were based on the new Rwanda national curriculum and covered the five core subjects: Kinyarwanda, English, Mathematics, Sciences, and Social Studies. There was one assessment per grade-subject, with students at the beginning of the year being assessed on the prior year's material. Each test aimed to cover the entire curriculum for the corresponding subject and year, with questions becoming progressively more difficult as a student advanced in the test. The questions were a combination of multiple choice and fill-in diagrams. The tests were piloted extensively; they have no ceiling effects while floor effects are limited.\footnote{Test scores are approximately normally distributed with a mean of close to 50 percent of questions answered correctly. A validation exercise of the test at baseline found its scores to be predictive of the national Primary Leaving Exam scores (both measured in school averages).} In each round, we randomly sampled a subset of students from each grade to take the test. In Year 1, both baseline and endline student samples were drawn from the official school register of enrolled students compiled by the head teacher at the beginning of the year. This ensured that the sampling protocol did not create incentives for strategic exclusion of students. In Year 2, students were assessed at the end of the year only, and were sampled from a listing that we collected in the second trimester.

Student samples were stratified by teaching \emph{streams} (subgroups of students taught together for all subjects). In Round 0, we sampled a minimum of 5 pupils per stream, and oversampled streams taught in at least one subject by a new recruit to fill available spaces, up to a maximum of 20 pupils per stream and 40 per grade.  In rare cases of grades with more than 8 streams, we sampled 5 pupils from all streams.  In Round 1, we sampled 10 pupils from each stream: 5 pupils retained from the baseline (if the stream was sampled at baseline) and 5 randomly sampled new pupils. We included the new students to alleviate concerns that teachers in P4P schools might teach (only) to previously sampled students. In Round 2, we randomly sampled 10 pupils from each stream using the listing for that year.\footnote{Consequently, the number of pupils assessed in Year 2 who have also been assessed in Year 1 is limited.  Because streams are reshuffled across years and because we were not able to match Year 2 pupil registers to Year 1 registers in advance of the assessment, it was not possible to sample pupils to maintain a panel across years while continuing to stratify by stream.}  

The tests were orally administered by trained enumerators. Students listened to an enumerator as he/she read through the instructions and test questions, prompting students to answer. The exam was timed for 50 minutes, allowing for 10 minutes per section. Enumerators administered the exam using a timed proctoring video on electronic tablets, which further ensured consistency in test administration. Individual student test results were kept confidential from teachers, parents, head teachers, and Ministry of Education officials, and have only been used for performance award and evaluation purposes in this study.

Responses were used to estimate a measure of student learning (for a given student in a given round and given subject in a given grade) based on a two-parameter IRT model. We use empirical Bayes estimates of student ability from this model as our measure of a student's learning level in a particular grade.

\subsection{Teacher inputs}\label{ss:data_inputs}

We collected data on several dimensions of teachers' inputs into the classroom.  This was undertaken in P4P schools only during Year 1, and in both P4P and FW schools in Year 2. This composite metric is based on three teacher input measures (presence, lesson preparation, and observed pedagogy), and one output measure (pupil learning)---the `4Ps'.  Here we describe the input components measured. 

To assess the three inputs, P4P schools received three unannounced surprise visits: two spot checks during Summer 2016, and one spot check in Summer 2017. During these visits, Sector Education Officers (SEOs) from the District Education Offices (in Year 1) or IPA staff (for logistical reasons, in Year 2) observed teachers and monitored their presence, preparation and pedagogy with the aid of specially designed tools.\footnote{Training of SEOs took place over two days. Day 1 consisted of an overview of the study and its objectives and focused on how to explain the intervention (in particular the 4Ps) to teachers in P4P schools using the enumerator manual in Online Appendix E. During Day 2, SEOs learned how to use the teacher monitoring tools and how to conduct unannounced school visits. SEOs were shown videos recorded during pilot visits. SEOs were briefed on the importance of not informing teachers or head teachers ahead of the visits. Field staff monitored the SEOs' adherence to protocol.}  FW schools also received an unannounced visit in Year 2, at the same time as the P4P schools. Table A.2 in Online Appendix A shows summary statistics for each of these three input measures over the three rounds of the study. 

\emph{Presence} is defined as the fraction of spot-check days that the teacher is present at the start of the school day. For the SEO to record a teacher present, the head teacher had to physically show the SEO that the teacher was in school. 

Lesson \emph{preparation} is defined as the planning involved with daily lessons, and is measured through a review of teachers' weekly lesson plans. Prior to any spot checks, teachers in grades 4, 5, and 6 in P4P schools were reminded how to fill out a lesson plan in accordance with REB guidelines. Specifically, SEOs provided teachers with a template to record their lesson preparation, focusing on three key components of a lesson---the lesson objective, the instructional activities, and the types of assessment to be used. A `hands-on' session enabled teachers to practice using this template. During the SEO's unannounced visit, he/she collected the daily lesson plans (if any had been prepared) from each teacher. Field staff subsequently used a lesson-planning scoring rubric to provide a subjective measure of quality.  Because a substantial share of upper-primary teachers did not have a lesson plan on a randomly chosen audit day, we used the presence of such a lesson plan as a summary measure in both the incentivized contracts and as an outcome for analysis.

\emph{Pedagogy} is defined as the practices and methods that teachers use in order to impact student learning. We collaborated with both the Ministry of Education and REB to develop a monitoring instrument to measure teacher pedagogy through classroom observation. Our classroom observation instrument measured objective teacher actions and skills as an input into scoring teachers' pedagogical performance.  Our rubric was adapted from the Danielson Framework for Teaching, which is widely used in the U.S. \citep{Dan07book}. The observer evaluated the teachers' effective use of 21 different activities over the course of a full 45-minute lesson. Based on these observations and a detailed rubric, the observer provided a subjective score, on a scale from zero to three, of four components of the lesson: communication of lesson objectives, delivery of material, use of assessment, and  student engagement. The teacher's incentivized score, as well the measure of pedagogy used in our analysis, is defined as the average of these ratings across the four domains.   

\subsection{Balance}

We use the baseline data described in this section to check whether the second-tier randomization produced an appropriately `balanced' experienced treatment assignment. Table \ref{t:balance} confirms that across a wide range of school, teacher, and student characteristics there are no statistically significant differences in means between the experienced P4P and FW treatment arms.\footnote{Since the teacher inputs described in Section \ref{ss:data_inputs} were collected \emph{after} the second-tier randomization, they are not included in Table \ref{t:balance}. See instead Table A.2 in Online Appendix A.}  

\begin{table}[!thbp]
\caption{Baseline characteristics and balance of experienced P4P assignment}
\label{t:balance}
\begin{footnotesize}
\begin{tabular}{l c c c}
\toprule
 & Control mean & Experienced P4P & \\ 
 & [St. Dev.] & ($p$-value) & Obs. \\ 
\midrule
\multicolumn{4}{l}{\emph{Panel A. School attributes}} \\ [1ex]
\multirow[t]{2}{0.25\linewidth}{Number of streams}   & 9.99 & -0.10 &  164\\
 & [4.48]  & (0.881)  \\[1ex]
\multirow[t]{2}{0.25\linewidth}{Number of teachers}   & 20.47 & 0.56 &  164\\
 & [8.49]  & (0.732)  \\[1ex]
\multirow[t]{2}{0.25\linewidth}{Number of new recruits}   & 1.94 & 0.13 &  164\\
 & [1.30]  & (0.505)  \\[1ex]
\multirow[t]{2}{0.25\linewidth}{Number of students}   & 410.06 & 1.42 &  164\\
 & [206.71]  & (0.985)  \\[1ex]
\multirow[t]{2}{0.25\linewidth}{Share female students}   & 0.58 & 0.00 &  164\\
 & [0.09]  & (0.777)  \\[2ex]
\multicolumn{4}{l}{\emph{Panel B. Upper-primary teacher recruit attributes}} \\ [2ex]
\multirow[t]{2}{0.25\linewidth}{Female}   & 0.36 & -0.02 &  242\\
 & [0.48]  & (0.770)  \\[1ex]
\multirow[t]{2}{0.25\linewidth}{Age}   & 25.82 & -0.25 &  242\\
 & [4.05]  & (0.616)  \\[1ex]
\multirow[t]{2}{0.25\linewidth}{DG share sent}   & 0.28 & -0.04 &  242\\
 & [0.33]  & (0.450)  \\[1ex]
\multirow[t]{2}{0.25\linewidth}{Grading task score}   & -0.24 & 0.12 &  242\\
 & [0.93]  & (0.293)  \\[2ex]
\multicolumn{4}{l}{\emph{Panel C. Pupil learning assessments}} \\ [2ex]
\multirow[t]{2}{0.25\linewidth}{English}   & -0.00 & 0.04 & 13826\\
 & [1.00]  & (0.551)  \\[1ex]
\multirow[t]{2}{0.25\linewidth}{Kinyarwanda}   & -0.00 & 0.05 & 13831\\
 & [1.00]  & (0.292)  \\[1ex]
\multirow[t]{2}{0.25\linewidth}{Mathematics}   & 0.00 & -0.00 & 13826\\
 & [1.00]  & (0.950)  \\[1ex]
\multirow[t]{2}{0.25\linewidth}{Science}   & -0.00 & 0.03 & 13829\\
 & [1.00]  & (0.607)  \\[1ex]
\multirow[t]{2}{0.25\linewidth}{Social Studies}   & -0.00 & 0.02 & 13829\\
 & [1.00]  & (0.670)  \\[1ex]
\bottomrule
\end{tabular}

\floatfoot{
The table provides summary statistics for attributes of schools, teachers (new recruits placed in upper primary only), and students collected at baseline.  The first column presents means in FW schools, (with standard deviations in brackets); the second column presents estimated differences between FW and P4P schools (with randomization inference $p$-values in parentheses).  
The sample in Panel B consists of new recruits placed in upper-primary classrooms at baseline, who undertook the lab-in-the-field exercises.
In Panel B, Grading Task IRT scores are standardized based on the distribution among incumbent teachers. In Panel C, student learning IRT scores are standardized based on the distribution in the experienced FW arm.
}
\end{footnotesize}
\end{table}

\section{Results}\label{s:results}

Our two-tiered experiment allows us to estimate impacts of pay-for-performance on the type of individuals applying to, and being placed in, primary teaching posts (the compositional margin), and on the activities undertaken by these new recruits (the effort margin). We report these results in Sections \ref{ss:composition_fx} and \ref{ss:ExperiencedP4P} respectively. Of course, the long-run effects of pay-for-performance will depend not only on selection-in, but also selection-\emph{out}, as well as the dynamics of the behavioral response on the part of teachers who stay. We address dynamic issues in Section \ref{ss:dynamics}, and postpone a substantive discussion of results until Section \ref{s:robustness}.  All statistical tests are conducted via randomization inference with 2,000 permutations of the experienced treatment.

\subsection{Compositional margin of pay-for-performance}\label{ss:composition_fx}

We study three types of compositional effect of pay-for-performance. These are impacts on: the quality of applicants; the observable skill and motivation of placed recruits on arrival; and the student learning induced by these placed recruits during their first and second year on the job.  

\paragraph{Quality of applicants} Motivated by the theoretical model sketched in Section \ref{ss:hypotheses}, we begin by testing for impacts of advertised P4P on the quality of applicants to a given district-by-qualification pool (Hypothesis I).
We focus on Teacher Training College final exam score since this is the only consistently measured quality-related characteristic we observe for all applicants. 

Our primary test uses a Kolmogorov-Smirnov (henceforth, KS) statistic to test the null that there is no difference in the distribution of TTC scores across advertised P4P and advertised FW labor markets. This test statistic can be written as
\begin{equation}\label{eq:KSstat}
    T^{KS} = \sup_{y} \left| \hat{F}_{P4P}(y) - \hat{F}_{FW}(y)\right| = \max_{i=1,\ldots,N} \left| \hat{F}_{P4P}(y_i) - \hat{F}_{FW}(y_i) \right|.
\end{equation}
Here, $\hat{F}_{P4P}(y)$ denotes the empirical cumulative distribution function of TTC scores among applicants who applied under advertised P4P, evaluated at some specific TTC score $y$. Likewise, $\hat{F}_{FW}(y)$ denotes the empirical cumulative distribution function of TTC scores among applicants who applied under advertised FW, evaluated at the same TTC score $y$.             
We test the statistical significance of this difference in distributions by randomization inference. To do so, we repeatedly sample from the set of potential (advertised) treatment assignments $\mathcal{T}^A$ and, for each such permutation, calculate the KS test statistic. The  $p$-value is then the share of such test statistics larger in absolute value than the statistic calculated from the actual assignment. 

\begin{figure}[!hbtp]
\includegraphics[width=0.65\textwidth]{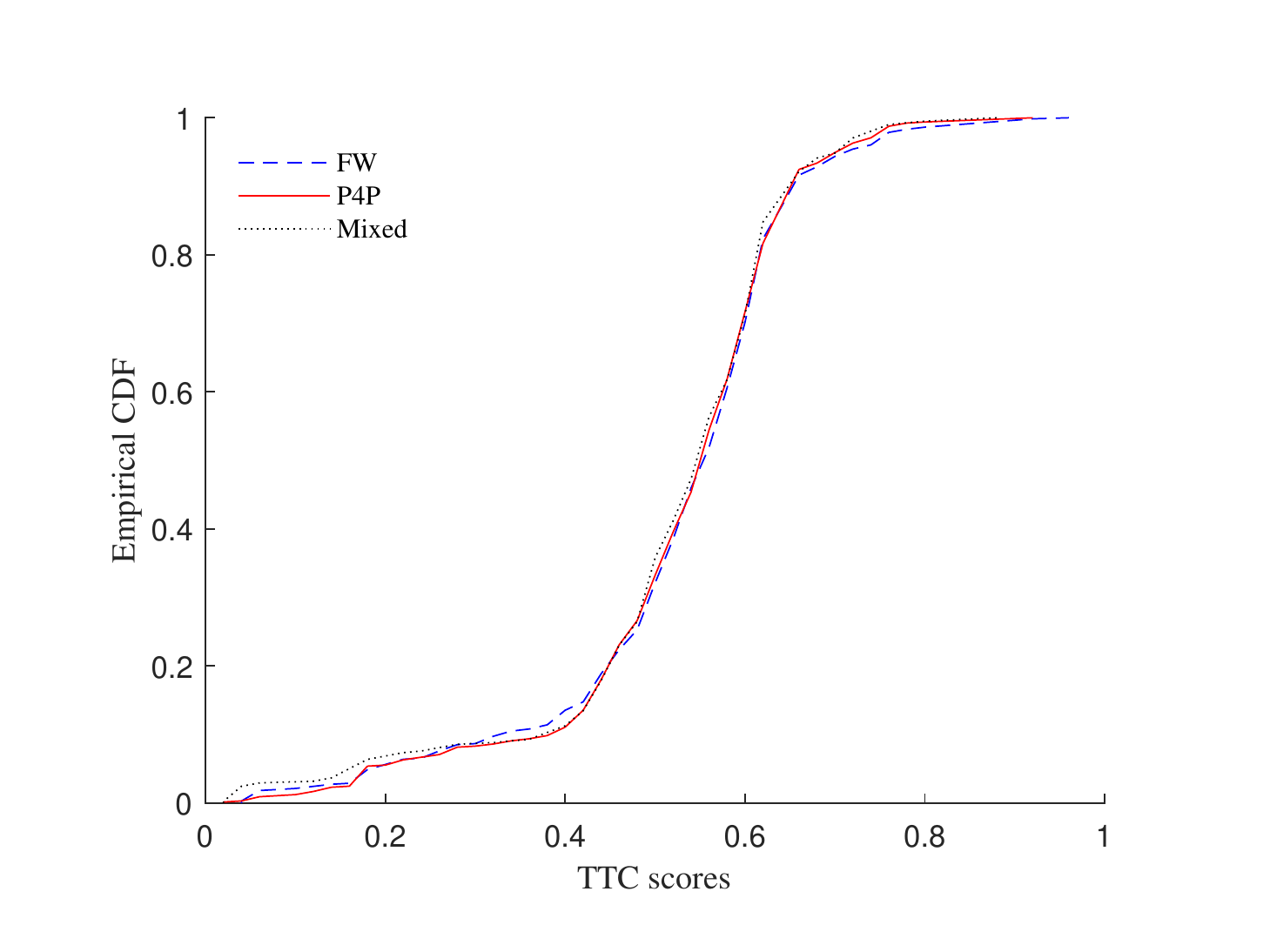}
\caption{Distribution of applicant TTC score, by advertised treatment arm \label{f:apps}}
\begin{footnotesize}
\floatfoot{
 KS test statistic is 0.026, with a $p$-value of 0.909. 
}
\end{footnotesize}
\end{figure}

Figure \ref{f:apps} depicts the distribution of applicant TTC score, by advertised treatment arm.  These distributions are statistically indistinguishable between advertised P4P and advertised FW. The KS test-statistic has a value of 0.026, with a $p$-value of 0.909. Randomization inference is well-powered, meaning that we can rule out even small effects on the TTC score distribution:  a 95 percent confidence interval based on inversion of the randomization inference test rules out additive treatment effects outside of the range $[-0.020, 0.020]$. (The OLS estimate of this additive treatment effect is -0.001, as reported in Table C.1 in Online Appendix C.) We therefore conclude that there was no meaningful impact of advertised P4P on the TTC final exam scores of applicants.\footnote{This conclusion is further substantiated by the battery of secondary tests in Online Appendix C.} 

Below, we move on to consider impacts of advertised P4P on the quality of applicants who were offered a post and chose to accept it---a subset that we term \emph{placed recruits}. It is worth emphasizing that we may find results here even though there is no evidence of an impact on the distribution of TTC score of applicants. This is because, for this well-defined set of placed recruits, we have access to far richer data: lab-in-the-field instruments measuring attributes on arrival, as well as measures of student learning in the first and second years on the job.

\paragraph{Skill and motivation of placed recruits}\label{ss:hires_fx} Along the lines suggested by \citet{Dalfin16}, we explore whether institutions can attract the most capable or the most intrinsically motivated into public service. We include multidimensional skill and motivation types in the theoretical model and test the resulting hypotheses (Hypotheses II and III) using the data described in Section \ref{ss:data_teachers}. Specifically, we use the Grading Task IRT score to measure a placed recruit's skill on arrival, and the framed Dictator Game share sent to capture  baseline intrinsic motivation. 

Our primary tests use these baseline attributes of placed recruits as outcomes.  For attribute $x$ of teacher $j$ with qualification $q$ in district $d$, we estimate a regression of the form 
        \begin{equation}\label{eq:baseline_chars}
        x_{jqd} = \tau_A T^A_{qd} + \gamma_q + \delta_d + e_{jqd},
        \end{equation}
where treatment $T^A_{qd}$ denotes the contractual condition under which a candidate applied.\footnote{Here and throughout the empirical specifications, we will define $T^A_{qd}$ as a \emph{vector} that includes indicators for both the P4P and mixed-treatment advertisement condition. However, for hypothesis testing, we are interested only in the coefficient on the pure P4P treatment. Defining treatment in this way ensures that only candidates who applied (and were placed) under the pure FW treatment are considered as the omitted category here, to which P4P recruits will be compared.} Our test of the null hypothesis is the $t$ statistic associated with coefficient $\tau_A$. We obtain a randomization distribution for this $t$ statistic under the sharp null of no effects for any hire by estimating equation \eqref{eq:baseline_chars} under the set of feasible randomizations of advertised treatments, $T^A \in \mathcal{T}^A$.

\begin{figure}[!hbtp]
\begin{minipage}{0.49\textwidth}
\begin{center}
\includegraphics[width=\textwidth]{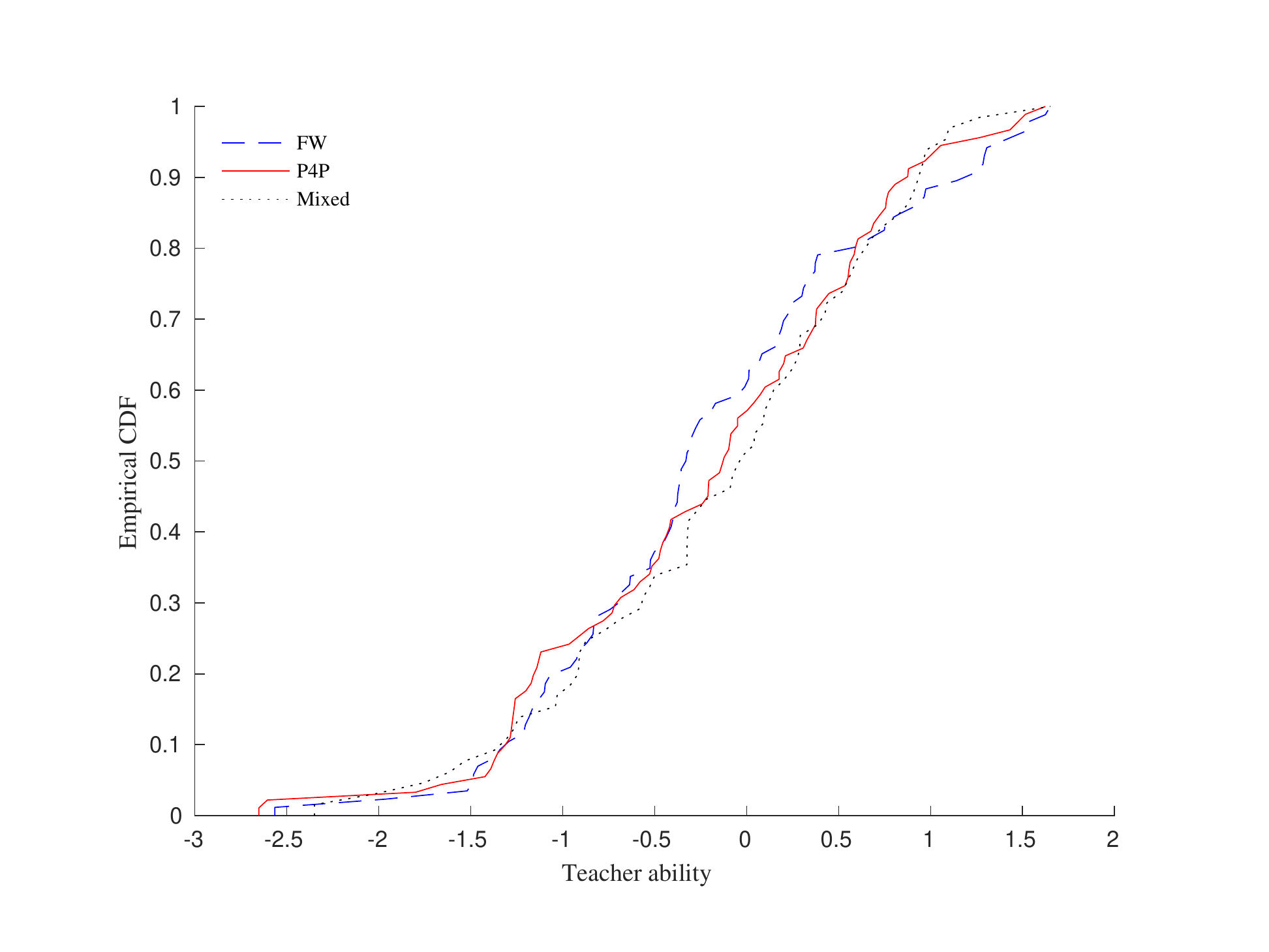} \\
{\footnotesize (a) Grading task score}
\end{center}
\end{minipage} %
\begin{minipage}{0.49\textwidth}
\begin{center}
\includegraphics[width=\textwidth]{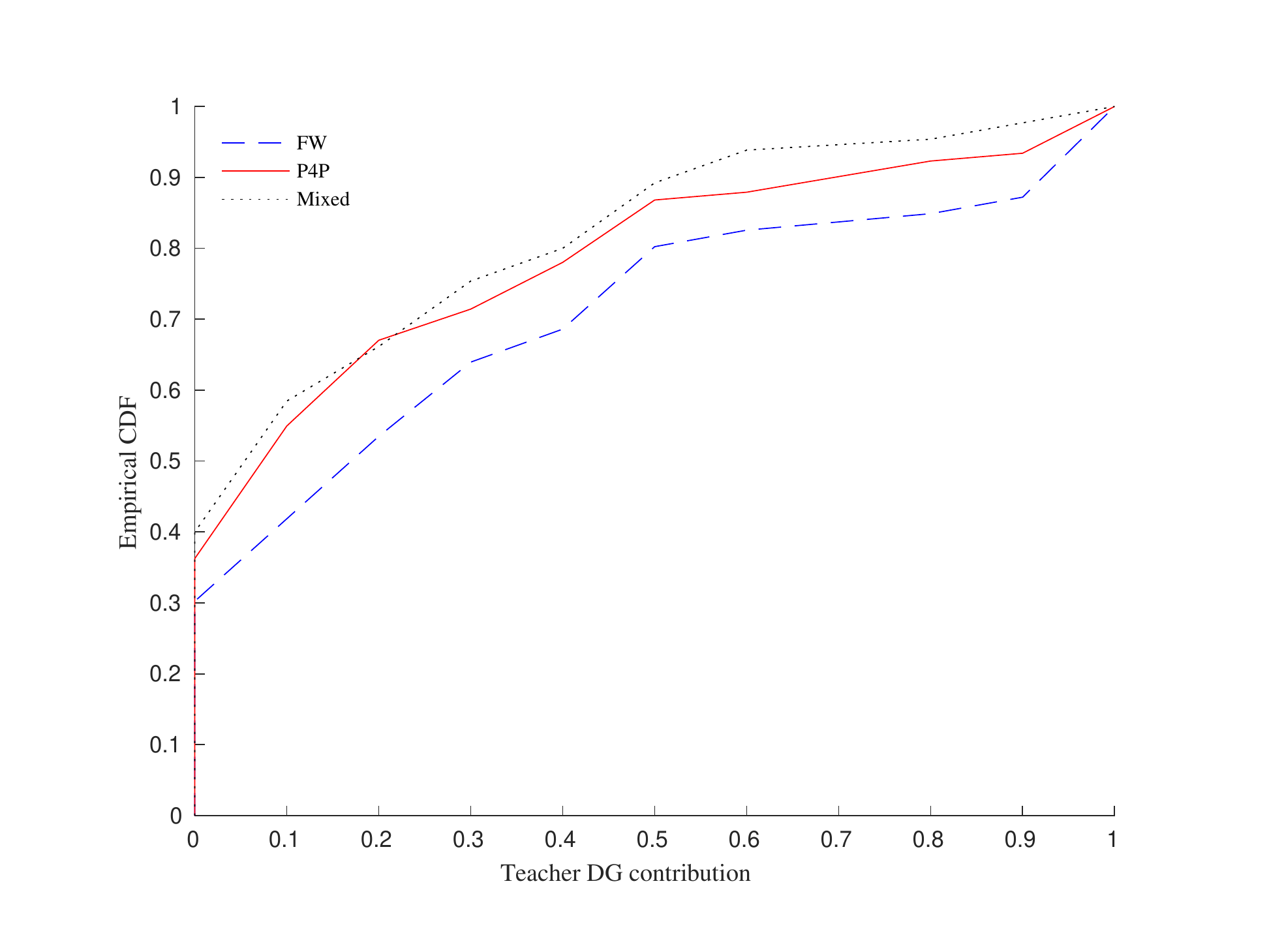} \\
{\footnotesize (b) Dictator Game contribution}
\end{center}
\end{minipage}

\caption{Distribution of placed recruit attributes on arrival, by advertised treatment arm}
\label{f:placedrecruits}
\begin{footnotesize}
\floatfoot{
In Figure 3a, the $t$ statistic for a difference in mean Grading Task IRT score across the P4P and FW treatments is $-0.184$, with a $p$-value of $0.367$. In Figure 3b, the $t$ statistic for a difference in mean DG share sent across the P4P and FW treatments is $-0.100$, with a $p$-value of $0.029$. 
}
\end{footnotesize}
\end{figure}

Before reporting these $t$ statistics, it is instructive to view the data graphically. Figure \ref{f:placedrecruits}a shows the distribution of Grading Task IRT score, and Figure \ref{f:placedrecruits}b the framed Dictator Game share sent, by advertised treatment arm and measured on placed recruits' arrival in schools. A difference in the distributions across treatment arms is clearly visible for the measure of intrinsic motivation but not for the measure of skill. Our regression results tell the same story. In the Grading Task IRT score specification, our estimate of $\tau_A$ is $-0.184$, with a (randomization inference) $p$-value of $0.367$. In the Dictator Game share sent specification, our estimate of $\tau_A$ is $-0.100$, with a $p$-value of $0.029$. It follows that we cannot reject the sharp null of no advertised P4P treatment effect on the measured skill of placed recruits, but we can reject the sharp null of no advertised P4P treatment effect on their measured intrinsic motivation (at the 5 percent level). Teachers recruited under advertised P4P allocated approximately 10 percentage points \emph{less} to the students on average.

We chose not to include additional teacher attributes in the theoretical model, or in the list of pre-specified hypotheses to avoid multiple hypothesis testing concerns. Notwithstanding this decision, we did collect additional data on placed recruits at baseline, meaning that we can use our two-tiered experimental design to conduct further \emph{exploratory} analysis of the impact of advertised P4P. Specifically, we estimate regressions of the form given in equation \eqref{eq:baseline_chars} for four additional teacher attributes: age, gender, risk aversion, and an index capturing the Big Five personality traits.\footnote{Here we follow \citet{DalFinRos13qje} who measure the risk preferences and Big Five personality traits of applicants for civil service jobs in Mexico, and \citet{CallenNBER} who study the relevance of Big Five personality traits for the performance of health workers in Pakistan.} Results are reported in Table A.3 in Online Appendix A, with details of the variable construction provided in the table-note. We are unable to reject the sharp null of no advertised P4P treatment effect for any of these exploratory outcomes.      

\paragraph{Student learning induced by placed recruits}\label{ss:learning_fx} The skill and motivation of placed recruits on arrival are policy relevant insofar as these attributes translate into teacher effectiveness.  To assess this, we combine experimental variation in the advertised contracts to which recruits applied, with the second-stage randomization in experienced contracts under which they worked.  This allows us to estimate the impact of advertised P4P on the student learning induced by these recruits, holding constant the experienced contract---a pure compositional effect (Hypothesis IV).

Our primary test is derived from estimates on student-subject-year level data.  The advertised treatment about which a given student's performance is informative depends on the identity of the placed recruit teaching that particular subject via qualification type and district. We denote this by $T^A_{qd}$ for teacher $j$ with qualification type $q$ in district $d$, and suppress the dependence of the teacher's qualification $q$ on the subject $b$, stream $k$, school $s$, and round $r$, which implies that $q=q(bksr)$. The experienced treatment is assigned at the school level, and is denoted by $T^E_s$. We pool data across the two years of intervention to estimate a specification of the type
    \begin{equation}
    \label{eq:assessments_pooled}
    z_{ibksr} = \tau_A T^A_{qd} + \tau_E T^E_s + \lambda_I I_j + \lambda_E T_s^E I_j + \rho_{br} \bar{z}_{ks,r-1} + \delta_d + \psi_r + e_{ibksr}
    \end{equation}
for the learning outcome of student $i$ in subject $b$, stream $k$, school $s$, and round $r$.  We define $j=j(bksr)$ as an identifier for the teacher assigned to that subject-stream-school-round. The variable $I_j$ is an indicator for whether the teacher is an incumbent, and the index $q=q(j)$ denotes the qualification type of teacher $j$ if that teacher is a recruit (and is undefined if the teacher is an incumbent, so that $T^A_{qd}$ is always zero for incumbents). Drawing on the pseudo-panel of student outcomes, the variable $\bar{z}_{ks,r-1}$ denotes the vector of average outcomes in the once-lagged assessment among students placed in that stream, and its coefficient, $\rho_{br}$, is subject- and round-specific. The coefficient of interest is $\tau_A$: the average of the within-year effect of advertised P4P on pupil learning in Year 1 and the within-year effect of advertised P4P on pupil learning in Year 2.\footnote{We focus on \emph{within-year} impacts because there is not a well-defined cumulative treatment effect. Individual students receive differing degrees of exposure to the advertised treatments depending on their path through streams (and hence teachers) over Years 1 and 2.}  

The theoretical model of Online Appendix B, as well as empirical evidence from other contractual settings \citep{EinFinRyaSchCul13aer}, suggests that pay-for-performance may induce selection on the \emph{responsiveness} to performance incentives. If so, then the impact of advertised treatment will depend on the contractual environment into which recruits are placed. Consequently, we also estimate a specification that allows advertised treatment effects to differ by experienced treatment, including an interaction term between the two treatments. This interacted model takes the form
    \begin{align}\label{eq:assessments_interacted}
    z_{ibksr}  = \tau_A T^A_{qd} + \tau_E T^E_s & + \tau_{AE} T^A_{qd} T^E_s 
    + \lambda_I I_j \notag \\
   & + \lambda_E T^E_s I_j + \rho_{bgr} \bar{z}_{ks,r-1} + \delta_d  + \psi_r +e_{ibksr}.
    \end{align}
Here, the compositional effect of advertised P4P among recruits placed in FW schools is given by $\tau_A$ (a comparison of on-the-job performance across groups $a$ and $b$, as defined in Figure \ref{f:twobytwo}). Likewise, the compositional effect of advertised P4P among recruits placed in P4P schools is given by $\tau_A + \tau_{AE}$ (a comparison of groups $c$ and $d$). 
If $\tau_{AE}$ is not zero, then this interacted model yields the more policy relevant estimands \citep{MurRomWut20nber}. Noting the distinction between estimands and test statistics \citep{ImbRub15book}, we pre-specified the pooled coefficient $\tau_A$ from equation \eqref{eq:assessments_pooled} as the primary test statistic for the presence of compositional effects. Our simulations, using blinded data, show that this pooled test is better powered under circumstances where the interaction term, $\tau_{AE}$, is small.  

We estimate equations \eqref{eq:assessments_pooled} and \eqref{eq:assessments_interacted} by a linear mixed effects model, allowing for normally distributed random effects at the student-round level.\footnote{In our pre-analysis plan, simulations using the blinded data indicated that the linear mixed effects model with a student-round normal random effects would maximize statistical power. We found precisely this in the unblinded data. For completeness, and purely as supplementary analysis, we also present estimates and hypotheses tests via ordinary least squares. See Table A.4 in Online Appendix A. These OLS estimates are generally larger in magnitude and stronger in statistical significance.} Randomization inference is used throughout.  To do so, we focus on the distribution of the estimated $z$-statistic (i.e., the coefficient divided by its estimated standard error), which allows rejections of the sharp null of no effect on any student's performance to be interpreted, asymptotically, as rejection of the non-sharp null that the coefficient is equal to zero \citep{DicRom17jasa}.  Inference for $\tau_A$ is undertaken by permutation of the advertised treatment, $T^A \in\mathcal{T^A}$, while inference for $\tau_E$ likewise proceeds by permuting the experienced treatment $T^E\in\mathcal{T^E}$. To conduct inference about the interaction term, $\tau_{AE}$ in equation \eqref{eq:assessments_interacted}, we simultaneously permute both dimensions of the treatment, considering pairs $(T^A,T^E)$ from the set $\mathcal{T^A}\times \mathcal{T^E}$.

\begin{table}[!hbtp]
\caption{Impacts on student learning, linear mixed effects model\label{t:learning_fx}}
\begin{footnotesize}
\begin{tabular}{lccc}
\toprule
    & \multicolumn{1}{c}{Pooled} & \multicolumn{1}{c}{Year 1} & \multicolumn{1}{c}{Year 2} \\ 
\midrule
\multicolumn{4}{l}{\emph{Model A: Direct effects only}} \\ [2ex] 
\multirow[t]{3}{0.25\linewidth}{Advertised P4P ($\tau_A$)} 
    & 0.01 
    & -0.03  
    & 0.04 \\
	& [-0.04, 0.08] 
	& [-0.06, 0.03] 
	& [-0.05, 0.16] \\
	& (0.75) %
	& (0.20) %
	& (0.31)  \\[2ex]
\multirow[t]{3}{0.25\linewidth}{Experienced P4P ($\tau_E$)} 
    & 0.11 
    & 0.06 
    & 0.16 \\ 
	& [0.02, 0.21] 
	& [-0.03, 0.15] 
	& [0.04, 0.28] \\
	& (0.02)  %
	& (0.17)  %
	& (0.00)  \\[2ex]
\multirow[t]{3}{0.25\linewidth}{Experienced P4P $\times$ Incumbent ($\lambda_E$)} %
	& -0.06 %
	& -0.05 %
	& -0.09 \\ 
	& [-0.20, 0.07] 
	& [-0.19, 0.11] 
	& [-0.24, 0.06] \\
	& (0.36) %
	& (0.54) %
	& (0.27) \\[2ex]
\multicolumn{4}{l}{\emph{Model B:  Interactions between advertised and experienced contracts}} \\ [2ex] 
\multirow[t]{3}{0.25\linewidth}{Advertised P4P ($\tau_A$)} %
	& 0.01 %
	& -0.02  %
	& 0.03 \\
	& [-0.05, 0.14] %
	& [-0.06, 0.07] %
	& [-0.05, 0.21] \\
	& (0.46) %
	& (0.62) %
	& (0.22) \\[2ex]
\multirow[t]{3}{0.25\linewidth}{Experienced P4P ($\tau_E$)} 
    & 0.12 %
    & 0.06  %
    & 0.18 \\ 
	& [0.05, 0.25] %
	& [-0.01, 0.19] %
	& [0.08, 0.33] \\
	& (0.01) %
	& (0.10) %
	& (0.00) \\[2ex]
\multirow[t]{3}{0.25\linewidth}{Advertised P4P $\times$ Experienced P4P ($\tau_{AE}$)} %
	& -0.03 % 
	& -0.01 %
	& -0.04 \\ 
	& [-0.17, 0.09] %
	& [-0.15, 0.10] %
	& [-0.22, 0.13] \\
	& (0.51) %
	& (0.65) %
	& (0.58) \\[2ex]
\multirow[t]{3}{0.25\linewidth}{Experienced P4P $\times$ Incumbent ($\lambda_E$)} %
	& -0.08 %
	& -0.05 %
	& -0.11  \\ 
	& [-0.31, 0.15] %
	& [-0.30, 0.18] %
	& [-0.36, 0.14] \\
	& (0.43) %
	& (0.56) %
	& (0.38) \\[2ex]
Observations %
	& 154594 %
	&  70821 %
	&  83773 \\
\bottomrule
\end{tabular}
\floatfoot{
For each estimated parameter, or combination of parameters, the table reports the point estimate (stated in standard deviations of student learning), 95 percent confidence interval in brackets, and $p$-value in parentheses. Randomization inference is conducted on the associated $z$ statistic. The measure of student learning is based on the empirical Bayes estimate of student ability from a two-parameter IRT model, as described in Section \ref{ss:data_assessments}. 
}
\end{footnotesize}
\end{table}
    
Results are presented in Table \ref{t:learning_fx}. Pooling across years, the compositional effect of advertised P4P is small in point-estimate terms, and statistically indistinguishable from zero (Model A, first row). We do not find evidence of selection on responsiveness to incentives; if anything, the effect of P4P is stronger among recruits who applied under advertised FW contracts, although the difference is not statistically significant and the 95 percent confidence interval for this estimate is wide (Model B, third row). The effect of advertised P4P on student learning does, however, appear to strengthen over time. By the second year of the study, the within-year compositional effect of P4P was 0.04 standard deviations of pupil learning. OLS estimates of this effect are larger, at 0.08 standard deviations, with a $p$-value of 0.10, as shown in Table A.4. 

\begin{figure}[!hb]
\begin{minipage}{0.49\textwidth}
\begin{center}
\includegraphics[width=\textwidth]{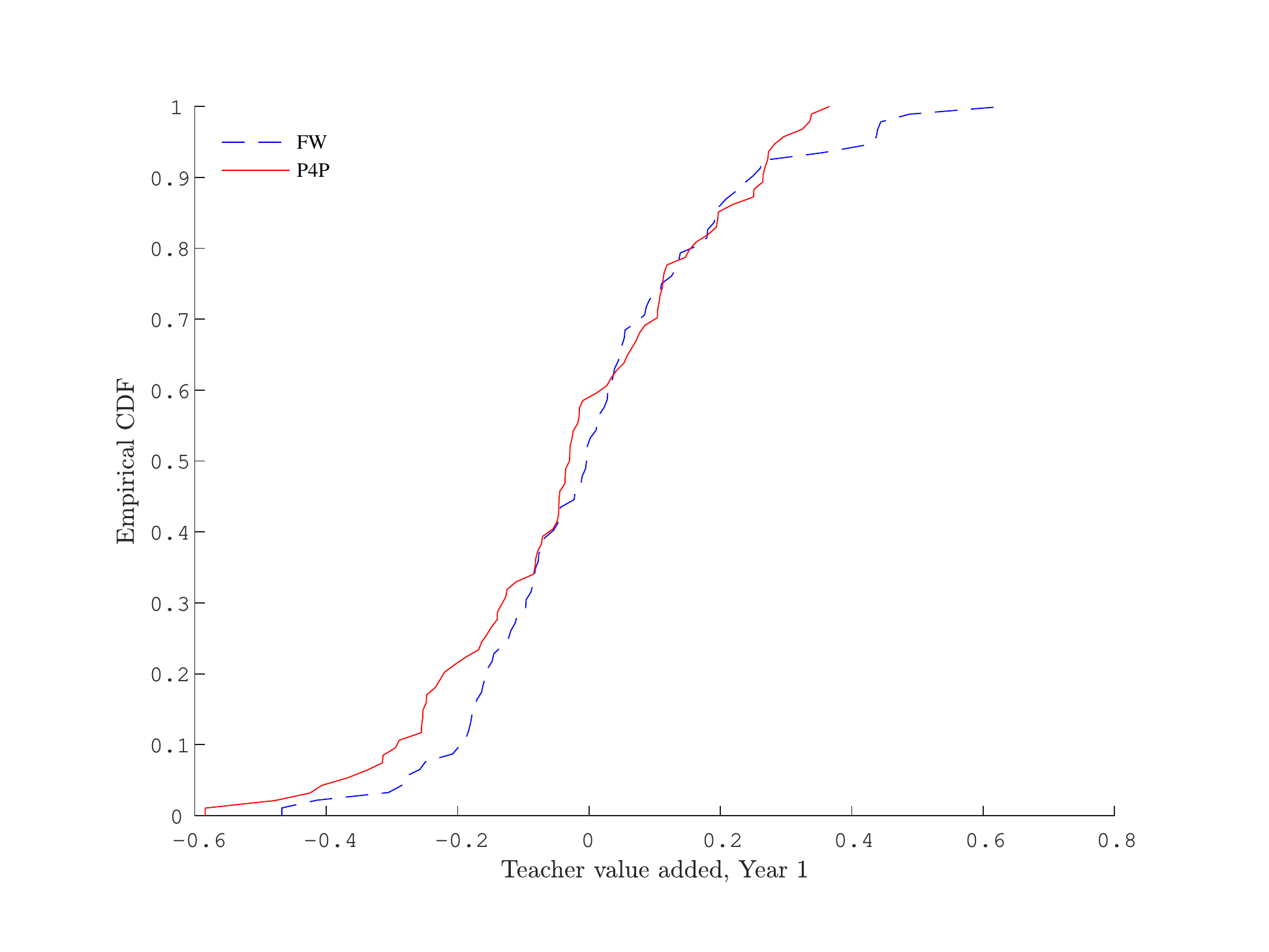} \\
{\footnotesize (a) Year 1}
\end{center}
\end{minipage} %
\begin{minipage}{0.49\textwidth}
\begin{center}
\includegraphics[width=\textwidth]{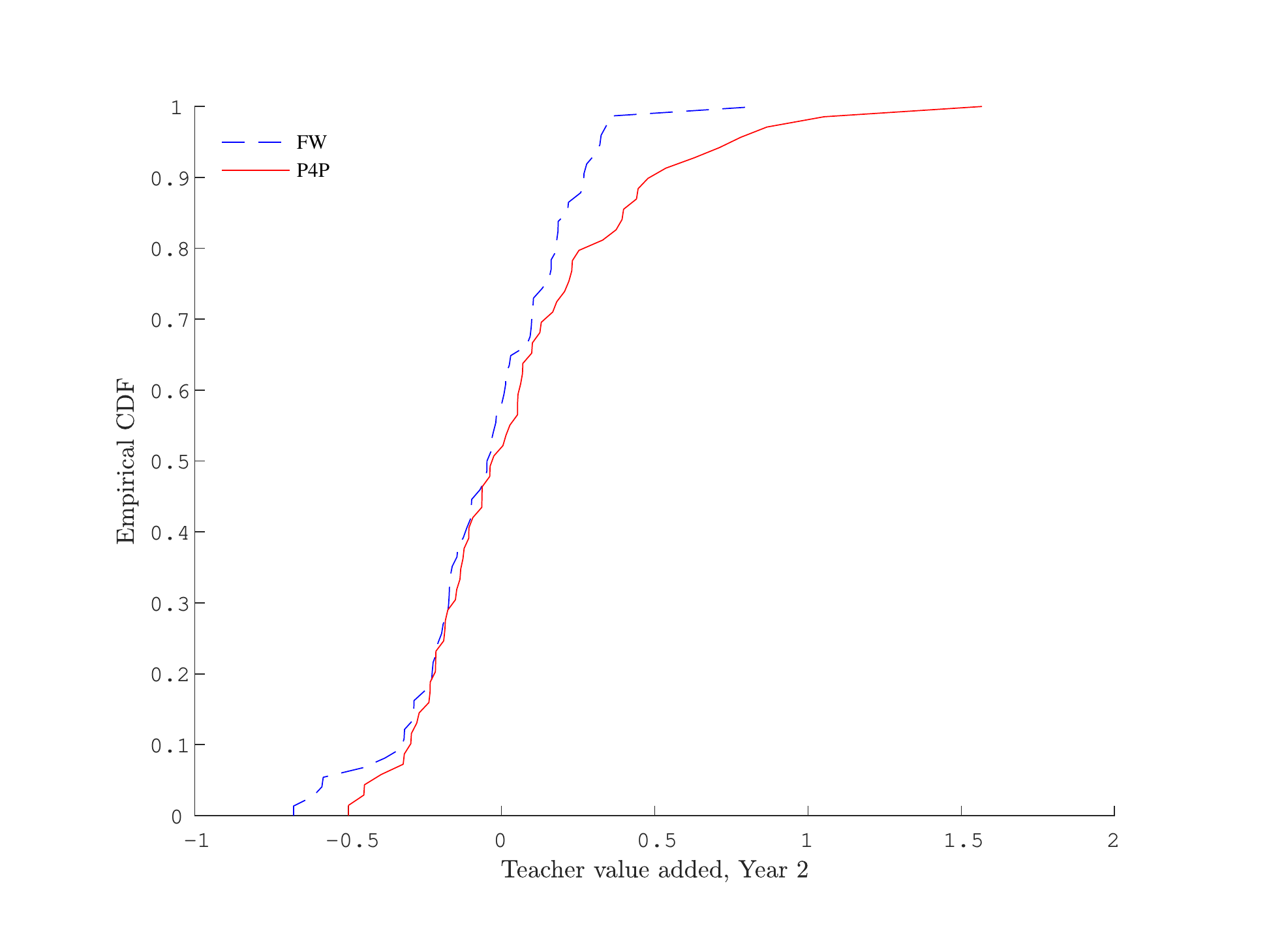} \\
{\footnotesize (b) Year 2}
\end{center}
\end{minipage}

\caption{Teacher value added among recruits, by advertised treatment and year}
\label{f:TVA_Recruits_byTx}
\begin{footnotesize}
\floatfoot{
The figures plot distributions of teacher value added under advertised P4P and advertised FW in Years 1 and 2. Value-added models estimated with school fixed effects. Randomization inference $p$-value for equality in distributions between P4P and FW applicants, based on one-sided KS test, is 0.796 using Year 1 data; 0.123 using Year 2 data; and 0.097 using pooled estimates of teacher value added (not pre-specified). 
}
\end{footnotesize}
\end{figure}

For the purposes of interpretation, it is useful to recast the data in terms of teacher value added. As detailed in Online Appendix D, we do so by estimating a teacher valued-added (TVA) model that controls for students' lagged test scores, as well as school fixed effects, with the latter absorbing differences across schools attributable to the experienced P4P treatment. This TVA model gives a sense of magnitude to the student learning estimates in Table \ref{t:learning_fx}. Applying the Year 2 point estimate for the effect of advertised P4P would raise a teacher from the 50th to above the 73rd percentile in the distribution of (empirical Bayes estimates of) teacher value added for placed recruits who applied under FW. The TVA model also reveals the impact of advertised P4P on the distribution of teacher effectiveness. Figure \ref{f:TVA_Recruits_byTx}b shows that the distribution of teacher value added among recruits in their second year on the job is better, by first order stochastic dominance, under advertised P4P than advertised FW. This finding is consistent with the view that a contract that rewards the top quintile of teachers attracts individuals who deliver greater learning. 

\subsection{Effort margin of pay-for-performance}\label{ss:ExperiencedP4P}

Having studied the type of individuals applying to, and being placed in, upper-primary posts, we now consider the activities undertaken by these new recruits.

\paragraph{Student learning induced by placed recruits} We start by using the two-tiered experimental variation to estimate the impact of experienced P4P on the student learning induced by the placed recruits, holding constant the advertised contract---a pure effort effect (Hypothesis V). Our primary test uses the specification in equation  \eqref{eq:assessments_pooled}, again estimated by a linear mixed effects model. The coefficient of interest is now $\tau_E$. To investigate possible `surprise effects' from the re-randomization, we also consider the interacted specification of equation \eqref{eq:assessments_interacted}. In this model, $\tau_E$ gives the effect of experienced P4P among recruits who applied under FW contractual conditions (a comparison of groups a and c, as defined in Figure 1), while $\tau_E+\tau_{AE}$ gives the effect of experienced P4P among recruits who applied under P4P contractual conditions (a comparison of groups b and d). If recruits are disappointed, because it is groups b and c who received the ‘surprise’, $\tau_E$ should be smaller than $\tau_E+\tau_{AE}$.\footnote{We are grateful to a referee for highlighting a further interpretation: $\tau_E$ in Model B is the policy-relevant estimate of experienced P4P at the start of any unexpected transition to P4P, while $\tau_E+\tau_{AE}$ is the policy-relevant estimate for that effect slightly further into a transition---the effect of P4P on a cohort anticipating P4P.} 

Results are presented in Table \ref{t:learning_fx}. Pooling across years, the within-year effect of experienced P4P is $0.11$ standard deviations of pupil learning (Model A, second row). The randomization inference $p$-value is $0.02$, implying that we can reject the sharp null of no experienced P4P treatment effect on placed recruits at the 5 percent level. We do not find evidence of disappointment caused by the re-randomization. The interaction term is insignificant (Model B, third row) and, in point-estimate terms, $\tau_E$ is larger than $\tau_E+\tau_{AE}$. As was the case for the compositional margin, the effort effect of experienced P4P on student learning appears to strengthen over time. By the second year of the study, the within-year effort effect of P4P was $0.16$ standard deviations of pupil learning.\footnote{Across all specifications, the interaction term between experienced P4P and an indicator for incumbent teachers is negative, though statistically insignificant, and smaller in magnitude than the direct effect of experienced P4P, implying a weaker---though still positive---effect of P4P on incumbents in point-estimate terms.}

To put this in perspective, we compare the magnitude of this effort effect to impacts in similar studies in the US, and beyond. \citet{SojMykWes14jhr} study pay-for-performance schemes in Minnesota, typically based on a composite metric of subjective teacher evaluation and student performance, and find an effect of 0.03 standard deviations of pupil learning. \citet{DeeWyc15jpam} study a high-stakes incentive over a composite metric in Washington, DC, and find effects consistent with those of \citet{SojMykWes14jhr} in terms of the implied magnitude of effects on pupil learning. \citet{GleMur16chapter} review a range of studies, including several in Benin, China, India and Kenya that employ incentives for either students or teachers based solely on student performance; effect sizes are larger, typically above 0.2 standard deviations of pupil learning. Our effort effect falls within this range and is of a comparable magnitude to the impact in \citet{DufHanRya12}, who study incentives for teacher attendance in India.

\paragraph{Dimensions of the composite performance metric}\label{s:4Ps}

The results in Table \ref{t:learning_fx} speak to the obvious policy question, namely whether there are impacts of advertised and experienced P4P contracts on student learning. For completeness, and to gain an understanding into mechanisms, we complete our analysis by studying whether there are impacts on the \emph{contracted} metrics which are calculated at teacher-level (Hypothesis VI). For these tests, we use the following specifications:
\begin{equation}\label{eq:spec_inputs}
 m_{jqsdr} = \tau_A T^A_{qd} + \tau_E T^E_{s}  + \lambda_I I_j + \lambda_E T_s^E I_j  + \gamma_q + \delta_d  + \psi_r + e_{jqsdr}
\end{equation}
%and 
\begin{equation}\label{eq:spec_inputs_split}
m_{jqsdr} = \tau_A T^A_{qd} + \tau_E T^E_{s} + \tau_{AE} T^A_{qd} T^E_{s} + \lambda_I I_j + \lambda_E T_s^E I_j+ \gamma_q + \delta_d + \psi_r  + e_{jqsdr},
\end{equation}
for the metric of teacher $j$ with qualification $q$ in school $s$ of district $d$, as observed in post-treatment round $r$. As above, the variable $I_j$ is an indicator for whether the teacher is an incumbent (recall that $T^A_{qd}$ is always zero for incumbents).\footnote{Note that any attribute of recruits themselves, even if observed at baseline, suffers from the `bad controls' problem, as the observed values of this covariate could be an outcome of the advertised treatment. These variables are therefore not included as independent variables.} A linear mixed effects model with student-level random effects is no longer applicable; outcomes are constructed at the teacher-level, and given their rank-based construction, normality does not seem a helpful approximation to the distribution of error terms. As stated in our pre-analysis plan, we therefore estimate equations \eqref{eq:spec_inputs} and \eqref{eq:spec_inputs_split} with a round-school random-effects estimator to improve efficiency. The permutations of treatments used for inferential purposes mirror those above.  

\begin{table}[!hpt]
\caption{Estimated effects on dimensions of the composite 4P performance metric\label{t:inputs}}
\begin{footnotesize}
\begin{tabular}{lccccc}
\toprule
    & \makecell[b]{Summary \\ metric} & Preparation & Presence & Pedagogy & \makecell[b]{Pupil \\ learning} \\ 
\midrule
\multicolumn{6}{l}{\emph{Model A: Direct effects only}} \\ [2ex] 
\multirow{2}{0.25\linewidth}{Advertised P4P ($\tau_A$)}    %
  &  -0.04  
  &  0.07  
  &  0.00  
  &  0.03  
  &  -0.02 \\ 
  &  [-0.09, 0.01]  
  &  [-0.13, 0.32]  
  &  [-0.05, 0.07]  
  &  [-0.06, 0.10]  
  &  [-0.08, 0.02] \\ 
  &  (0.11)  
  &  (0.40)  
  &  (0.93)  
  &  (0.42)  
  &  (0.27)  \\[2ex] 
\multirow{2}{0.25\linewidth}{Experienced P4P ($\tau_E$)}    %
  &  0.23  
  &  0.02  
  &  0.08  
  &  0.10  
  &  0.09 \\ 
  &  [0.19, 0.28]  
  &  [-0.13, 0.16]  
  &  [0.02, 0.14]  
  &  [-0.00, 0.21]  
  &  [0.03, 0.15] \\ 
  &  (0.00)  
  &  (0.84)  
  &  (0.01)  
  &  (0.05)  
  &  (0.00)  \\[2ex] 
\multirow{2}{0.25\linewidth}{Experienced P4P $\times$ Incumbent ($\lambda_E$)}    %
  &  0.03 
  &  0.07 
  &  -0.01 
  &  0.07 
  &  -0.00\\ 
  &  [-0.01, 0.07]  
  &  [-0.03, 0.18]  
  &  [-0.06, 0.05]  
  &  [-0.01, 0.16]  
  &  [-0.04, 0.03] \\ 
  &  (0.10)  
  &  (0.17)  
  &  (0.70)  
  &  (0.11)  
  &  (0.86)  \\[2ex] 
\multicolumn{6}{l}{\emph{Model B:  Interactions between advertised and experienced contracts}} \\ [2ex] 
\multirow{2}{0.25\linewidth}{Advertised P4P ($\tau_A$)}    %
  &  -0.03  
  &  0.16  
  &  -0.01  
  &  0.12  
  &  -0.01 \\ 
  &  [-0.12, 0.05]  
  &  [-0.11, 0.48]  
  &  [-0.16, 0.17]  
  &  [-0.27, 0.55]  
  &  [-0.12, 0.11] \\ 
  &  (0.42)  
  &  (0.19)  
  &  (0.86)  
  &  (0.44)  
  &  (0.91)  \\[2ex] 
\multirow{2}{0.25\linewidth}{Experienced P4P ($\tau_E$)}    %
  &  0.22  
  &  -0.00  
  &  0.08  
  &  0.17  
  &  0.08 \\ 
  &  [0.15, 0.29]  
  &  [-0.26, 0.25]  
  &  [-0.01, 0.16]  
  &  [-0.05, 0.38]  
  &  [0.00, 0.16] \\ 
  &  (0.00)  
  &  (0.97)  
  &  (0.07)  
  &  (0.12)  
  &  (0.04)  \\[2ex] 
\multirow{2}{0.25\linewidth}{Advertised P4P $\times$ Experienced P4P ($\tau_{AE}$)}    %
  &  -0.02  
  &  -0.11  
  &  0.02  
  &  -0.11  
  &  -0.03 \\ 
  &  [-0.11, 0.07]  
  &  [-0.45, 0.23]  
  &  [-0.12, 0.16]  
  &  [-0.45, 0.24]  
  &  [-0.15, 0.08] \\ 
  &  (0.65)  
  &  (0.53)  
  &  (0.69)  
  &  (0.53)  
  &  (0.64)  \\[2ex] 
\multirow{2}{0.25\linewidth}{Experienced P4P $\times$ Incumbent ($\lambda_E$)}    %
  &  0.05 
  &  0.09 
  &  -0.01 
  &  0.00 
  &  0.00\\ 
  &  [-0.01, 0.10]  
  &  [-0.07, 0.26]  
  &  [-0.09, 0.07]  
  &  [-0.13, 0.14]  
  &  [-0.05, 0.06] \\ 
  &  (0.07)  
  &  (0.27)  
  &  (0.82)  
  &  (0.96)  
  &  (0.90)  \\[2ex] 
Observations  %
  &      3996  
  &      2514  
  &      3455  
  &      2136  
  &      3049  \\  
FW recruit mean    
  &    0.49  
  &    0.65  
  &    0.89  
  &    1.98  
  &    0.48  \\  
(SD)    %
  &    (0.22)  
  &    (0.49)  
  &    (0.31)  
  &    (0.57)  
  &    (0.27)  \\  
FW incumbent mean    %
  &    0.37  
  &    0.50  
  &    0.87  
  &    2.05  
  &    0.45  \\  
(SD)    %
  &    (0.24)  
  &    (0.50)  
  &    (0.33)  
  &    (0.49)  
  &    (0.28)  \\  
\bottomrule
\end{tabular}

\end{footnotesize}
\floatfoot{
For each estimated parameter, the table reports the point estimate, 95 percent confidence interval in brackets, and $p$-value (or for FW means, standard deviations) in parentheses.  Randomization inference is conducted on the associated $t$ statistic. All estimates are pooled across years, but outcomes are observed in the FW arm during the second year only. Outcomes are constructed at teacher-round-level as follows: \emph{preparation} is a binary indicator for existence of a lesson plan on a randomly chosen spot-check day; \emph{presence} is the fraction of spot-check days present at the start of the school day; \emph{pedagogy} is the classroom observation score, measured on a four-point scale; and \emph{pupil learning} is the  Barlevy-Neal percentile rank. The \emph{summary metric} places 50 percent weight on learning and 50 percent on teacher inputs, and is measured in percentile ranks.
}
\end{table}

Results are reported in Table \ref{t:inputs} and, to the extent available, are based on pooled data.\footnote{As discussed in Section \ref{ss:data_inputs}, FW schools only received unannounced visits to measure teacher inputs in Year 2.} Consistent with the pooled results in Table \ref{t:learning_fx}, we see a positive and significant impact of experienced P4P on both the summary metric and the learning sub-component. The specifications with teacher inputs as dependent variables suggest that this impact on student learning is driven, at least in part, by improvements in teacher presence and pedagogy. Teacher presence was 8 percentage points higher among recruits who experienced the P4P contract compared to recruits who experienced the FW contract; an impact that is statistically significant at the 1 percent level and sizeable in economic terms given that baseline teacher presence was already nearly 90 percent. Recruits who experienced P4P were 0.10 points more effective in their classroom practices than recruits receiving FW, although it is possible that this improvement occurred only during the observation. We find no evidence of impacts on lesson planning.

\subsection{Dynamic effects}\label{ss:dynamics}

Our two-tiered experiment was designed to evaluate the impact of pay-for-performance and, in particular, to quantify the relative importance of a compositional margin at the recruitment stage versus an effort margin on the job. The hypotheses specified in our pre-analysis plan refer to selection-in and incentives among placed recruits. Since within-year teacher turnover was limited by design and within-year changes in teacher skill and motivation are likely small, the total effect of P4P in Year 1 can plausibly only be driven by a change in the type of teachers recruited and/or a change in effort resulting from the provision of extrinsic incentives. 

Interpreting the total effect of P4P in Year 2 is more complex, however. First, we made no attempt to discourage \emph{between}-year teacher turnover, and so there is the possibility of a further compositional margin at the retention stage (c.f. \citeauthor{MurSun11jpe} 2011). Experienced P4P may have selected-out the low skilled \citep{Laz00aer} or, more pessimistically, the highly intrinsically motivated. Second, given the longer time frame, teacher characteristics could have changed. Experienced P4P may have eroded a given teacher's intrinsic motivation (as hypothesized in the largely theoretical literature on motivational crowding out) or, more optimistically, encouraged a given teacher to improve her classroom skills. In this section, we conduct an exploratory analysis of these dynamic effects.\footnote{We emphasize that this material is exploratory; the hypotheses tested in this section were not part of our pre-analysis plan. That said, the structure of the analysis in this section does follow a related pre-analysis plan (intended for a companion paper) which we uploaded to our trial registry on October 3, 2018, \emph{prior} to unblinding of our data.} 

\paragraph{Retention effects}\label{ss:retention}

We begin by exploring whether experienced P4P affects retention rates among recruits. Specifically, we look for an impact on the likelihood that a recruit is still employed at midline in February 2017 at the start of the Year 2; i.e. after experiencing pay-for-performance in Year 1, although before the performance awards were announced. To do so, we use a linear probability model of the form
\begin{equation}\label{eq:retentionprob}
\Pr[employed_{iqd2} =1] = \tau_E T_s^E + \gamma_q + \delta_d, % + e_{iqd},
\end{equation}
where $employed_{iqd2}$ is an indicator for whether teacher $i$ with subject-family qualification $q$ in district $d$ is still employed by the school at the start of Year 2, and $\gamma_q$ and $\delta_d$ are the usual subject-family qualification and district indicators.

As the first column of Table \ref{t:retention} reports, our estimate of $\tau_E$ is zero with a randomization inference $p$-value of 0.94. There is no statistically significant impact of experienced P4P on retention of recruits; the retention rate is practically identical---at around 80 percent---among recruits experiencing P4P and those experiencing FW.  

\begin{table}[!hbtp]
    \caption{Retention of placed recruits} 
    \label{t:retention}
    \begin{footnotesize}
    \begin{tabular}{l S S S}
\toprule & \multicolumn{1}{c}{(1)} & \multicolumn{1}{c}{(2)} & \multicolumn{1}{c}{(3)} \\ 
\midrule
Experienced P4P & 0.00 & -0.04 & -0.08 \\ 
  & (0.94) & (0.42) & (0.24) \\ 
Interaction &  & -0.05 & 0.15 \\ 
  &   & (0.39) & (0.37) \\ 
\midrule
Heterogeneity by\ldots & & \multicolumn{1}{c}{Grading Task} & \multicolumn{1}{c}{Dictator Game} \\ 
Observations  & \multicolumn{1}{c}{   249} & \multicolumn{1}{c}{   238} & \multicolumn{1}{c}{   238} \\ 
\bottomrule
\end{tabular}

    \end{footnotesize}
    \floatfoot{
   For each estimated parameter, the table reports the point estimate and $p$-value in parentheses. Randomization inference is conducted on the associated $t$ statistic. In each column, the outcome is an indicator for whether the teacher is still employed at the start of Year 2. The mean of this dependent variable for FW recruits is 0.80. In the second column, the specification includes an interaction of experienced treatment with the teacher's baseline Grading Task IRT score (not de-meaned); in the third column, the interaction is with the teacher's share sent in the baseline framed Dictator Game (again not de-meaned). All specifications include controls for districts and subjects of teacher qualification.
   }
\end{table}

It is worth noting that there is also no impact of experienced P4P on \emph{intentions} to leave in Year 3. In the endline survey in November 2017, we asked teachers the question:``How likely is it that you will leave your job at this school over the coming year?''. Answers were given on a 5-point scale. For analytical purposes we collapse these answers into a binary indicator coded to 1 for `very likely' or `likely' and 0 otherwise, and estimate specifications analogous to equations (\ref{eq:spec_inputs}) and (\ref{eq:spec_inputs_split}). As the second column of Table A.5 in Online Appendix A shows, there is no statistically significant impact of experienced P4P on recruits' self-reported likelihood of leaving in Year 3. Our estimate of $\tau_E$ is $-0.06$ with a randomization inference $p$-value of 0.39.

Of course, a retention rate of 80 percent implies 20 percent attrition from Year 1 to Year 2, which is non-negligible. And the fact that retention \emph{rates} are similar does not rule out the possibility of an impact of experienced P4P on the \emph{type} of recruits retained. To explore this, we test whether experienced P4P induces differentially skilled recruits to be retained. Here, we use teachers' performance on the baseline Grading Task in the primary subject they teach to obtain an IRT estimate of their ability in this subject, denoted $z_i$, and estimate an interacted model of the form
\begin{equation}\label{eq:placed_skill}
\Pr[employed_{iqd2} =1] = \tau_E T_s^E + \zeta T_s^E z_i + \beta z_i + \gamma_q  + \delta_d .
\end{equation}
Inference for the key parameter, $\zeta$, is undertaken by performing randomization inference for alternative assignments of the school-level experienced treatment indicator. As the second column of Table \ref{t:retention} reports, our estimate of $\zeta$ is $-0.05$, with a randomization inference $p$-value of $0.39$. There is not a significant difference in selection-out on baseline teacher skill across the experienced treatments. Hence, there is no evidence that experienced P4P induces differentially skilled recruits to be retained. 

We also test whether experienced P4P induces differentially intrinsically motivated recruits to be retained. Here, we use the contribution sent in the framed Dictator Game played by all recruits at baseline, denoted $x_i$, and re-estimate the interacted model in equation (\ref{eq:placed_skill}), replacing $z_i$ with $x_i$. As the third column of Table \ref{t:retention} reports, our estimate of $\zeta$ in this specification is $0.15$, with a randomization inference $p$-value of $0.37$. There is not a significant difference in selection-out on baseline teacher intrinsic motivation across the experienced treatments. Hence, there is also no evidence that experienced P4P induces differentially intrinsically motivated recruits to be retained. 

\paragraph{Changes in retained teacher characteristics}\label{ss:endline}

To assess whether experienced P4P changes within-retained-recruit teacher skill or intrinsic motivation from baseline to endline, we estimate the following ANCOVA specification
\begin{equation}\label{eq:placed_motivation_diff}
y_{isd2} = \tau_E T^E_{s} + \rho y_{isd0} + \gamma_q + \delta_d + e_{isd}, 
\end{equation}
where $y_{iqsd2}$ is the characteristic (raw Grading Task score or framed Dictator Game contribution) of retained recruit $i$ with qualification $q$ in school $s$ and district $d$ at endline (round 2), and $y_{iqsd0}$ is this characteristic of retained recruit $i$ at baseline (round 0). As the first column of Table \ref{t:round2chars} reports, our estimate of $\tau_E$ in the Grading Task specification is $0.68$, with a randomization inference $p$-value of 0.57. Our estimate of $\tau_E$ in the Dictator Game specification is $-0.04$, with a randomization inference $p$-value of 0.06. Both estimates are small in magnitude and, in the case of the Dictator Game share sent, we reject the sharp null only at the 10 percent level. Hence, to the extent that contributions in the Dictator Game are positively associated with teachers' intrinsic motivation, we find no evidence that the \emph{rising} effects of experienced P4P from Year 1 to Year 2 are driven by \emph{positive} changes in our measures of within-retained-recruit teacher skill or intrinsic motivation.\footnote{Although repeated play of lab experimental games may complicate interpretation in some contexts, several factors allay this concern here.  First, unlike strategic games, the `Dictator Game' has no second `player' about whom to learn. Second, the two rounds of play were fully two years apart.} 

\begin{table}
\caption{Characteristics of retained recruits at endline}
\label{t:round2chars}
\begin{footnotesize}
\begin{tabular}{l S S }
\toprule & \multicolumn{1}{c}{Grading Task} & \multicolumn{1}{c}{Dictator Game} \\ 
\midrule
Experienced P4P & 0.68 & -0.04 \\ 
  & (0.57) & (0.06) \\[2ex] 
Observations & \multicolumn{1}{c}{   170} & \multicolumn{1}{c}{   169} \\ 
\bottomrule
\end{tabular}

\floatfoot{
For each estimated parameter, the table reports the point estimate and $p$-value in parentheses. Randomization inference is conducted on the associated $t$ statistic. In the first column, the outcome is the Grading Task score of the teacher at endline on a (raw) scale from $0$ to $30$; in the second column, it is the teacher's share sent in the framed Dictator Game played at endline. All specifications include the outcome measured at baseline and controls for district and subject-of-qualification.
}
\end{footnotesize}

\end{table}

Before moving on, it is worth noting that the Dictator Game result could be interpreted as weak evidence that the experience of P4P contracts crowded out the intrinsic motivation of recruits. We do not have any related measures observed at both baseline and endline with which to further probe \emph{changes} in motivation. However, we do have a range of related measures at endline: job satisfaction, likelihood of leaving, and positive/negative affect.\footnote{We follow \cite{Bloomqje15} in using the Maslach Burnout Index to capture job satisfaction and the Clark-Tellgen Index of positive and negative affect to capture the overall attitude of teachers.} As Table A.5 shows, there is no statistically significant impact of experienced P4P on any of these measures. 

Further substantiating this point, Table A.6 in Online Appendix A shows the distribution of answers to the endline survey question: ``What is your overall opinion about the idea of providing high-performing teachers with bonus payments on the basis of objective measures of student performance improvement?''\footnote{We follow the phrasing used in the surveys run by \cite{MurSun11eer}.} The proportion giving a favorable answer exceeds 75 percent in every study arm. In terms of Figure \ref{f:twobytwo}, group $a$ (recruits who both applied for and experienced FW) had the most negative view of pay-for-performance, while group $c$ (who applied for FW but experienced P4P) had the most positive view. Hence it seems that it was the idea, rather than the reality, of pay-for-performance that was unpopular with (a minority of) recruits.\footnote{Consistent with our failure to find `surprise effects' in student learning, there is no evidence that the re-randomization resulted in hostility toward pay-for-performance; if anything the reverse.} 

\section{Discussion}\label{s:robustness}

\paragraph{Compositional margin} To recap from Section \ref{ss:composition_fx}, we find no evidence of an advertised treatment impact on the measured quality of applicants for upper-primary teaching posts in study districts, but we do find evidence of an advertised treatment impact on the measured intrinsic motivation of individuals who are placed into study schools. We draw three conclusions from these results. 

First, potential applicants were aware of, and responded to, the labor market intervention. The differences in distributions across advertised treatment arms in Figure \ref{f:placedrecruits}b (Dictator Game share sent) and Figure \ref{f:TVA_Recruits_byTx}b (teacher valued added in Year 2) show that the intervention changed behavior. Since these differences are for placed recruits not applicants, it could be that this behavior change was on the labor demand rather than supply side. In Figure A.2 in Online Appendix A, we plot the empirical probability of hiring as a quadratic function of the rank of an applicant's TTC score within the set of applicants in their district.
It is clear from the figure that the predicted probabilities are similar across P4P and FW labor markets. We also test formally whether the probability of hiring, as a function of  CV characteristics (TTC score, age and gender), is the same under both P4P and FW advertisements.\footnote{Note that this is a sufficient but not necessary test of the absence of a demand-side response. It is sufficient because districts do not interview applicants, so CVs give us the full set of characteristics that could determine hiring. It is not necessary, however, because we observe hires rather than offers. The probability that an offer is accepted could be affected by the advertised contract associated with that post, even if applicants apply to jobs of both types and even if DEOs do not take contract offer types into account when selecting the individuals to whom they would like to make offers.} We find no statistically significantly differences across advertised treatment arms. 

Second, the supply-side response was, if anything, beneficial for student learning. The P4P contract negatively selected-in the attribute measured by the baseline Dictator Game. However, Table D.1 in Online Appendix D shows that the rank correlation between the baseline DG share sent by recruits and their teacher value added is small and not statistically significant. Consistent with this, our primary test rules out meaningful negative effects of advertised P4P on student learning. In fact, our supplementary analyses---the OLS estimates in Table A.4 and the distributions of teacher value added in Figure \ref{f:TVA_Recruits_byTx}---point to \emph{positive} effects on learning by recruits' second year on the job. It therefore appears that only positively selected attribute(s) mattered, at least in the five core subjects that we assessed. 

Finally, districts would struggle to achieve this compositional effect directly via the hiring process. The positively selected attribute(s) were not evident  in the metrics observed at baseline---either in TTC scores, or in the Grading Task scores that districts could in principle adopt.\footnote{An alternative explanation for the null KS test on applicant TTC scores is that individuals applied everywhere. If this were true, we would expect to see most candidates make multiple applications, and a rejection of the null in a KS test on \emph{placed recruits'} TTC scores (if the supply-side response occurred at  acceptance rather than application). We do not see either in the data.} This suggests that there is not an obvious demand-side policy alternative to contractually induced supply-side selection.

\paragraph{Effort margin}  To recap from Section \ref{ss:ExperiencedP4P}, we find evidence of a positive impact of experienced P4P on student learning, which is considerably larger (almost tripling in magnitude) in recruits' second year on the job. In light of Section \ref{ss:dynamics}, we draw the following conclusions from these results. 

The additional learning achieved by recruits working under P4P, relative to recruits working under FW, is unlikely to be due to selection-out---the compositional margin famously highlighted by \citet{Laz00aer}. Within-year teacher turnover was limited by design. Between-year turnover did happen but cannot explain the experienced P4P effect. In Online Appendix D, we show that the rank correlation between recruits' baseline Grading Task IRT score and their teacher value added is positive. However, in Section \ref{ss:dynamics} we reported that, if anything, selection-out on baseline teacher skill runs the wrong way to explain the experienced P4P effect. 

Neither is the experienced P4P effect likely to be due to within-teacher changes in skill or motivation. We find no evidence that recruits working under P4P made greater gains on the Grading Task from baseline to endline than did recruits working under FW. As already noted, recruits' Dictator Game share sent is not a good predictor of teacher value added. But even if it were, we find no evidence that recruits working under P4P contributed more from baseline to endline than did recruits working under FW, if anything the reverse.  

Instead, the experienced P4P effect is most plausibly driven by teacher effort. This conclusion follows  from the arguments above and the direct evidence that recruits working under P4P provided greater inputs than did recruits working under FW. Specifically, the P4P contract encouraged recruits to be present in school more often and to use better pedagogy in the classroom, behaviors that were incentivized components of the 4P performance metric.

\paragraph{Total effect} The total effect of the P4P contract combines both the advertised and experienced impacts: $\tau_A + \tau_E$. By the second year of the study, the within-year total effect of P4P is $0.04 + 0.16= 0.20$ standard deviations of pupil learning, which is statistically significant at the one percent level.
Roughly four fifths of the total effect can thus be attributed to increased teacher effort, while the remainder arises from supply-side selection during recruitment. At a minimum, our results suggest that in relation to positive effort-margin effects, fears of pay-for-performance causing motivational crowd-out among new public-sector employees may be overstated.  

Our estimates raise the question of why this effect is so much stronger in Year 2 compared to Year 1, particularly on the effort margin. One interpretation is that this is because it takes time for recruits to settle into the job and for the signal to noise ratio in our student learning measures to improve \citep{StaRoc10jep}. Consistent with this interpretation, we note that the impact of experienced P4P on incumbents did not increase in the second year.  
This interpretation suggests that Year 2 effects are the best available estimates of longer-term impacts.

\section{Conclusion}\label{s:conclusion}

This two-tier, two-year, randomized controlled trial featuring extensive data on teachers---their skills and motivations before starting work, multiple dimensions of their on-the-job performance, and whether they left their jobs---offers new insights into the compositional and effort margins of pay-for-performance. We found that potential applicants were aware of, and responded to, the first-tier labor market intervention. This supply-side response to advertised P4P was, if anything, beneficial for student learning. We also found a positive impact of experienced P4P that appears to stem from increased teacher effort, rather than selection-out or changes in measured skill or intrinsic motivation. 

Given these encouraging results, it is natural to ask whether it would be feasible and cost effective to implement this P4P contract at scale. We worked closely with the government to design a contract that was contextually feasible and well-grounded in theory. A composite P4P metric was used to avoid narrowly emphasizing any single aspect of teacher performance and, when measuring learning, we followed the pay-for-percentile approach that aims to give all teachers a fair chance, regardless of the composition of the students they teach. We also took care to ensure that the P4P contract, if successful, could be built into the growth path of teacher wages. While a larger bonus might have elicited stronger impacts, the expected value of the P4P bonus was set at three percent of teacher salaries to be commensurate with annual teacher salary increments (and discretionary pay in other sectors under Rwanda's \emph{imihigo} system of performance contracts for civil servants). 

The fact that we compared a P4P contract with an expenditure-equivalent fixed wage alternative that is equal in magnitude to annual teacher salary increments means that it is reasonable to think about cost effectiveness primarily in terms of measurement. For pupil learning, the minimum requirement for the P4P contract we study is a system of repeated annual assessments across grades and key subjects.\footnote{Such a system may  soon be a part of Rwanda's `comprehensive assessment' program. See, e.g., https://www.newtimes.co.rw/opinions/mineducs-new-guide-student-assessment-triggers-debate.} Measurement of the other aspects of performance---teacher presence, preparation, and pedagogy---can in principle be conducted by head teachers or district staff (who are increasingly being asked to monitor teacher performance) at modest cost.     

There are nonetheless limitations of our work. Inasmuch as the impacts on either the compositional or effort margin might differ after five or ten years, there is certainly scope for further study of this topic in low- and middle-income countries. For instance, it would be interesting to explore whether long-term P4P commitments influence early-career decisions to train as a teacher; our study restricts attention to employment choices by individuals who have already received TTC degrees. 

Another set of issues relate to unintended consequences of pay-for-performance. We found that advertised P4P attracted teachers with lower intrinsic motivation, as measured by the share sent in the framed baseline Dictator Game. It is possible that the students taught by these more self-regarding teachers became more self-regarding themselves or otherwise developed different soft skills. We also found that experienced P4P improved performance on three of the four incentivized dimensions of the composite metric: teacher presence and pedagogy, and pupil learning. It is conceivable that the students in P4P schools may have been impacted by `multi-tasking' as teachers focused on these dimensions to the detriment of others. Since we did not measure aspects of student development beyond test score gains, it would be interesting to explore these issues in future work.

Rwanda's labor market has a characteristic that is unusual for low- and middle-income countries: it has no public sector pay premium, and consequently many of those qualified to teach choose not to, making it more similar to high- income country labor markets in this regard. Whether the positive effects we find in Rwanda of a multidimensional, pay-for-percentile contract---improving performance without dampening employee satisfaction---will generalize to settings where public-sector wage premiums differ remains an open question, for the education sector and beyond.

\cleardoublepage 

\bibliography{STARSbib}

%--------------------------------------------------------------------------%
%--  APPENDIX MATERIALS BEGIN HERE --%
%--------------------------------------------------------------------------%

\cleardoublepage
\appendix
\noappendicestocpagenum \addappheadtotoc
% Below, adding 'Appendix' to section header but NOT to TOC entry or to calls to the section label in text.
\makeatletter
\def\@seccntformat#1{Appendix\ \csname the#1\endcsname\quad}
\def\@subseccntformat#1{\csname the#1\endcsname\quad}
\makeatother
%%  And adding Appendix counter to table and figure numbers
%\setcounter{table}{0} \setcounter{figure}{0}
\renewcommand\thetable{\Alph{section}.\arabic{table}}
\renewcommand\thefigure{\Alph{section}.\arabic{figure}}
\renewcommand{\thepage}{\Alph{section}.\arabic{page}}

\counterwithin{figure}{section}
\counterwithin{table}{section}
\counterwithin{page}{section}

%--------------------------------------------------------------------------%
% Supplemental figures and tables
%--------------------------------------------------------------------------%
\cleardoublepage
\begin{center}
\Large{\textbf{Online Appendix}}\\
\vskip2ex
\Large{Recruitment, effort, and retention effects of performance contracts for civil servants: Experimental evidence from Rwandan primary schools} 
\vskip3ex
\large{Clare Leaver, Owen Ozier, Pieter Serneels, and Andrew Zeitlin}
\end{center}    

\vfill
\thispagestyle{empty}
\pagebreak

\section{Supplemental figures and tables }\label{app:figures}
\setcounter{page}{1} 

\begin{figure}[!hb]
\caption{Study profile}\label{f:StudyProfile}
\centering 
\forestset{
    qtree/.style={
        baseline,
        for tree={
            draw, 
            parent anchor=south,
            child anchor=north,
            align=center,
            inner sep=2pt,
            forked edge, 
            edge=->,
            font=\footnotesize
        }%
    }%
}

\begin{forest}
qtree,
[{$~$\textbf{Study sample definition} $~$\\ \scriptsize{6 Districts} \\ \scriptsize{18 Labor markets enrolled} }, name=districts
    [{$~$\textbf{Randomization of labor markets to advertised contracts}$~$} 
        [Advertised P4P,name=P4P1]
        [Advertised FW,name=FW1 
            [{$~$\textbf{Applications placed at District Education Offices}$~$ \\ %
            \scriptsize{1,962 qualified applications} \\  %  1,963 qualified applications
            }  %
            ,name=apps % 
                [$~$\textbf{Teachers placed into schools and assigned to classes}$~$  
                    [ {$~$\textbf{Baseline schools enrolled} $~$\\ \scriptsize{164 schools enrolled in study} % \scriptsize{164 schools enrolled in study} %
                        }
                        [{$~$\textbf{Randomization of schools to experienced contracts}$~$}
                            [{\textbf{Experienced P4P contracts} 
                            \\ \scriptsize{85 schools } %{85 schools} %
                            \\ \scriptsize{176 new recruits at baseline (131 upper primary)} % {176 new recruits at baseline (134 upper primary)} %
                            \\ \scriptsize{1,608 incumbent and other teachers at baseline} % {1,608 incumbent and other teachers at baseline} 
                            \\ \scriptsize{(657 upper primary of these 1,608)} % {(682 upper primary of these 1,608)} 
                            \\ \scriptsize{7,229 pupils assessed} % {7,229 pupils assessed} %
                                }%
                                ,name=P4P2
                                [{$~$\textbf{Year 1 teacher inputs measured}$~$ \\ %
                                    \scriptsize{Presence, preparation, pedagogy}%
                                    },tier=align y1inputs
                                    [{\textbf{Year 1 endline} %
                                    \\  \scriptsize{7,495 pupils assessed} % {7,495 pupils assessed} %
                                    },tier=align y1endline
                                        [{$~$\textbf{Year 2 teacher inputs measured}$~$}, tier=align y2inputs
                                        [{\textbf{Year 2 endline} %
                                            \\ \scriptsize{8,910 pupils assessed} % {8,910 pupils assessed} %
                                            }, tier=agn y2endline]
                                        ]
                                    ]
                                ]
                            ]
                            [{\textbf{Experienced FW contracts} %
                                \\ \scriptsize{79 schools} % 79 schools
                                \\ \scriptsize{153 new recruits at baseline (125 upper primary)} % {153 new recruits at baseline (126 upper primary)} 
                                \\ \scriptsize{1,459 incumbent and other teachers at baseline} % {1,459 incumbent and other teachers at baseline} 
                                \\ \scriptsize{(595 upper primary of these 1,459)} % {(618 upper primary of these 1,459)} 
                                \\ \scriptsize{6,602 pupils assessed} % {6,602 pupils assessed} %
                                }
                                    [{\textbf{Year 1 endline} %
                                        \\ \scriptsize{6,815 pupils assessed} % {6,815 pupils assessed}%
                                        } ,tier=align y1endline
                                        [{$~$\textbf{Year 2 teacher inputs measured}$~$}, tier=align y2inputs
                                        [{\textbf{Year 2 endline} %
                                            \\ \scriptsize{7,964 pupils assessed} % {7,964 pupils assessed} %
                                            }, tier=agn y2endline]
                                        ]
                                    ]
                            ]
                        ]
                    ]
                ]
            ]
        ]
        [Advertised mixed, name=M1]
    ]
]
\draw[->] (P4P1) to (apps) ;
\draw[->] (FW1) to (apps) ;
\draw[->] (M1) to (apps) ;
\end{forest}
\end{figure}

\begin{figure}[!hbtp]
\caption{Probability of hiring as a function of TTC score, by advertised treatment arm}
\label{f:HiringRule}
\begin{center}
  \includegraphics[width=0.5\textwidth]{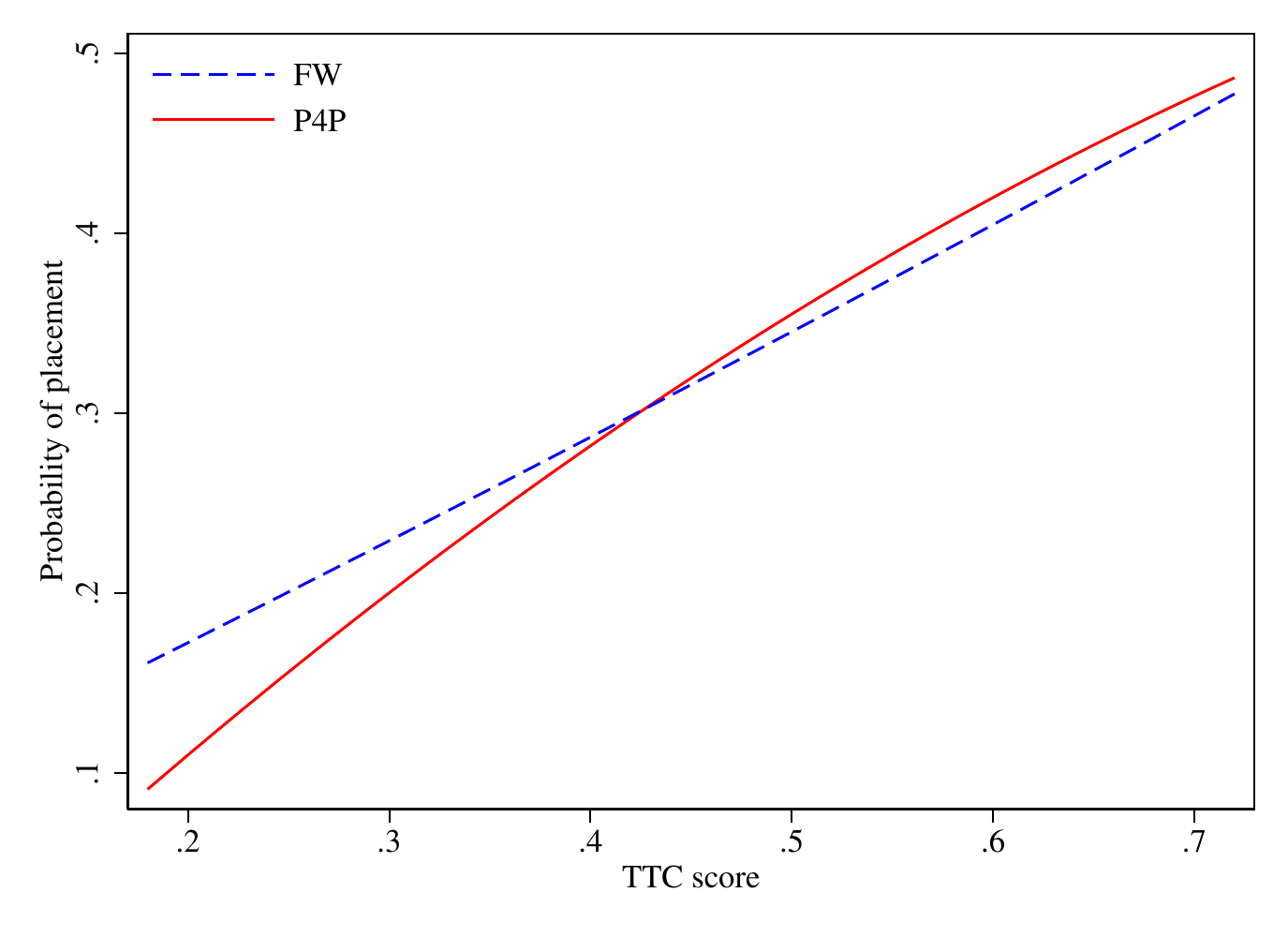}
\end{center}

\begin{footnotesize}
\floatfoot{
\emph{Note}:  The figure illustrates estimated hiring probability as a (quadratic) function of the rank of an applicant's TTC final exam score within the set of applicants in their district.
}
\end{footnotesize}

\end{figure}

\cleardoublepage

\begin{landscape}
\begin{footnotesize}
\begin{singlespace} 
	\begin{longtable}{$p{0.25\textwidth} ^p{0.5\textwidth} ^p{0.25\textwidth} ^p{0.15\textwidth}} %
 	%  Header for first page, including caption.
    \caption{Summary of hypotheses, outcomes, samples, and specifications}
    \label{t:specsummary}
    \\
 	\toprule 
	Outcome & Sample & Test statistic & Randomization inference \\
	\midrule %
    \endfirsthead

    \multicolumn{4}{c}{Table \ref{t:specsummary}, continued}
    \\
 	\toprule 
	Outcome & Sample & Test statistic & Randomization inference \\
	\midrule %
	\endhead

	%  Footer for pages that are not last
	\\
	\midrule
    \multicolumn{4}{r}{\emph{Continues\ldots}}
    \endfoot

    % Footer for final page
    \\ 
    \bottomrule
	\multicolumn{4}{p{1.2\textwidth}}{\footnotesize{\emph{Note}: Primary tests of each family of hypotheses appear first, preceded by a superscript $^*$; those that appear subsequently under each family without the superscript $^*$ are secondary hypotheses. Under inference, $\mathcal{T}^A$ refers to randomization inference involving the permutation of the \emph{advertised} contractual status of the recruit \emph{only}; $\mathcal{T}^E$ refers to randomization inference that includes the permutation of the \emph{experienced} contractual status of the school; $\mathcal{T}^A \times \mathcal{T}^E$ indicates that randomization inference will permute both treatment vectors to determine a distribution for the relevant test statistic. Test statistic is a studentized coefficient or studentized sum of coefficients (a $t$ statistic), except where otherwise noted (as in Hypothesis I); in linear mixed effects estimates of equation \eqref{eq:assessments_pooled} and \eqref{eq:assessments_interacted}, which are estimated by maximum likelihood, this is a $z$ rather than $t$ statistic, but we maintain notation to avoid confusion with the test score outcome, $z_{jbksr}$.}}
	\endlastfoot 

	%  Body text
	\\
	\multicolumn{4}{l}{\textsc{Hypothesis I: Advertised P4P induces differential application qualities}} \\
	% \rowcolor{LightCyan} \rowstyle{\itshape}
	$^*$TTC exam scores & Universe of applications & KS test of eq. \eqref{eq:KSstat} & $\mathcal{T}^A$  \\
	District exam scores & Universe of applications & KS test of eq. \eqref{eq:KSstat} & $\mathcal{T}^A$ \\ 
	TTC exam scores & Universe of applications &  $t_A$ in eq. \eqref{eq:ttest_applicants} &  $\mathcal{T}^A$  \\
	TTC exam scores & Applicants in the top $\hat{H}$ number of applicants, where $\hat{H}$ is the predicted number of hires based on subject and district, estimated off of FW applicant pools &  $t_A$ in eq. \eqref{eq:ttest_applicants} & $\mathcal{T}^A$ \\
    TTC exam scores & Universe of application, weighted by probability of placement &  $t_A$ in eq. \eqref{eq:ttest_applicants} &  $\mathcal{T}^A$  \\
	Number of applicants & Universe of applications & $t_A$ in eq. \eqref{eq:n_applicants} & $\mathcal{T}^A$ \\
	\\ 
	\multicolumn{4}{l}{\textsc{Hypothesis II: Advertised P4P affects the observable skills of placed recruits in schools}} \\
	% \rowcolor{LightCyan} % \rowstyle{\itshape}
    $^*$Teacher skills assessment IRT model EB score & Placed recruits &  $t_A$ in eq. \eqref{eq:baseline_chars} &  $\mathcal{T}^A$  \\
	%$^*$TTC exam scores & Placed recruits &  Coefficient $\tau_A$ in eq. \eqref{eq:placed_skill} &  $\mathcal{T}^A$ \\
	\\
	\multicolumn{4}{l}{\textsc{Hypothesis III: Advertised P4P induces differentially  `intrinsically' motivated recruits to be placed in schools}} \\
	% \rowcolor{LightCyan} % \rowstyle{\itshape}
    $^*$Dictator-game donations & Placed recruits &  $t_A$ in eq. \eqref{eq:baseline_chars} & $\mathcal{T}^A$ \\ 
	Perry PSM instrument & Placed recruits retained through Year 2 &  $t_A$ in eq. \eqref{eq:baseline_chars} & $\mathcal{T}^A$ \\
	% $^*$Taste for competition (net of prediction based on teacher skill) & Placed recruits at baseline & As above \\
	% $^*$ Risk aversion \\ 
	% $^*$ Overconfidence & Placed recruits at baseline & As above \\
	% $^*$ Locus of control \\
	% $^*$ Self esteem \\
	% $^*$ Extroversion? \\
	% $^*$ Gender \\ 
	\\
	\multicolumn{4}{l}{\textsc{Hypothesis IV:  Advertised P4P induces the selection of higher-(or lower-) value-added teachers}} \\
	% \rowcolor{LightCyan} % \rowstyle{\itshape} 
    $^*$Student assessments (IRT EB predictions) & Pooled Year 1 \& Year 2 students & $t_A$ in eq. \eqref{eq:assessments_pooled} & $\mathcal{T}^A$  \\
    Student assessments & Pooled Year 1 \& Year 2  students & \makecell[lt]{$t_A$ and $t_{A+AE}$; \\ $t_{AE}$ in eq. \eqref{eq:assessments_interacted} } & \makecell[lt]{$\mathcal{T}^A$ \\ $\mathcal{T}^A \times \mathcal{T}^E $}\\	
    Student assessments & Year 1 students & $t_A$ in eq. \eqref{eq:assessments_pooled} & $\mathcal{T}^A$  \\ 
	Student assessments & Year 2 students & $t_A$ in eq. \eqref{eq:assessments_pooled} & $\mathcal{T}^A$  \\
	\\ 
	\multicolumn{4}{l}{\textsc{Hypothesis V: Experienced P4P creates incentives which contribute to higher (or lower) teacher value-added}} \\
	$^*$Student assessments (IRT EB predictions) & Pooled Year 1 \& Year 2 students & $t_E$ in eq. \eqref{eq:assessments_pooled} & $T^E$ \\
	Student assessments & Pooled Year 1 \& Year 2  students & \makecell[lt]{$t_E$ and $t_{E+AE}$; \\ $t_{AE}$ in eq. \eqref{eq:assessments_interacted} } & \makecell[lt]{$\mathcal{T}^E$ \\ $\mathcal{T}^A \times \mathcal{T}^E $}\\	%
	Student assessments & Year 1 students & $t_E$ in eq. \eqref{eq:assessments_pooled} & $\mathcal{T}^E$	\\
	Student assessments & Year 2 students &  $t_E$ in eq. \eqref{eq:assessments_pooled} &  $\mathcal{T}^E$	\\
	\\
	\multicolumn{4}{l}{\textsc{Hypothesis VI: Selection and incentive effects are apparent in the 4P performance metric}} \\ % P4P contracts attract individuals who are more responsive to P4P contracts}} \\
	% \rowcolor{LightCyan} % \rowstyle{\itshape} 
    $^*$Composite 4P metric & Teachers, pooled Year 1 (experienced P4P only) \&  Year 2  & $t_A$ in eq. \eqref{eq:spec_inputs} & $\mathcal{T}^A$ \\
    Composite 4P metric & Teachers, pooled Year 1 (experienced P4P only) \&  Year 2 & 
    \makecell[lt]{$t_A$ and $t_{A+AE}$; \\
    $t_E$ and $t_{E+AE}$; \\
    $t_{AE}$ in eq. \eqref{eq:spec_inputs_split}} & \makecell[lt]{$\mathcal{T}^A$ \\ $\mathcal{T}^E$\\ $\mathcal{T}^A \times \mathcal{T}^E $}\\	%
	Barlevy-Neal rank & As above &  &  \\
	Teacher attendance & As above & & \\
	Classroom observation & As above &  & \\ 
	Lesson plan (indicator) & As above &  & \\ 
	\\
	\end{longtable}

\end{singlespace} 
\end{footnotesize}
\end{landscape}

\begin{table}[!hbpt]
\caption{Measures of teacher inputs in P4P schools}
\label{t:teacherinputs}

\begin{footnotesize}
\begin{center} 
\begin{tabular}{l c c c}
\toprule
 & Mean  & St~Dev  & Obs  \\  
\midrule 
\addlinespace[2ex] 
\multicolumn{4}{l}{\textbf{Year 1, Round 1}} \\
Teacher present  & 0.97 & (0.18) & 640 \\  
Has lesson plan  & 0.53 & (0.50) & 569 \\  
Classroom observation: Overall score  & 2.01 & (0.40) & 631 \\  
\hspace{0.5em}Lesson objective  & 2.00 & (0.71) & 631 \\  
\hspace{0.5em}Teaching activities  & 1.94 & (0.47) & 631 \\  
\hspace{0.5em}Use of assessment  & 1.98 & (0.50) & 629 \\  
\hspace{0.5em}Student engagement  & 2.12 & (0.56) & 631 \\  
\addlinespace[2ex] 
\multicolumn{4}{l}{\textbf{Year 1, Round 2}} \\
Teacher present  & 0.97 & (0.18) & 629 \\  
Has lesson plan  & 0.53 & (0.50) & 587 \\  
Classroom observation: Overall score  & 2.27 & (0.41) & 628 \\  
\hspace{0.5em}Lesson objective  & 2.22 & (0.76) & 627 \\  
\hspace{0.5em}Teaching activities  & 2.18 & (0.46) & 627 \\  
\hspace{0.5em}Use of assessment  & 2.23 & (0.48) & 627 \\  
\hspace{0.5em}Student engagement  & 2.46 & (0.49) & 628 \\  
\addlinespace[2ex] 
\multicolumn{4}{l}{\textbf{Year 2, Round 1}} \\
Teacher present  & 0.91 & (0.29) & 675 \\  
Has lesson plan  & 0.79 & (0.41) & 568 \\  
Classroom observation: Overall score  & 2.37 & (0.34) & 520 \\  
\hspace{0.5em}Lesson objective  & 2.45 & (0.68) & 520 \\  
\hspace{0.5em}Teaching activities  & 2.28 & (0.43) & 518 \\  
\hspace{0.5em}Use of assessment  & 2.25 & (0.47) & 519 \\  
\hspace{0.5em}Student engagement  & 2.49 & (0.45) & 520 \\  
\bottomrule
\end{tabular}

\end{center}
\floatfoot{
\emph{Note}:  Descriptive statistics for upper-primary teachers only. Overall score for the classroom observation is the average of four components: lesson objective, teaching activities, use of assessment, and student engagement, with each component scored on a scale from 0 to 3.
}
\end{footnotesize}

\end{table}

\begin{landscape}
\begin{table}[!hbpt]
\caption{Impacts of advertised P4P on characteristics of placed recruits}\label{t:selection_nonPAP}

\begin{footnotesize}
\begin{center}
\begin{tabular}{p{0.1\linewidth}*{6}{C{0.15\linewidth}}}
\toprule
 & \multicolumn{2}{c}{Primary outcomes} & \multicolumn{4}{c}{Exploratory outcomes} \\ 
\cmidrule(lr){2-3} \cmidrule(lr){4-7} 
    & Teacher skills & DG contribution & Age & Female & Risk aversion & Big Five \\ 
\midrule
\multirow{3}{0.9\linewidth}{Advertised P4P}    &  -0.184  &  -0.100  &  -0.161  &  0.095  &  0.010  &  -0.007 \\ 
  &  [-0.836, 0.265]  &  [-0.160, -0.022]  &  [-1.648, 1.236]  &  [-0.151, 0.255]  &  [-0.125, 0.208]  &  [-0.270, 0.310] \\ 
  &  (0.367)  &  (0.029)  &  (0.782)  &  (0.325)  &  (0.859)  &  (0.951)  \\[2ex] 
Observations  &     242  &     242  &     242  &     242  &     242  &     241  \\  
\bottomrule
\end{tabular}

\end{center}
\floatfoot{
\emph{Note}:  The table reports the point estimate of $\tau_A$, together with the 95 percent confidence interval in brackets, and the randomization inference $p$-value in parentheses, from the specification in equation \eqref{eq:baseline_chars}. The primary  outcomes are described in detail in Section \ref{ss:data_teachers}. In the third column, the outcome is placed recruit age, measured in years. In the fourth column, the outcome is coded to 1 for female recruits and 0 for males. In the fifth column, the outcome is a binary measure of risk aversion constructed from placed recruits' responses in a hypothetical lottery choice game \citepApp{EckGross08handb,ChetEcketal10jru}. It is coded to 1 when the respondent chooses either of the two riskiest of the five available lotteries, and 0 otherwise (53 percent of the sample make one of these choices). In the final column, the outcome is an index of the Big Five personality traits constructed from the 15 item version, validated by \citetApp{LangJohnetalBehRes2011} and following \citetApp{DohmenFalk10EJ}.
}
\end{footnotesize}
\end{table}
\end{landscape}

\begin{table}[!hbpt]
\caption{Impacts on student learning, OLS model}\label{t:learning_ols}

\begin{footnotesize}
\begin{center}
\begin{tabular}{lccc}
\toprule
    & \multicolumn{1}{c}{Pooled} & \multicolumn{1}{c}{Year 1} & \multicolumn{1}{c}{Year 2} \\ 
\midrule
\multicolumn{4}{l}{\emph{Model A: Direct effects only}} \\ [2ex] 
\multirow[t]{3}{0.25\linewidth}{Advertised P4P ($\tau_A$)} 
    & 0.03 
    & -0.03  
    & 0.08 \\
	& [-0.04, 0.14] 
	& [-0.10, 0.08] 
	& [-0.03, 0.24] \\
	& (0.37) %
	& (0.51) %
	& (0.10)  \\[2ex]
\multirow[t]{3}{0.25\linewidth}{Experienced P4P ($\tau_E$)} 
    & 0.13 
    & 0.10 
    & 0.17 \\ 
	& [0.03, 0.24] 
	& [0.00, 0.20] 
	& [0.04, 0.32] \\
	& (0.01)  %
	& (0.05)  %
	& (0.02)  \\[2ex]
\multirow[t]{3}{0.25\linewidth}{Experienced P4P $\times$ Incumbent ($\lambda_E$)} %
	& -0.09 %
	& -0.10 %
	& -0.09 \\ 
	& [-0.31, 0.15] 
	& [-0.32, 0.16] 
	& [-0.34, 0.16] \\
	& (0.44) %
	& (0.40) %
	& (0.48) \\[2ex]
\multicolumn{4}{l}{\emph{Model B:  Interactions between advertised and experienced contracts}} \\ [2ex] 
\multirow[t]{3}{0.25\linewidth}{Advertised P4P ($\tau_A$)} %
	& 0.04 %
	& -0.03  %
	& 0.12 \\
	& [-0.07, 0.23] %
	& [-0.14, 0.13] %
	& [-0.03, 0.33] \\
	& (0.41) %
	& (0.59) %
	& (0.10) \\[2ex]
\multirow[t]{3}{0.25\linewidth}{Experienced P4P ($\tau_E$)} 
    & 0.14 %
    & 0.10  %
    & 0.17 \\ 
	& [0.03, 0.26] %
	& [-0.02, 0.22] %
	& [0.02, 0.35] \\
	& (0.01) %
	& (0.11) %
	& (0.03) \\[2ex]
\multirow[t]{3}{0.25\linewidth}{Advertised P4P $\times$ Experienced P4P ($\tau_{AE}$)} %
	& -0.03 % 
	& 0.01 %
	& -0.06 \\ 
	& [-0.22, 0.17] %
	& [-0.18, 0.21] %
	& [-0.32, 0.18] \\
	& (0.72) %
	& (0.97) %
	& (0.60) \\[2ex]
\multirow[t]{3}{0.25\linewidth}{Experienced P4P $\times$ Incumbent ($\lambda_E$)} %
	& -0.09 %
	& -0.09 %
	& -0.09  \\ 
	& [-0.52, 0.36] %
	& [-0.47, 0.40] %
	& [-0.56, 0.51] \\
	& (0.62) %
	& (0.62) %
	& (0.68) \\[2ex]
Observations %
	& 154594 %
	&  70821 %
	&  83773 \\
\bottomrule
\end{tabular}
\end{center}
\floatfoot{
\emph{Note}:  For each estimated parameter, or combination of parameters, the table reports the point estimate (stated in standard deviations of student learning), 95 percent confidence interval in brackets, and $p$-value in parentheses. Randomization inference is conducted on the associated $t$ statistic. The measure of student learning is based on the empirical Bayes estimate of student ability from a two-parameter IRT model, as described in Section \ref{ss:data_assessments}.
}
\end{footnotesize}
\end{table}

\begin{table}[!hbpt]
\caption{Teacher endline survey responses}\label{t:endline}
\begin{footnotesize}
\begin{center}
\begin{tabular}{lcccc}
\toprule
    & Job satisfaction & Likelihood of leaving & Positive affect & Negative affect \\ 
\midrule
\multicolumn{5}{l}{\emph{Model A: Direct effects only}} \\ [2ex] 
Advertised P4P    &  -0.04  &  -0.07  &  -0.06  &  -0.02 \\ 
  &  [-0.41, 0.48]  &  [-0.27, 0.08]  &  [-0.44, 0.33]  &  [-0.29, 0.32] \\ 
  &  (0.82)  &  (0.36)  &  (0.74)  &  (0.86)  \\[2ex] 
Experienced P4P    &  0.05  &  -0.06  &  -0.00  &  0.09 \\ 
  &  [-0.25, 0.36]  &  [-0.18, 0.06]  &  [-0.28, 0.28]  &  [-0.14, 0.33] \\ 
  &  (0.72)  &  (0.39)  &  (0.99)  &  (0.47)  \\[2ex] 
Experienced P4P $\times$ Incumbent    &  -0.00  &  0.04  &  0.04  &  -0.07 \\ 
  &  [-0.45, 0.48]  &  [-0.13, 0.21]  &  [-0.45, 0.52]  &  [-0.50, 0.37] \\ 
  &  (0.99)  &  (0.61)  &  (0.84)  &  (0.70)  \\[2ex] 
\multicolumn{5}{l}{\emph{Model B:  Interactions between advertised and experienced contracts}} \\ [2ex] 
Advertised P4P    &  -0.10  &  -0.01  &  0.02  &  -0.33 \\ 
  &  [-0.57, 0.55]  &  [-0.26, 0.18]  &  [-0.52, 0.44]  &  [-0.75, 0.30] \\ 
  &  (0.67)  &  (0.93)  &  (0.89)  &  (0.20)  \\[2ex] 
Experienced P4P    &  0.08  &  -0.07  &  -0.02  &  -0.25 \\ 
  &  [-0.42, 0.54]  &  [-0.27, 0.14]  &  [-0.56, 0.47]  &  [-0.67, 0.17] \\ 
  &  (0.75)  &  (0.50)  &  (0.93)  &  (0.23)  \\[2ex] 
Advertised P4P $\times$ Experienced P4P    &  0.13  &  -0.13  &  -0.16  &  0.64 \\ 
  &  [-0.66, 0.85]  &  [-0.42, 0.14]  &  [-0.81, 0.43]  &  [0.04, 1.28] \\ 
  &  (0.71)  &  (0.34)  &  (0.59)  &  (0.03)  \\[2ex] 
Experienced P4P $\times$ Incumbent    &  -0.03  &  0.05  &  0.06  &  0.27 \\ 
  &  [-0.90, 0.90]  &  [-0.28, 0.37]  &  [-0.86, 0.90]  &  [-0.54, 1.09] \\ 
  &  (0.92)  &  (0.69)  &  (0.84)  &  (0.40)  \\[2ex] 
Observations  &    1483  &    1492  &    1474  &    1447  \\  
FW recruit mean (SD)    &  5.42  &  0.26  &  0.31  &  0.00 \\ 
     &  (0.90)  &  (0.44)  &  (0.93)  &  (0.99) \\ 
FW incumbent mean (SD)    &  5.26  &  0.29  &  -0.05  &  0.00 \\ 
     &  (1.10)  &  (0.46)  &  (1.00)  &  (1.04) \\ 
\bottomrule
\end{tabular}

\end{center}
\floatfoot{
\emph{Note}: For each estimated parameter, or combination of parameters, the table reports the point estimate (stated in standard deviations of student learning), 95 percent confidence interval in brackets, and $p$-value in parentheses. Randomization inference is conducted on the associated $t$ statistic. Outcomes are constructed as follows: \emph{job satisfaction} is scored on a 7-point scale with higher numbers representing greater satisfaction; \emph{likelihood of leaving} is a binary indicator coded to 1 if the teacher responds that they are likely or very likely to leave their job at the current school over the coming year; \emph{positive affect} and \emph{negative affect} are standardized indices derived from responses on a 5-point Likert scale.
}
\end{footnotesize}
\end{table}

\begin{table}
\caption{Teacher attitudes toward pay-for-performance at endline}\label{t:p4p_opinions}
\begin{footnotesize}
\begin{center}
\begin{tabular}{l *{5}{p{0.1\textwidth}}}
\toprule
 & Very unfavorable  & Somewhat unfavorable  & Neutral  & Somewhat favorable  & Very favorable  \\
\midrule
Recruits applying under FW (64) & 4.7\% & 4.7\% & 7.8\% & 10.9\% & 71.9\% \\
---Experiencing FW (33) & 6.1\% & 9.1\% & 9.1\% & 3.0\% & 72.7\% \\
---Experiencing P4P (31) & 3.2\% & 0.0\% & 6.5\% & 19.4\% & 71.0\% \\[2ex]
Recruits applying under P4P (60) & 5.0\% & 3.3\% & 8.3\% & 1.7\% & 81.7\% \\
---Experiencing FW (32) & 6.2\% & 0.0\% & 6.2\% & 0.0\% & 87.5\% \\
---Experiencing P4P (28) & 3.6\% & 7.1\% & 10.7\% & 3.6\% & 75.0\% \\[2ex]
Incumbent teachers (1,113) & 5.0\% & 7.5\% & 7.2\% & 9.9\% & 70.4\% \\
---Experiencing FW (537) & 5.2\% & 8.6\% & 8.0\% & 8.6\% & 69.6\% \\
---Experiencing P4P (576) & 4.9\% & 6.6\% & 6.4\% & 11.1\% & 71.0\% \\
\bottomrule
\end{tabular}

\end{center}
\floatfoot{
\emph{Note}: The table reports the distribution of answers to the following question on the endline teacher survey: ``What is your overall opinion about the idea of providing high-performing teachers with bonus payments on the basis of objective measures of student performance improvement?'' Figures in parentheses give the number of respondents in each treatment category.  
}
\end{footnotesize}
\end{table}

%--------------------------------------------------------------------------%
% THEORY
%--------------------------------------------------------------------------%
\clearpage 

\onehalfspacing 

\section{Theory}\label{s:theory}
\setcounter{page}{1} 

This appendix sets out a simple theoretical framework, adapted from \citetApp{LeavLemSCur}, that closely mirrors the experimental design described in Section \ref{s:design}. We used this framework as a device to organize our thinking when choosing what hypotheses to test in our pre-analysis plan. We did not view the framework as a means to deliver sharp predictions for one-tailed tests.

\subsection*{The model}

We focus on an individual who has just completed teacher training, and who must decide whether to apply for a teaching post in a public school, or a job in a generic `outside sector'.\footnote{ \citetApp{LeavLemSCur} focus on a teacher who chooses between three alternatives: (i) accepting an offer of a job in a public school on a fixed wage contract, (ii) declining and applying for a job in a private school on a pay-for-performance contract, and (iii) declining and  applying for a job in an outside sector on a different performance contract.}

\paragraph{Preferences}
The individual is risk neutral and cares about compensation $w$ and effort $e$. Effort costs are sector-specific. The individual's payoff in the education sector is $w - (e^2 - \tau \, e)$, while her payoff in the outside sector is $w - e^2$. The parameter $\tau \geq 0 $ captures the individual's \emph{intrinsic motivation} to teach, and can be thought of as the realization of a random variable. The individual observes her realization $\tau$ perfectly, while (at the time of hiring) employers observe nothing.

\paragraph{Performance metrics} 
Irrespective of where the individual works, her effort generates a performance metric $m = e\, \theta + \e$. The parameter $\theta \geq 1$ is the individual's \emph{ability}, and can also be thought of as the realization of a random variable. The individual observes her realization of $\theta$ perfectly, while (at the time of hiring) employers observe nothing. Draws of the error term $\e$ are made from $U\left[\underline{\e},\overline{\e}\right]$, and are independent across employments.  

\paragraph{Compensation schemes}

\noindent Different compensation schemes are available depending on advertised treatment status. In the advertised P4P treatment, individuals choose between: (i) an education contract of the form, $w^G + B$ if $m \geq \overline{m}$, or $w^G$ otherwise; and (ii) an outside option of the form $w^0$ if $m\geq\underline{m}$, or 0 otherwise. In the advertised FW treatment, individuals choose between: (i) an education contract of the form $w^F$; and (ii) the same outside option. In our experiment, the bonus $B$ was valued at RWF 100,000, and the fixed-wage contract exceeded the guaranteed income in the P4P contract by RWF 20,000 (i.e. $w^F-w^G= 20,000$). 

\paragraph{Timing}
The timing of the game is as follows. 

\begin{enumerate}
    \item Outside options and education contract offers are announced.
    \item Nature chooses type $(\tau,\theta)$.
    \item Individuals observe their type $(\tau,\theta)$, and choose which sector to apply to.
    \item Employers hire (at random) from the set of applicants.
    \item \emph{Surprise} re-randomization occurs.
    \item Individuals make effort choice $e$.
    \item Individuals' performance metric $m$ is realized, with $\e \sim U[\underline{\e},\bar{\e}]$.
    \item Compensation paid in line with (experienced) contract offers.
\end{enumerate}

\begin{figure}[!tbp]
\caption{Compensation schemes in the numerical example}\label{f:contracts}
\centering
\includegraphics[width=.55\textwidth]{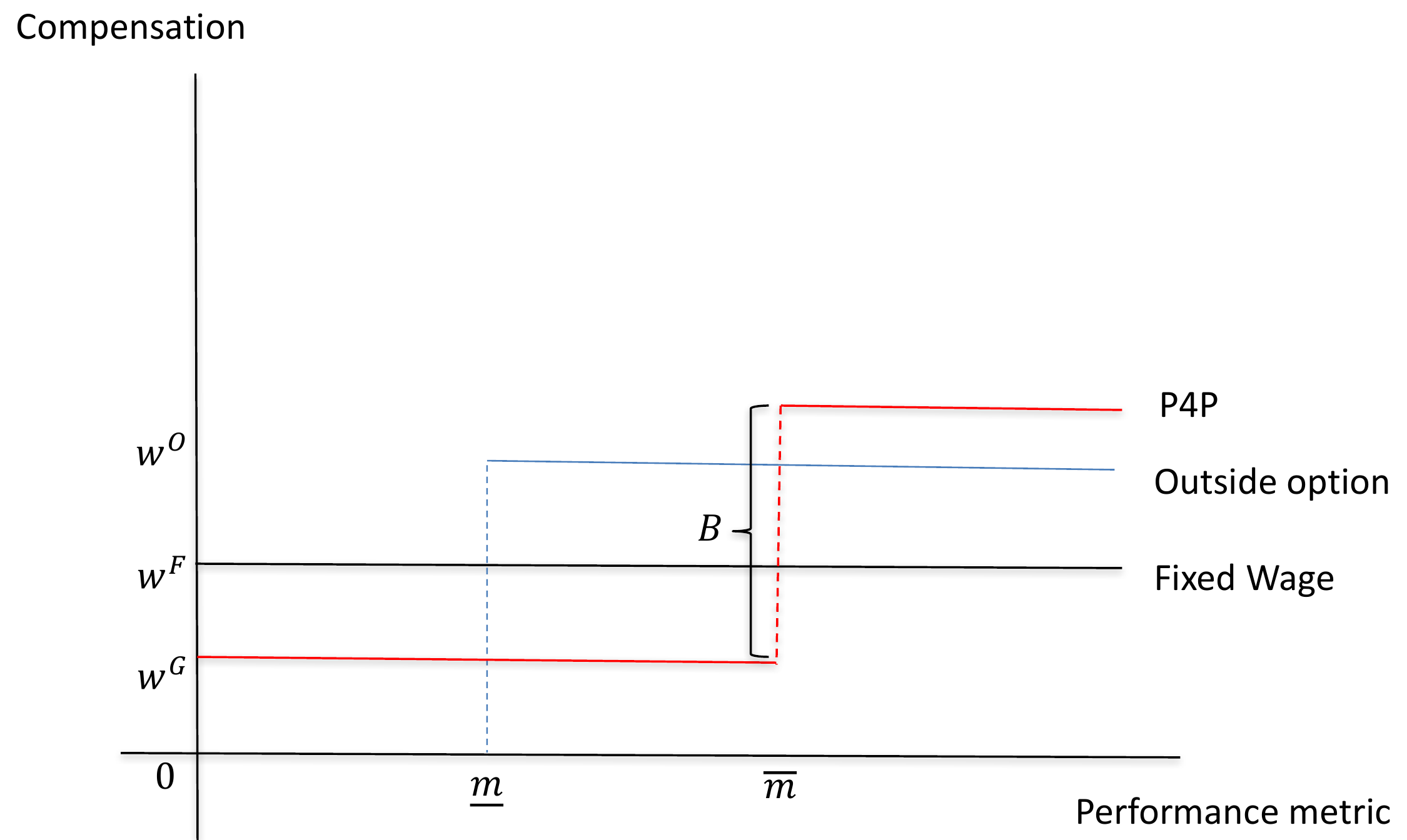}
\end{figure}

\paragraph{Numerical example}
To illustrate how predictions can be made using this framework, we draw on a numerical example. First, in terms of the compensation schemes, we assume that $w^O=50$, $B=40$, $w^G=15$, $\underline{m}=1$, and $\overline{m}=4.5$ (as illustrated in Figure \ref{f:contracts}). These five parameters, together with $\underline{\e} =-5$ and $\overline{\e}=5$, pin down effort and occupational choices by a \emph{given} $(\tau,\theta)$-type. If, in addition, we make assumptions concerning the distributions of $\tau$ and $\theta$, then we can also make statements about the expected intrinsic motivation and expected ability of applicants, and the expected performance of placed recruits. Here, since our objective is primarily pedagogical, we go for the simplest case possible and assume that $\tau$ and $\theta$ are drawn independently from uniform distributions. Specifically, $\tau$ is drawn from $U\left[0,10\right]$, and $\theta$ is drawn from $U\left[1,5\right]$. 

%------------------------------------------------------------------%
%------------------------------------------------------------------%

\subsection*{Analysis}

As usual, we solve backwards, starting with effort choices.

\paragraph{Effort incentives}

Effort choices under the three compensation schemes are:
\begin{align*}
e^F &=  \tau/2 \\
e^P &=  \frac{\theta \, B}{2(\bar{\e} - \underline{\e})} + \tau/2 \\
e^O &= \frac{\theta \, w^O}{2(\bar{\e} - \underline{\e})},
\end{align*}
where we have used the fact that $\e$ is drawn from a uniform distribution. Intuitively, effort incentives are higher under P4P than under FW, i.e. $e^P>e^F$. 

\begin{figure}
    \caption{Decision rules under alternative contract offer treatments}
    \label{f:selectionexample}
    \centering
    \includegraphics[width=.45\textwidth]{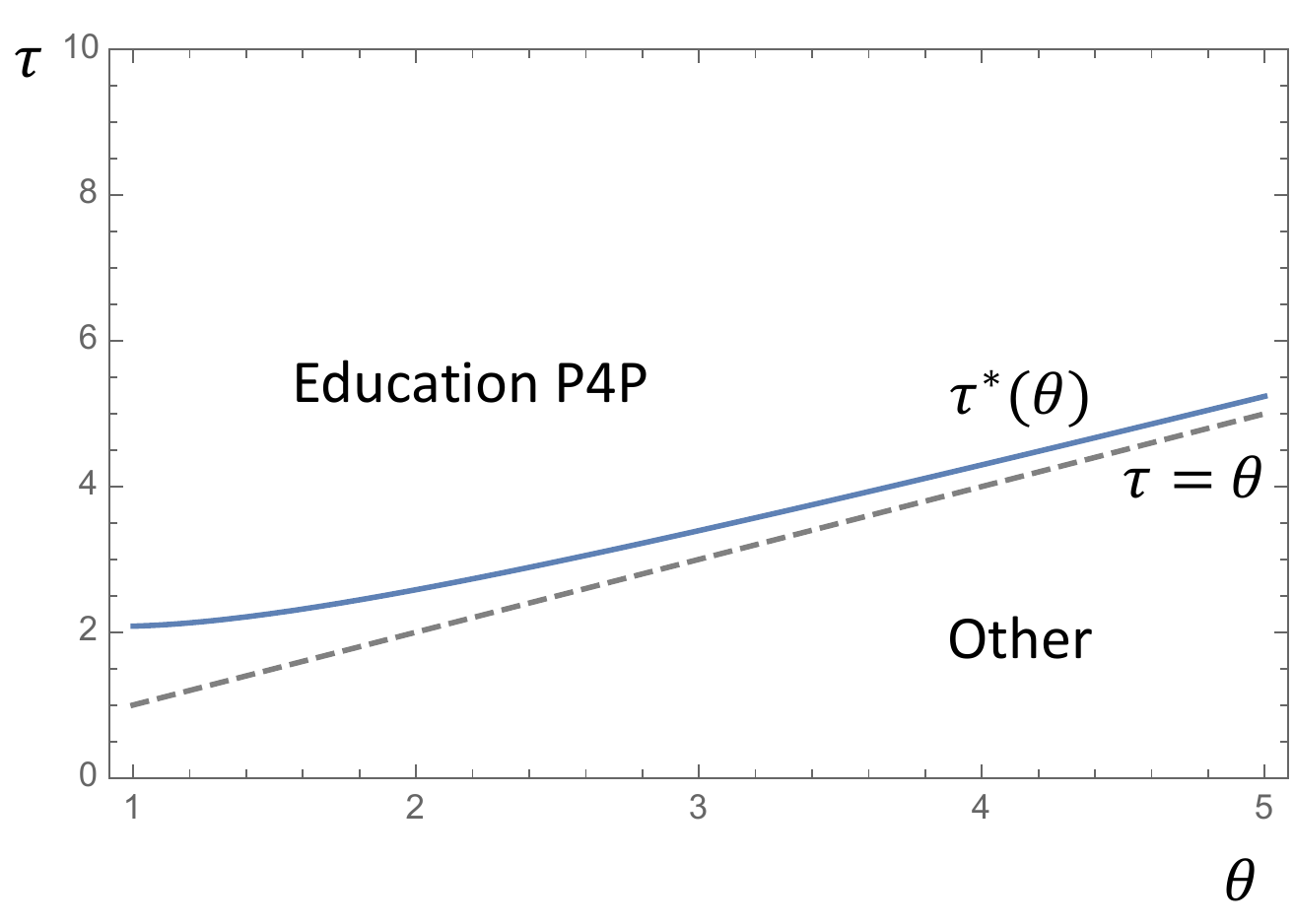}
    \includegraphics[width=.45\textwidth]{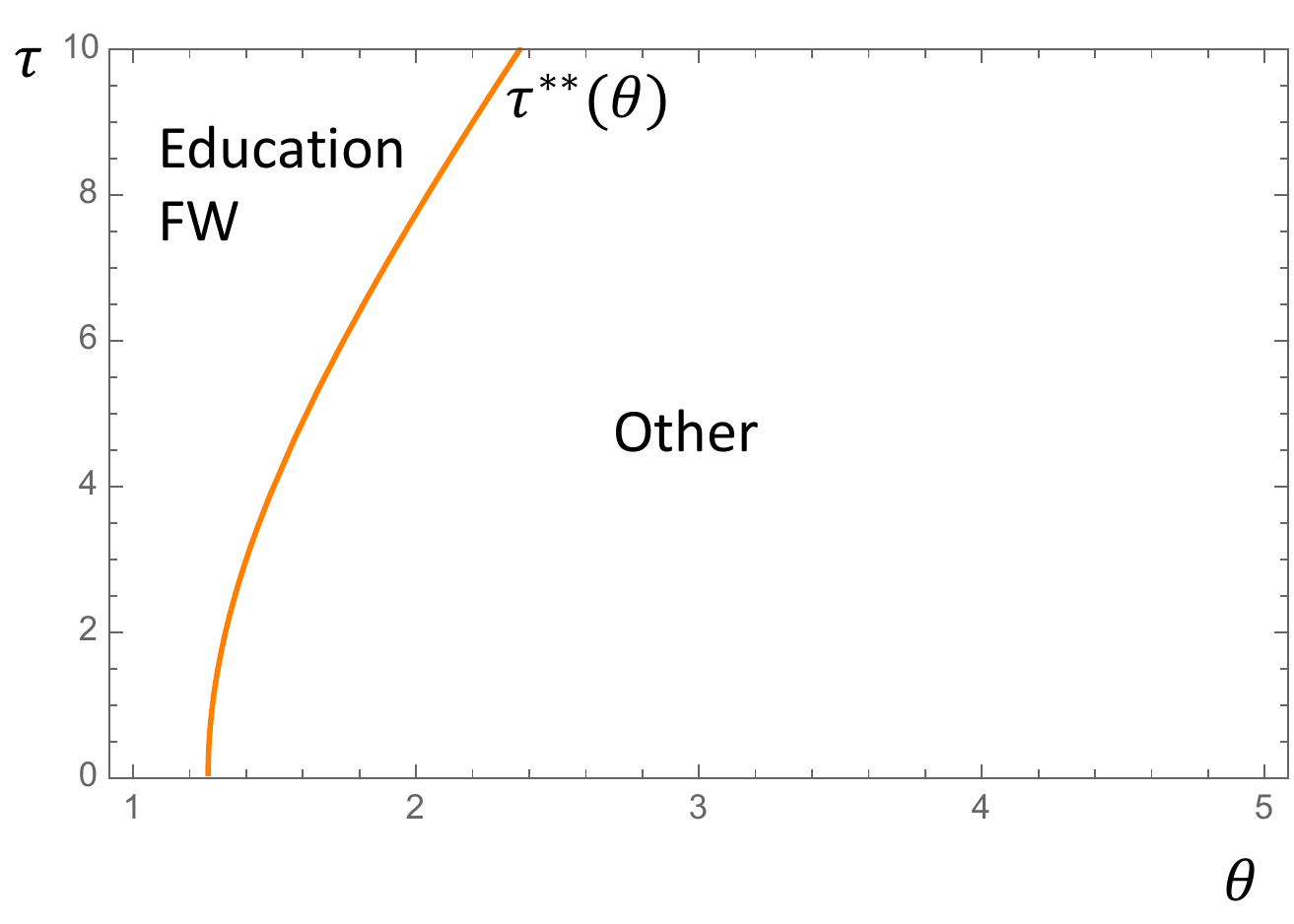}
\end{figure}

\paragraph{Supply-side selection.} The individual applies for a teaching post advertised under P4P if, given her $(\tau,\theta)$ type, she expects to receive a higher payoff teaching in a school on the P4P contract than working in the outside sector. We denote the set of such $(\tau,\theta)$ types by $\mathcal{T}^P$. Similarly, the individual applies for a teaching post advertised under FW if, given her $(\tau,\theta)$ type, she expects to receive a higher payoff teaching in a school on the FW contract than working in the outside sector. We denote the set of such $(\tau,\theta)$ types by $\mathcal{T}^F$. Figure \ref{f:selectionexample} illustrates these sets for the numerical example. Note that the function $\tau^*(\theta)$ traces out motivational types who, given their ability, are just indifferent between applying to the education sector under advertised P4P and applying to the outside sector, i.e.:
    \begin{equation*}
        \Pr\left[\theta e^P + \e > \overline{m}\right]\, B + w^G - (e^P)^2 + \tau^* e^P =  \Pr\left[\theta e^O + \e > \underline{m}\right]\, w^O - (e^O)^2.
    \end{equation*}
Similarly, the function $\tau^{**}(\theta)$ traces out motivational types who, given their ability, are just indifferent between applying to the education sector under advertised FW and applying to the outside sector, i.e.:
    \begin{equation*}
        w^F - (e^F)^2 + \tau^{**} = \Pr\left[\theta e^O + \e > \underline{m}\right]\cdot w^O - (e^O)^2.
    \end{equation*}
In the numerical example, we see a case of positive selection on intrinsic motivation and negative selection on ability under both the FW and P4P treatments. But there is \emph{less} negative selection on ability under P4P than under FW.

\subsection*{Empirical implications}

We used this theoretical framework when writing our pre-analysis plan to clarify what hypotheses to test. We summarize this process for Hypotheses I and VI below.

\paragraph{Hypothesis I: Advertised P4P induces differential application qualities.} Define $1_{\left\{(\tau,\theta)\in \mathcal{T}^F\right\}}$  and $1_{\left\{(\tau,\theta)\in \mathcal{T}^P\right\}}$ as indicator functions for the application event in the advertised FW and P4P treatments respectively. The difference in expected intrinsic motivation and expected ability across the two advertised treatments, can be written as:
\begin{equation*}
\mathrm{E}\left[\tau \cdot 1_{\left\{(\tau,\theta)\in \mathcal{T}^F\right\}}\right] - \mathrm{E}\left[\tau \cdot 1_{\left\{(\tau,\theta)\in \mathcal{T}^P\right\}}\right]
\end{equation*}
and
\begin{equation*}
\mathrm{E}\left[\theta \cdot 1_{\left\{(\tau,\theta)\in \mathcal{T}^F\right\}}\right] - \mathrm{E}\left[\theta \cdot 1_{\left\{(\tau,\theta)\in \mathcal{T}^P\right\}}\right].
\end{equation*}
In the numerical example, both differences are negative: expected intrinsic motivation and expected ability are higher in the P4P treatment than in the FW treatment.

\paragraph{Hypothesis VI: Selection and incentive effects are apparent in the composite 4P performance metric.} 
We start with the selection effect. Maintaining the assumption of no demand-side selection treatment effects, and using the decomposition in \citetApp{LeavLemSCur}, we can write the difference in expected performance across sub-groups $a$ and $b$ (i.e. placed recruits who experienced FW) as:
\begin{equation*}
\mathrm{E}[m^a]-\mathrm{E}[m^b]=\underbrace{\mathrm{E}\left[(\theta \,e^F-\theta \,e^F)\cdot 1_{\left\{(\tau,\theta)\in \mathcal{T}^F\right\}}\right]}_{\text{incentive effect = 0}}+\underbrace{\mathrm{E}\left[\theta \, e^F\cdot\left(1_{\left\{(\tau,\theta)\in \mathcal{T}^F\right\}}-1_{\left\{(\tau,\theta)\in \mathcal{T}^P\right\}}\right)\right]}_{\text{selection effect}}.
\end{equation*}
Similarly, the difference in expected performance across sub-groups c and d (i.e. placed recruits who experienced P4P) can be written as:
\begin{equation*}
\mathrm{E}[m^c]-\mathrm{E}[m^d]=\underbrace{\mathrm{E}\left[(\theta \,e^P-\theta \,e^P)\cdot 1_{\left\{(\tau,\theta)\in \mathcal{T}^F\right\}}\right]}_{\text{incentive effect = 0}}+\underbrace{\mathrm{E}\left[\theta \, e^P\cdot\left(1_{\left\{(\tau,\theta)\in \mathcal{T}^F\right\}}-1_{\left\{(\tau,\theta)\in \mathcal{T}^P\right\}}\right)\right]}_{\text{selection effect}}.
\end{equation*}
In the numerical example, both differences are negative, and the second is larger than the first. 

Turning to the incentive effect, we can write the difference in expected performance across sub-groups a and c (i.e. placed recruits who applied under advertised FW) as:
\begin{equation*}
\mathrm{E}[m^a]-\mathrm{E}[m^c]=\underbrace{\mathrm{E}\left[(\theta \,e^F-\theta \,e^P)\cdot 1_{\left\{(\tau,\theta)\in \mathcal{T}^F\right\}}\right]}_{\text{incentive effect}}+\underbrace{\mathrm{E}\left[\theta \, e^F\cdot\left(1_{\left\{(\tau,\theta)\in \mathcal{T}^F\right\}}-1_{\left\{(\tau,\theta)\in \mathcal{T}^F\right\}}\right)\right]}_{\text{selection effect=0}}.
\end{equation*}
Similarly, the difference in expected performance across sub-groups b and d (i.e. placed recruits who applied under advertised P4P) can be written as:
\begin{equation*}
\mathrm{E}[m^b]-\mathrm{E}[m^d]=\underbrace{\mathrm{E}\left[(\theta \,e^F-\theta \,e^P)\cdot 1_{\left\{(\tau,\theta)\in \mathcal{T}^P\right\}}\right]}_{\text{incentive effect}}+\underbrace{\mathrm{E}\left[\theta \, e^P\cdot\left(1_{\left\{(\tau,\theta)\in \mathcal{T}^P\right\}}-1_{\left\{(\tau,\theta)\in \mathcal{T}^P\right\}}\right)\right]}_{\text{selection effect=0}}.
\end{equation*}
In the numerical example, both differences are negative, and the second is larger than the first. Hypothesis IV and V focus on one component of the performance metric---student performance---and follow from the above. 

%--------------------------------------------------------------------------%
% HIRING PATTERNS
%--------------------------------------------------------------------------%
\cleardoublepage
\section{Applications}\label{s:hiring}
\setcounter{page}{1}

Here, we report results from secondary tests of Hypothesis I: advertised P4P induces differential application qualities, and also provide a robustness check of our assumption that district-by-subject-family labour markets are distinct.

\subsection*{Secondary tests} \label{ss:labor supply}

Our pre-analysis plan included a small number of secondary tests of Hypothesis I (see Table \ref{t:specsummary}). Three of these tests use estimates from TTC score regressions of the form
\begin{equation}\label{eq:ttest_applicants}
y_{iqd} = \tau_A T^A_{qd} + \gamma_q + \delta_d + e_{iqd},\quad \mathrm{with \, weights \, } w_{iqd},
\end{equation}
where $y_{iqd}$ denotes the TTC exam score of applicant teacher $i$ with qualification $q$ in district $d$ and treatment $T^A_{qd}$ denotes the contractual condition under which a candidate applied. The weighted regression parameter $\tau_A$ estimates the difference in (weighted) mean applicant skill induced by advertised P4P. The fourth test is for a difference in the number of applicants by treatment status, conditional on district and subject-family indicators. Here, we use a specification of the form
\begin{equation}\label{eq:n_applicants}
     \log N_{qd} = \tau_A  T^A_{qd} + \gamma_q + \delta_d + e_{qd},
\end{equation}
 where $q$ indexes subject families and $d$ indexes districts; $N_{qd}$ measures the number of qualified applicants in each district.\footnote{`Qualified' here means that the applicant has a TTC degree. In addition to being a useful filter for policy-relevant applications, since only qualified applicants can be hired, in some districts' administrative data this is also necessary in order to determine the subject-family under which an individual has applied.} Although our pre-analysis plan proposes a fifth test---a KS test of equation \eqref{eq:KSstat} using district exam scores---we did not do this because our sample of these scores was incomplete.   
 
To undertake inference about these differences in means, we use randomization inference, sampling repeatedly from the set of potential (advertised) treatment assignments $\mathcal{T^A}$.  Following \citetApp{ChuRom13annstat}, we studentize this parameter by dividing it by its  (cluster-robust, clustered at the district-subject level) standard error to control the asymptotic rejection probability against the null hypothesis of equality of means. These are two-sided tests.\footnote{We calculated $p$-values for two-sided tests as provided in \citetApp{Ros10book} and in the `Standard Operating Procedures' of Donald Green's Lab at Columbia \citepApp{LinGreCop16sop}.}  The absolute value of the resulting test statistic, $|t_A|$, is compared to its randomization distribution in order to provide a test of the hypothesis that $\tau_A=0$.   

Results are in Table \ref{t:AppDistributions}. The first column restates the confidence interval and $p$-value from the KS test for comparison purposes. The second column reports results for the TTC score regression where all observations are weighted equally (i.e. a random hiring rule, as assumed in the theory). Our estimate of $\tau_A$ is  $-0.001$. The randomization inference $p$-value is $0.984$, indicating that we cannot reject the sharp null of no impact of advertised P4P. The third column reports results for the TTC score regression with weights $w_{iqd}=\hat{p}_{iqd}$, where $\hat{p}_{iqd}$ is the estimated probability of being hired as a function of district and subject indicators, as well as a fifth-order polynomial in TTC exam scores, estimated using FW applicant pools only (i.e. the status quo mapping from TTC scores to hiring probabilities). The fourth column reports results for the TTC score regression with weights $w_{iqd}=1$ for the top $\hat{H}$ teachers in their application pool, and zero otherwise (i.e. a meritocratic hiring rule based on TTC scores alone). Here, we test for impacts on the average ability of the top $\hat{H}$ applicants, where $\hat{H}$ is the predicted number hired in that district and subject based on outcomes in advertised FW district-subjects. Neither set of weights changes the conclusion from the second column: we cannot reject the sharp null of no impact of advertised P4P. 
The final column reports results for the (logged) application volume regression. Our estimate of $\tau_A$ is $-0.040$. The randomization inference $p$-value of $0.811$, indicating that we cannot reject the sharp null of no impact of advertised P4P on application volumes.

\begin{table}
\begin{footnotesize}
\begin{center}
\caption{Secondary tests of impacts on teacher ability in application pool}
\label{t:AppDistributions}

\begin{tabular}{p{0.1\linewidth}*{5}{C{0.15\linewidth}}}
\toprule
    & KS & Unweighted & Empirical weights & Top & Number of Applicants \\ 
\midrule
\multirow{3}{0.9\linewidth}{Advertised P4P}    &  n.a.  &  -0.001  &  -0.001  &  -0.009  &  -0.040 \\ 
  &  [-0.020, 0.020]  &  [-0.040, 0.036]  &  [-0.038, 0.032]  &  [-0.025, 0.008]  &  [-0.306, 0.292] \\ 
  &  (0.909)  &  (0.984)  &  (0.948)  &  (0.331)  &  (0.811)  \\[2ex] 
Observations  &    1715  &    1715  &    1715  &    1715  &      18  \\  
\bottomrule
\end{tabular}

\end{center}

\floatfoot{
\emph{Note}: The first column shows the confidence interval in brackets, and the $p$-value in parentheses, from the primary KS test discussed in Section \ref{ss:composition_fx}. The second column reports the (unweighted OLS) point estimate of $\tau_A$ from the applicant TTC exam score specification in \eqref{eq:ttest_applicants}. The third and fourth columns report the point estimate of $\tau_A$ from the same specification with the stated weights. The fifth column reports the point estimate of $\tau_A$ from the number of applicants per labor market specification in \eqref{eq:n_applicants}, with the outcome $N_{qd}$ in logs.
}
\end{footnotesize}

\end{table}

\subsection*{Robustness}\label{ss:CrossApps}

To illustrate the implications of cross-district applications, consider an individual living in, say, Ngoma with the TTC qualification of TSS. On the assumption that this individual is willing to travel only to the neighbouring district of Rwamagana, she could be impacted by the contractual offer of P4P in her home `Ngoma-TSS' market and/or the contractual offer of P4P in the adjacent `Rwamagana-TSS' market. That is, she might apply in both markets, or in Rwamagana instead of Ngoma---what we term \emph{a cross-district labor-supply effect}. The former behavior would simply make it harder to detect a selection effect at the application stage (although not at the placement stage since only one job can be accepted). But the latter cross-district labor-supply effect would be more worrying. We would not find a selection effect where none existed---without a direct effect of advertised P4P on a given market, there cannot be cross-district effects by this posited mechanism---but we might overstate the magnitude of any selection effect.  

Our random assignment provides us with an opportunity to test for the presence cross-district labor-supply effects. To do so, we construct an \emph{adjacency matrix}, defining two labor markets as adjacent if they share a physical border and the same TTC subject-family qualification. We then construct a count of the number of adjacent markets that are assigned to Advertised P4P, and an analogous count for `mixed' treatment status. Conditional on the number of adjacent markets, this measure of the local saturation of P4P is randomly determined by the experimental assignment of districts to advertised contractual conditions.  A regression of labor-market outcomes in a given district on both its own advertised contractual status (direct effect) and this measure of local saturation, conditional on the number of neighboring labor markets, provides an estimate of cross-district labor-supply effects and, by randomization inference, a test for their presence.  

\begin{table}[!hbtp]
\caption{Cross-district effects in teacher labor market outcomes}
\label{t:CrossFx}
\begin{footnotesize}
\begin{center}
\begin{tabular}{p{0.3\linewidth}*{2}{C{0.25\linewidth}}}
\toprule
    & TTC scores & Number of applicants \\ 
\midrule
\multirow{2}{0.9\linewidth}{Advertised P4P}    &  0.032  &  -0.085 \\ 
  &  [-0.050, 0.103]  &  [-0.469, 0.972] \\ 
  &  (0.297)  &  (0.900)  \\[1ex] 
\multirow{2}{0.9\linewidth}{Adjacent P4P markets}    &  0.027  &  -0.047 \\ 
  &  [-0.022, 0.087]  &  [-0.833, 0.573] \\ 
  &  (0.115)  &  (0.710)  \\[1ex] 
Observations  &    1715  &      18  \\  
\bottomrule
\end{tabular}

\end{center}

\floatfoot{
\emph{Note}: The table shows point estimates for the direct and local saturation effects of P4P contracts, with confidence intervals in brackets and randomization inference $p$-values in parentheses. In the first column, the unit of analysis is the application and the outcome is the TTC score of the applicant. In the second column, the unit of analysis is the labor market and the outcome is the number of applications, in logs.  All specifications control for the total number of adjacent markets.
}
\end{footnotesize}
\end{table}

Table \ref{t:CrossFx} shows results of this analysis for two key labor-market outcomes---applicant TTC scores analyzed at the application level, and the number of applications per labor-market analyzed at the labor-market level.  The direct effects of advertised P4P on each of these outcomes are presented for comparison and remain qualitatively unchanged relative to the estimates in Table \ref{t:AppDistributions}, which did not allow for saturation effects.  Estimated saturation effects of neighboring P4P markets are modest in estimated size and statistically insignificant for both outcomes. This suggests that saturation effects were of limited consequence in our setting.   

%--------------------------------------------------------------------------%
% TEACHER VALUE ADDED
%--------------------------------------------------------------------------%
\cleardoublepage 
\section{Test-score constructs}
\label{s:tva}
\setcounter{page}{1}

\subsection*{Barlevy-Neal metric}

At the core of our teacher evaluation metric is a measure of the learning gains that teachers bring about, measured by their students' performance on assessments.  (See Section \ref{s:data} for a description of assessment procedures; throughout, we use students' IRT-based predicted abilities to capture their learning outcomes in a given subject and round.)  To address concerns over dysfunctional strategic behavior, our objective was to follow Barlevy and Neal's  \emph{pay-for-percentile} scheme as closely as was practically possible \citepApp[henceforth BN]{BarNea12aer}.

The logic behind the BN scheme is that it creates a series of `seeded tournaments' that incentivize teachers to promote learning gains at all points in the student performance distribution. In short, a teacher expects to be rewarded equally for enabling a weak student to outperform his/her comparable peers as for enabling a strong student to outperform his/her comparable peers. Roughly speaking, the implemented BN scheme works as follows. Test all students in the district in each subject at the start of the year. Take student $i$ in stream $k$ for subject $b$ at grade $g$ and find that student's percentile rank in the district-wide distribution of performance in that subject and grade at baseline. Call that percentile (or interval of percentiles if data is sparse) student $i$'s baseline bin.\footnote{In setting such as ours where the number of students is modest, there is a tradeoff in determining how wide to make the percentile bins. As these become very narrowly defined, they contain few students, and the potential for measurement error to add noise to the results increases. But larger bins make it harder for teachers to demonstrate learning gains in cases where their students start at the bottom of a bin.  In practice, we use vigintiles of the district-subject distribution.} Re-test all students in each subject at the end of the year. Establish student $i$'s end-of-year percentile rank within the comparison set defined by his/her baseline bin. This metric constitutes student $i$'s contribution to the performance score of the teacher who taught that subject-stream-grade that school year. Repeat for all students in all subjects-streams-grades taught by that teacher in that school year, and take the average to give the BN performance metric at teacher level.

We adapt the student test score component of the BN scheme to allow for the fact that we observe only a sample of students in each round in each school-subject-stream-grade. (This was done for budgetary reasons and is a plausible feature of the cost-effective implementation of such a scheme at scale, in an environment in which centrally administered standardized tests are not otherwise taken by all students in all subjects.) To avoid gaming behavior---and in particular, the risk that teachers would distort effort toward those students sampled at baseline---we re-sampled (most) students across rounds, and informed teachers in advance that we would do so.

Specifically, we construct \emph{pseudo-baseline bins} as follows. Students sampled for testing at the end of the year are allocated to district-wide comparison bins using empirical CDFs of start of year performance (of different students). To illustrate, suppose there are 20 baseline bins within a district, and that the best baseline student in a given school-stream-subject-grade is in the (top) bin 20. Then the best endline student in the same school-stream-subject-grade will be assigned to bin 20, and will be compared against all other endline students within the district who have also been placed in bin 20. 

To guard against the possibility that schools might selectively withhold particular students selected from the exam, all test takers were drawn from beginning-of-year administrative registers of students in each round.  Any student who did not take the test was assigned the minimum theoretically possible score.  This feature of our design parallels similar incentives to mitigate incentives for selective test-taking in \citetApp{GleIliKre10aejapplied}.

Denote by $z_{ibkgdr}$ the IRT estimate of the ability of student $i$ in subject $b$, stream $k$, grade $g$, district $d$, and round $r$.  We can outline the resulting algorithm for producing the student learning component of the assessment score for rounds $r\in\{1,2\}$ in the following steps:
\begin{enumerate}
    \item \emph{Create baseline bins}. 
        \begin{itemize}
            \item Separately for each subject and grade, form a within-district ranking of the students sampled at round $r-1$ on the basis of $z_{ibkgd,r-1}$.  Use this ranking to place these round $r-1$ students into $B$ baseline bins.
            \item For each subject-grade-school-stream within a school, calculate the empirical CDF of these baseline bins.\footnote{There are 40 subject-grade-school streams (out of a total of 4,175) for which no baseline students were sampled. In such cases, we use the average of the CDFs for the same subject in other streams of the same school and grade (if available) or in the school as a whole to impute baseline learning distributions for performance award purposes.}  
        \end{itemize}

    \item \emph{Place end-of-year students into pseudo-baseline bins}.
        \begin{itemize}
            \item Form a within subject-stream-grade-school percentile ranking of the students sampled at round $r$ on the basis of $z_{ibkgdr}$. In practice, numbers of sampled students varies for a given stream between baseline and endline, so we use percentile ranks rather than simple counts.  Assign the lowest possible learning level to students who were sampled to take the test but did not do so. 
            \item Map percentile-ranked students at endline onto baseline bins through the empirical CDF of baseline bins. For example, if there are 20 bins and the best round 1 student in that subject-stream-grade-school was in the top bin, then the best round 2 student in that subject-stream-grade-school will be placed in pseudo-bin 20. 
        \end{itemize}
    
    \item \emph{BN performance metric at student-subject level}. Separately for each subject, grade, and district, form a within-psuedo-baseline bin ranking of the students sampled at round $r$ on the basis of $z_{ibgdr}$. This is the BN performance metric at student-subject level, which we denote by $\pi_{ibkgdr}$. It constitutes student $i$'s contribution to the performance score of the teacher who taught subject $b$ stream $k$ at grade $g$ for school year $r$. 
    
    \item \emph{BN performance metric at teacher-level}. For each teacher, compute the weighted average of the $\pi_{ibkgt}$ for all the students in the subject-stream-grades that they taught in round $r$ school year. This is the BN performance metric at teacher-level. Weights $w_{ik}$ are given by the (inverse of the) probability that student $i$ was sampled in stream $k$: the number of sampled students in that stream divided by the number of students enrolled in the same stream.  Note these weights are determined by the number of students \emph{sampled} for the test, \emph{not} the number of students who actually took the test (which may be smaller), since our implementation of the BN metric includes, with the penalty described above, students who were sampled for but did not sit the test.\footnote{Our endline sampling frame covered all grades, streams, and subjects.  In practice, out of 4,200 school-grade-stream-subjects in the P4P schools, we have data for a sample of students in all but five of these, which were missed in the examination.}
\end{enumerate}

To construct the BN performance metric at teacher-level for the second performance round, $r=2$, we must deal with a further wrinkle, namely the fact that we did not sample students at the start of the year. We follow the same procedure as above except that at Step 2 we use  the set of students who were sampled for and actually sat the round 1 endline exam, and can be linked to an enrollment status in a specific stream round 2, to create the baseline bins and CDFs for that year.

\subsection*{Teacher value added}

This section briefly summarizes how we construct the measure of teacher value added for the placed recruits, referred to at the end of Section \ref{ss:composition_fx}. 

We adapt the approach taken in prior literature, most notably \citetApp{KanSta08nber} and \citetApp{BauDas20AEJEP}. Denoting as in equations \eqref{eq:assessments_pooled} and \eqref{eq:assessments_interacted} the learning outcomes of student $i$ in subject $b$, stream $k$ of grade $g$, taught by teacher $j$ in school $s$ and round $r$ by $z_{ibgjsr}$, we express the data-generating process as:
\begin{equation}\label{eq:tva_dgp}
z_{ibgjsr} = \rho_{bgr} \bar{z}_{ks,{r-1}} + \mu_{bgr} + \lambda_s + \theta_j + \eta_{jr} + \e_{ibgjr},
\end{equation}   
This adapts a standard TVA framework to use the full pseudo-panel of student learning measures.  Our sampling strategy implies that most students are not observed in consecutive assessments, as discussed in Section \ref{ss:data_assessments}.  We proxy for students' baseline abilities using the vector of means of lagged learning outcomes in all subjects, $\bar{z}_{ks,r-1}$, where the parameter $\rho_{bgr}$ allows these lagged mean outcomes to have distinct own- and cross-subject associations with subsequent learning for all subjects, grades, and rounds.  In a manner similar to including means instead of fixed effects \citepApp{Mun78ecta,Cha82jectx}, these baseline peer means block any association between teacher ability (value added) and the baseline learning status of sampled students.

In equation \eqref{eq:tva_dgp}, the parameter $\theta_j$ is the time-invariant effect of teacher $j$: her value added. We allow for fixed effects by subject-grade-rounds, $\mu_{bgr}$, and schools $\lambda_s$, estimating these within the model. We then form empirical Bayes estimates of TVA as follows.
\begin{enumerate}
\item Estimate the variance of the TVA, teacher-year, and student-level errors, $\theta_j, \eta_{jr}, \e_{ibgjr}$ respectively, from equation \eqref{eq:tva_dgp}.  Defining the sum of these errors as $v_{ibgjr}=\theta_j + \eta_{jr} + \e_{ibr}$: the last variance term can be directly estimated by the variance of student test scores around their teacher-year means: $\hat\sigma^2_\e = \Var(v_{ibgjr} - \bar{v}_{jr})$; the variance of TVA can be estimated from the covariance in teacher mean outcomes across years: $\hat\sigma^2_\theta = \Cov(\bar{v}_{jr},\bar{v}_{j,r-1})$, where this covariance calculation is weighted by the number of students taught by each teacher; and the variance of teacher-year shocks can be estimated as the residual, $\hat\sigma^2_{\eta}=\Var(v_{ibgjr}) -\hat\sigma_\theta^2 - \hat\sigma_\e^2 $.
\item Form a weighted average of teacher-year residuals $\bar{v}_{jr}$ for each teacher.  
\item Construct the empirical Bayes estimate of each teacher's value added by multiplying this weighted average of classroom residuals, $\bar{v}_j$, by an estimate of its reliability:
	\begin{equation}\label{eq:tva_estimate}
	\widehat{VA}_j = \bar{v}_j \left( \frac{\hat\sigma_\theta^2}{\Var(\bar{v}_j)} \right)
	\end{equation}
where $\Var(\bar{v}_j) = \hat\sigma_\theta^2+ \left(\sum_r h_{jr} \right)^{-1}$, with $h_{jr} = \Var(\bar{v}_{jr}|\theta_j)^{-1} = \left(\hat\sigma^2_\eta + \frac{\hat\sigma_\e^2}{n_{jr}}  \right)^{-1}$.  
\end{enumerate}

Following this procedure, we obtain a distribution of (empirical Bayes estimates of) teacher value added for placed recruits who applied under advertised FW. The Round 2 point estimate from the student learning model in Equation \eqref{eq:assessments_pooled} would raise a teacher from the 50th to above the 76th percentile in this distribution. Figure \ref{f:TVA_Recruits_byTx} plots the distributions of (empirical Bayes estimates of) $\theta_j + \eta_{jr}$ separately for $r=1,2$, and for recruits applying under advertised FW and advertised P4P.

It is of interest to know whether the measures of teacher ability and intrinsic motivation that we use in Section \ref{ss:composition_fx} are predictive of TVA. This is undertaken in Table \ref{t:predict_tva}, where  TVA is the estimate obtained pooling across rounds and treatments.\footnote{We obtain qualitatively similar results for the FW sub-sample, where TVA cannot be impacted by treatment with P4P.} Interestingly, the measure of teacher ability that we observe among recruits at baseline, Grading Task IRT score, \emph{is} positively correlated with TVA (rank correlation of $0.132$, with a $p$-value of $0.039$). It is also correlated with TTC final exam score (rank correlation of $0.150$, with a $p$-value of $0.029$). However, neither the measure that districts have access to at the time of hiring, TTC final exam score, nor the measure of intrinsic motivation that we observe among recruits at baseline, DG share sent, is correlated with TVA.

\begin{table}[!htbp]
    \begin{footnotesize}
    \begin{center}
    \caption{Rank correlation between TVA estimates, TTC scores, Grading Task IRT scores, and Dictator Game behavior among new recruits}
    \label{t:predict_tva}
    \begin{tabular}{l c c c} \\ 
\toprule
 & TVA  & TTC score  & Grading task \\
\midrule 
 TTC score  & -0.087 &    . &    .\\
         & (0.178) \\[2ex]
 Grading task  & 0.132 & 0.150 &    .\\
         & (0.039)  & (0.029) \\[2ex]
 DG share sent  & -0.078 & 0.062 & -0.047\\
         & (0.203)  & (0.349)  & (0.468) \\[2ex]
\bottomrule 
\end{tabular} 

    \end{center}

    \floatfoot{
    \emph{Note}:  The table provides rank correlations and associated $p$-values (in parentheses) for relationships between recruits' teacher value added and various measures of skill and motivation: TTC final exam scores, baseline Grading Task IRT scores, and baseline Dictator Game share sent. We obtain the empirical Bayes estimate of TVA from $\theta_j$ estimated in the school fixed-effects model in equation \eqref{eq:tva_dgp}.
    }
    \end{footnotesize}
\end{table}

%--------------------------------------------------------------------------%
% COMMUNICATIONS TO APPLICANTS AND TEACHERS
%--------------------------------------------------------------------------%

\cleardoublepage 
\section{Communication about the intervention}\label{s:communication}
\setcounter{page}{1}

\subsection*{Promotion to potential applicants}\label{ss:promotion}

\begin{singlespace}

The subsections below give details of the (translated) promotional materials that were used in November and December 2015.

\subsubsection*{Leaflets and posters in district offices}

\begin{figure}[!hbtp]
\centering
\caption{Leaflet advertising treatments\label{f:leaflet}}
\includegraphics[width=0.88\textwidth]{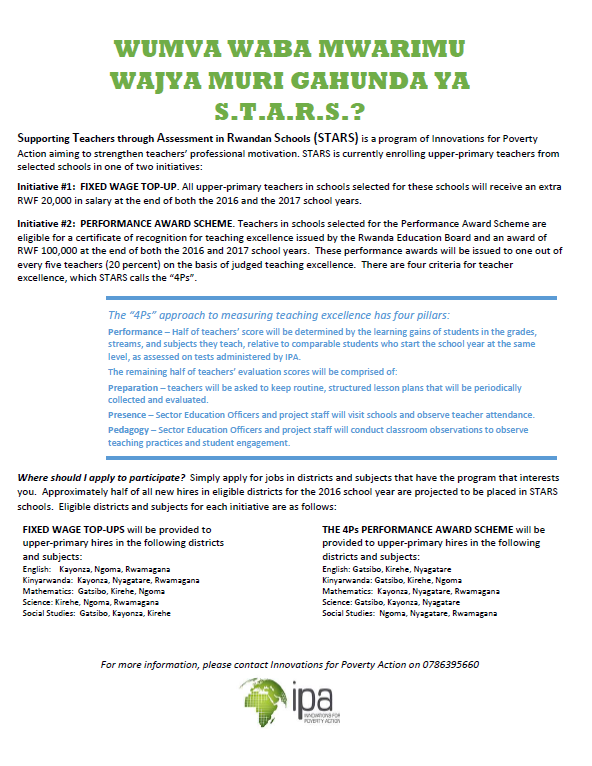}
\end{figure}

\begin{figure}[!hbtp]
\centering
\caption{Poster explaining the programme\label{f:posterREB}}
\includegraphics[width=0.40\textwidth]{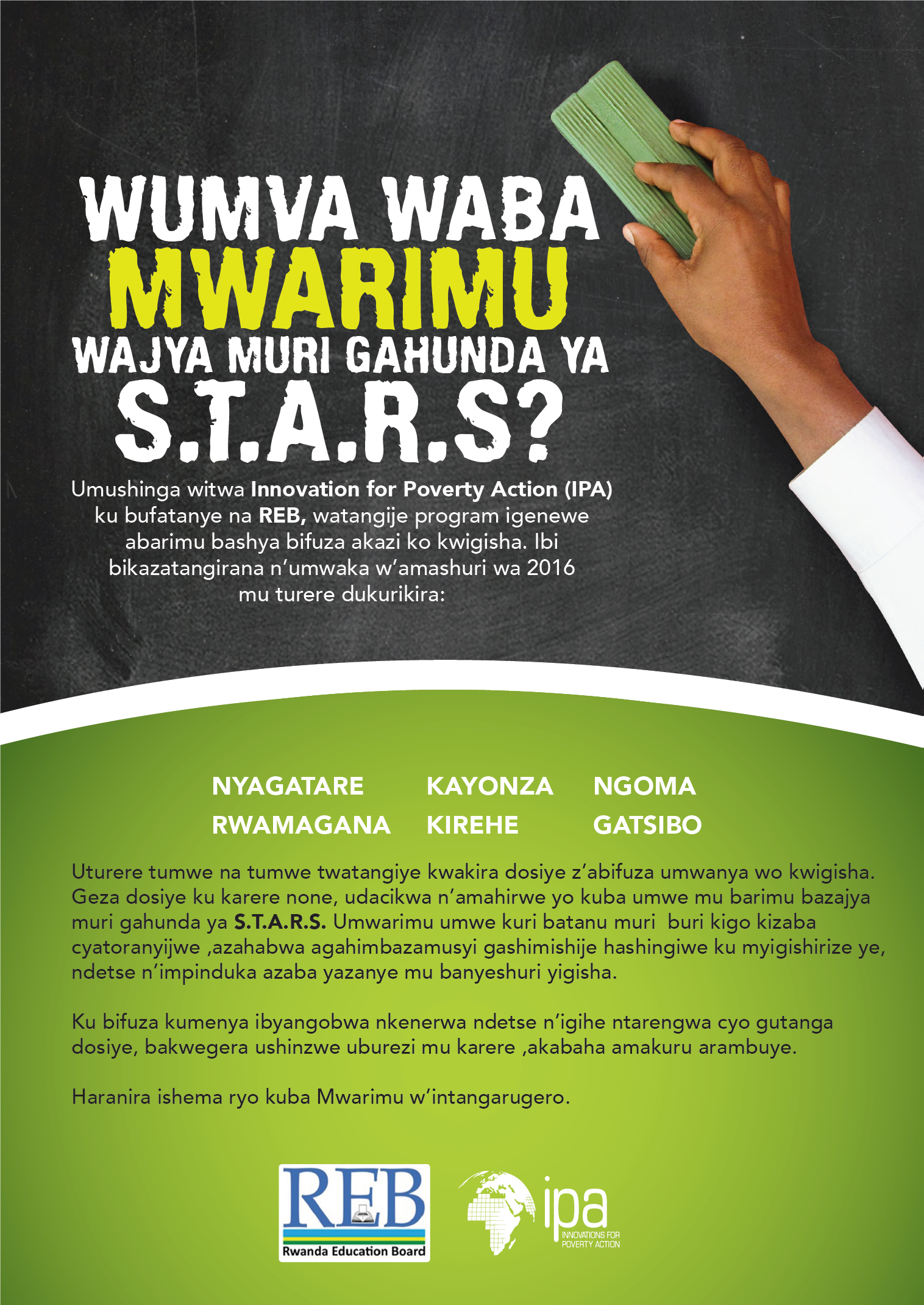}
\end{figure}

A help desk was set up in every District Education Office.  Staffers explained the advertised contracts to individuals interested in applying, and distributed the leaflet shown in Figure \ref{f:leaflet}, and stickers. Permanent posters, like the example shown in Figure \ref{f:posterREB} further summarised the programme. Staffers kept records of the number of visitors and most frequent questions, and reported back to head office.\footnote{The respective number of visitors were: Gatsibo 305, Kayonza 241, Kirehe 411, Ngoma 320, Nyagatara 350, and Rwamagama 447.}  

\subsubsection*{Radio Ads} 

Radio ads were broadcast on Radio Rwanda, the national public broadcaster, during November/December 2015 to promote awareness of the intervention. The scripts below were developed in partnership with a local advertising agency.

\paragraph{Radio script 1}

\emph{SFX: Noise of busy environment like a trading centre}

\begin{itemize}
    \item[] FVO: Hey, Have you seen how good Gasasira's children look? [\emph{This is a cultural reference implying that teachers are smart, respected individuals and nothing literal about how the child looks.}]
    \item[] MVO: Yeah! That's not surprising though, their parents are teachers.
    \item[] FVO: Hahahahah...[\emph{Sarcastic laugh as if to say, what is so great about that.}]
    \item[] MVO: Don’t laugh...haven’t you heard about the new programme in the district to recognize and reward good teachers? I wouldn’t be surprised if Gasasira was amongst those that have been recognised.
    \item[] ANNOUNCER: Innovations for Poverty Action in collaboration with REB and MINEDUC, is running the STARS program in the districts Kayonza, Ngoma, Rwamagama, Kirehe, Gatsibo, and Nyagatare for the 2016 academic year. Some \emph{new} teachers applying to these districts will be eligible for STARS which rewards the hardest working, most prepared and best performing teachers. Eligible districts are still being finalized---keep an eye out for further announcements!
\end{itemize}

\paragraph{Radio script 2} \emph{SFX: Sound of a street with traffic and cars hooting}

\begin{itemize}
    \item[] VO1: Mari, hey Mariko!....What's the rush, is everything OK?
    \item[] VO2: Oh yes, everything is fine. I am rushing to apply for a job and don't want to find all the places taken.
    \item[] VO1: Oh that's good. And you studied to be a teacher right?
    \item[] VO2: Exactly! Now I am going to submit my papers at the District Office and hope I get lucky on this new programme that will be recognizing good teachers!
    \item[] ANNOUNCER: Innovations for Poverty Action in collaboration with REB and MINEDUC, is running the STARS program in the districts Kayonza, Ngoma, Rwamagama, Kirehe, Gatsibo, and Nyagatare for the 2016 academic year. Some \emph{new} teachers applying to these districts will be eligible for STARS which rewards the hardest working, most prepared and best performing teachers. Eligible districts are still being finalized---keep an eye out for further announcements!
\end{itemize}

\paragraph{Radio script 3} \emph{SFX: Calm peaceful environment}

\begin{itemize}
    \item[] VO1: Yes honestly, Kalisa is a very good teacher!
    \item[] VO2: You are right, ever since he started teaching my son, the boy now understands maths!
    \item[] VO1: Yes and because of him other parents want to take their children to his school.
    \item[] VO2: Aaah!...That must be why he was selected for the programme that rewards good teachers.
    \item[] VO1: He definitely deserves it, he is an excellent teacher.
    \item[] ANNOUNCER: Innovations for Poverty Action in collaboration with REB and MINEDUC, is running the STARS program in the districts Kayonza, Ngoma, Rwamagama, Kirehe, Gatsibo, and Nyagatare for the 2016 academic year. Some \emph{new} teachers applying to these districts will be eligible for STARS which rewards the hardest working, most prepared and best performing teachers. Eligible districts are still being finalized---keep an eye out for further announcements!
\end{itemize}

\subsection*{Briefing in P4P schools}\label{ss:training} 
The subsections below provide extracts of the (translated) script that was used during briefing sessions with teachers in P4P schools in April 2016. The main purpose of these sessions was to explain the intervention and maximise understanding of the new contract.   

\subsubsection*{Introduction}
[Facilitator speaks.] You have been selected to participate in a pilot program that Rwanda Education Board (REB) and Innovations for Poverty Action (IPA) are undertaking together on paid incentives and teacher performance. As a participant in this study, you will be eligible to receive a competitive bonus based on your performance in the study. The top 20 percent of teachers in participating schools in your district will receive this bonus. All participants will be considered for this paid bonus. It is important to note that your employment status will not be affected by your participation in this study. It will not affect whether you keep your job, receive a promotion, etc. 

You will be evaluated on \textbf{four different categories}:
\begin{enumerate}
    \item \textbf{Presence}, which we will measure through whether you are present in school on days when we visit;
    \item \textbf{Preparation},  which we will measure through lesson planning;
    \item \textbf{Pedagogy},  which we will measure through teacher observation; and
    \item \textbf{Performance}, which we will measure through student learning assessments. You will receive additional information on each of these categories throughout this training. 
\end{enumerate}	
	
In your evaluation, the first three categories (presence, preparation, and pedagogy) will contribute equally to your ‘inputs’ score.  This will be averaged with your ‘performance’ score (based on student learning assessments) which will therefore be worth half of your overall score. [Teachers are then provided with a visual aid.]
  
The SEO will now tell you how we are going to measure each of these components of your performance. Before I do so, are there any questions?

\subsubsection*{Presence: Teacher attendance score}

[SEO now speaks.] I will now explain to you the first component of your performance score:  Teacher Presence.  
During this pilot program, I will visit your school approximately one time per term.  Sometimes I will come twice or more; you will not know in advance how many times I  plan to visit in any term. These visits to your school will be unannounced.  Neither your Head Teacher nor you will know in advance when I  plan to visit your school.
I will arrive approximately at the start of the school day. Teachers who are present at that time will be marked ‘present’; those that are not will be marked ‘late’ or ‘absent’.  The type of absence will be recorded.  Teachers who have excused reasons for not being present in school will be marked ‘excused’.  These reasons include paid leaves of absence, official trainings, and sick leaves that have been granted in advance by the Head Teacher.  If you are not present because you feel unwell but have not received advance permission from the Head Teacher, you will be marked as absent.  

It is in your best interest to be present every day, or in the case of emergency, notify the head teacher of your absence with an appropriate excuse before the beginning of classes. I  will also record what time you arrive to school. You will be marked for arriving on time and arriving late to work. It is in your best interest to arrive on time to school every day. 

\subsubsection*{Preparation: Lesson planning score}

Later in this session, you’ll be shown how to use a lesson planning form. Lesson planning is a tool to help you improve both your organization and teaching skills. The lesson planning form will help you to include the following components into your lesson:
\begin{itemize}
    \item A clear lesson objective to guide the lesson.
    \item Purposeful teaching activities that help students learn the skill.
    \item Strong assessment opportunities or exercise to assess students’ understanding of the skill.
\end{itemize}

This lesson planning form consists of three categories: lesson objective, teaching activities, assessment/exercises. You will be evaluated on these three categories. I  will not evaluate your lesson plans. Instead, I  will collect your lesson planning forms at the end of the study. An IPA education specialist will review your lesson plans and score them. They will compare your lesson plans to other teachers' plans in the district. Please be aware that these lesson plans will only be used for this study and will not be reviewed by any MINEDUC officials. They will use the following scoring scale, with 0 being the lowest score and 3 being the highest score. [Teachers are then provided with a visual aid.]
 
You will be responsible for filling out the lesson planning form to be eligible for the paid bonus. You will fill out a lesson plan for each day and each subject you teach. You will fill out the lesson planning form in addition to your MINEDUC lesson journal. Later in this session, you will have a chance to practice using the lesson planning form. You will also see examples of strong and weak lesson plans to help you understand our expectations. 

\subsubsection*{Pedagogy: Teacher observation score}

The third component that will affect your eligibility for the paid bonus is your observation score. I will observe your classrooms during the next few weeks at least once, and again next term. I will score your lesson in comparison to other teachers in your district using a rubric. During the observation, I will record all the activities and teaching strategies you use in your lesson. At the end of your lesson, I  will use my notes to evaluate your performance in the following four categories: 

\begin{itemize}
    \item Lesson objective, does your lesson have a clear objective?;
    \item Teaching activities, does your lesson include activities that will help students learn the lesson?;
    \item Assessment and exercises, does your lesson include exercises for students to practice the skill?; and 
    \item Student engagement, are students engaged during the lesson and activities?
\end{itemize}

I  will use a scoring rubric designed by IPA, Georgetown, and Oxford University to evaluate your performance in each category. You will receive a score from 0 (unsatisfactory) to 3 (exemplary) in each category. I  will observe your entire lesson, from beginning to end. I  will then evaluate your performance based on the observation. You will not know when I am coming to observe your lesson, so it is in your best interest to plan your lessons everyday as if I were coming to observe. After the lesson, I will share your results with the Head Teacher.  You will be able to obtain a copy of your scores, together with an explanation, from the Head Teacher.  

\subsubsection*{Performance: Student test scores}

[Field supervisor now speaks.] Half of your overall evaluation will be determined by the learning achievements of your students. We have devised a system to make sure that all teachers compete on a level playing field. If students in your school are not as well off as students in other schools, you do not have to worry:  we are rewarding teachers for how much their students can improve, not for where they start.

Here is how this works. We randomly selected a sample of your students to take a cumulative test, testing their knowledge of grade level content. These tests were designed based on the curriculum, to allow us to measure the learning of students for each subject separately.  The performance of each teacher will be measured by the learning outcomes of students in the subjects and streams that they themselves teach. (So, if you are a P4 Maths Teacher, your performance will not be affected by students’ scores in P4 English.  And if you teach P4 Math for Stream A but not Stream B, your performance measure will not depend on students’ scores in Stream B.) We will compare the marks for this test with those from other students in the same district, and place each student into one of ten groups, with Group 1 being the best performing, Group 2 being the next-best performing, and so on, down to Group 10.  In the district as a whole, there are equal numbers of students placed in each of these groups, but some of your students may be in the same group, and there may be some groups in which you do not have any students at all.

At the end of this school year, we will return to your school and we will sample 10 new pupils from every stream in Upper Primary school to take a new test.  This will be a random sample.  We do not know in advance who will be drawn, and students who participated in the initial assessment have the same chance of appearing in the end-of-year sample as anyone else. We will draw students for this assessment based on the student enrollment register.  If any student from that register is asked to participate in the test but is no longer enrolled at the school, they will receive a score of zero.  So, you should do your best to encourage students to remain enrolled and to participate in the assessment if asked. Once the new sample has taken the assessment, we will sort them into groups, with the best-performing student from the final assessment being placed into the group that was determined by the best-performing student in the initial assessment.  The second-best student from the final assessment will be placed into the second-highest group achieved from the initial assessment, and so on, until all students have been placed into groups. We will then compare your students’ learning levels with the learning levels of other students in the same group only.  Each of your students will receive a rank, with 1 being the best, 2 being the next, and so on, within their group.  (This means there will be a 1st-ranked student in Group 1, and another student ranked first in Group 2, and so on.)  The measure of your performance that we will use for your score is the average of these within-group ranks of the students whom you teach.  

This all means that you do not have to have the highest performing students in the District in order to be ranked well. It is possible to be evaluated very well even if, for example, all of your students are in Group 10, the lowest-performing group:  what matters is how they perform relative to other students at the same starting point.
I will now demonstrate how this works with some examples.  Please feel free to ask questions as we go along.

\paragraph*{Worked example 1}  

[Field supervisor sets up Student Test Scores Poster and uses the Student Test Scores Figures to explain this example step by step.] Let us see how the learning outcomes score works with a first example. For this example, suppose that we were to sample 5 students from your class in both the beginning-of-year and end-of-year assessments.  (In reality there will be at least ten, but this is to make the explanation easier.) Now, suppose in the initial assessment, we drew 5 students.  And those students’ scores on the assessment might mean that they are placed as follows:

\begin{itemize}
    \item One student in Group 1 (top);
    \item One student in Group 3;
    \item One student in Group 6;
    \item One student in Group 9; and
    \item One student in Group 10. 
\end{itemize}

Then, at the end of the school year, we will return and we will ask 5 new students to sit for a different assessment. These are unlikely to be the same students as before.  
Once they have taken the test, we will rank them, and we will put the best-performing of the new students into Group 1, the next-best-performing of the new students into Group 3, the next-best performing of the new students into Group 6, then Group 9, and Group 10.  So, the Groups into which the new students are placed are determined by the scores of the original students.

Finally, we will compare the actual scores of the new students to the other new students from schools in this district who have been placed into the same groups.  For example:

\begin{itemize}
    \item The new student placed into Group 1 might be ranked 1st within that group;
    \item The new student placed into Group 3 might be ranked 7th within that group;
    \item The new student placed into Group 6 might be ranked 4th within that group;
    \item The new student placed in Group 9 might also be ranked 4th within her group;
    \item The new student placed into Group 10 might be ranked 1st within his group.
\end{itemize}

Then, we add up these ranks to determine your score:  in this case, it is $1+7+4+4+1=17$.  That is pretty good!  Remember, the lower the sum of these ranks, the better.   And notice that even though the student in Group 10 did not have a very high score compared to everyone else in the district, he really helped your performance measure by doing very well within his group.

\paragraph{Worked example 2}  

Now, let us try a second example. Again let us suppose that we were to sample 5 students from your class in both the beginning-of-year and end-of-year assessments.  (Remember:  in reality there will be at least ten, but this is to make the explanation easier.)
Now, suppose in the initial assessment, we drew 5 students.  And those students' scores on the assessment might mean that they are placed as follows:

\begin{itemize}
    \item One student in Group 1 (top);
    \item TWO students in Group 3;
    \item One student in Group 4; and
    \item One student in Group 5. 
\end{itemize}

Notice that it is possible for two or more of your students to be in the same group. Then, at the end of the school year, we will return and we will ask 5 new students to sit for a different assessment.  Again, these are unlikely to be the same students as before. Now, suppose that one out of the five students that we ask for has dropped out of school, or fails to appear for the test.  They will still be counted, but their exam will be scored as if they answered zero questions correctly---the worst possible score. Once they have taken the test, we will rank them, and we will put the best-performing of the new students into Group 1, the \emph{two} next-best-performing of the new students into Group 3, the next-best performing of the new students into Group 4. The student who was not present for the test because they had dropped out of school is placed into Group 5. As in the previous example, notice that the groups into which the new students are placed are determined by the scores of the original students.

Finally, we will compare the actual scores of the new students to the other new students from schools in this district who have been placed into the same groups.  For example:
\begin{itemize}
    \item The new student placed into Group 1 might be ranked 1st within that group;
    \item The new students placed in Group 3 might be ranked 4th \& 7th in that group;
    \item The new student placed into Group 4 might be ranked 8th within that group;
    \item The new student placed in Group 5, who did not actually take the test, will be placed last in his group. If there are 40 students in the group from across the whole district, then this would mean that his rank in that group is 40th.
\end{itemize}
 
Then, we add up these ranks to determine your score:  in this case, it is $1+4+7+8+40=60$.  Notice three points. First, even though in this example, your students did better on the initial assessment than in the first example, this does not mean that you scored better overall.  All groups are counted equally, so that no school or teacher will be disadvantaged in this process. Second, notice that the student who dropped out was ranked worst out of the group to which he was assigned.  Since the lowest-performing student in the initial assessment was in Group 5, the student who had dropped out was compared with other students placed into Group 5. Since he received the worst possible score, he was ranked last (in this case, fortieth) within that group.  This was bad for the teacher’s overall performance rank. Third, teachers will be evaluated based on the same number of students. So even if a teacher would be teaching in several streams, resulting in more students taking the tests, his final score will be based on a random subsample of students, such that all teachers are evaluated on the same number of students. 

\end{singlespace}

%--------------------------------------------------------------------------%
% APPENDIX BIBLIOGRAPHY
%--------------------------------------------------------------------------%
\cleardoublepage
\singlespacing
\thispagestyle{empty}
\bibliographystyleApp{aea}
\bibliographyApp{STARSbib}

\end{document}